\documentclass[aps,pre,citeautoscript,twocolumn,superscriptaddress,nofootinbib,nobibnotes,10pt]{revtex4-1} 
\usepackage[utf8]{inputenc}%latin1
\usepackage{amsmath,amssymb}
\usepackage{mathrsfs} 
\usepackage{graphicx,color}
\usepackage{varwidth,xcolor}
\usepackage{placeins}

\DeclareGraphicsExtensions{.pdf,.PDF,.png,.PNG}

\newcommand{\abs}[1]{\lvert#1\rvert}%
\newcommand{\norm}[1]{\lVert#1\rVert}%
\newcommand{\ZT}[1]{\textquotedblleft#1\textquotedblright}%

\begin{document}

\title{Collective guiding of acoustically propelled nano- and microparticles for medical applications}

\author{Tobias Nitschke}
\affiliation{Institut f\"ur Theoretische Physik, Center for Soft Nanoscience, Westf\"alische Wilhelms-Universit\"at M\"unster, 48149 M\"unster, Germany}

\author{Joakim Stenhammar}
\affiliation{Division of Physical Chemistry, Lund University, SE-221 00 Lund, Sweden}

\author{Raphael Wittkowski}
\email[Corresponding author: ]{raphael.wittkowski@uni-muenster.de}
\affiliation{Institut f\"ur Theoretische Physik, Center for Soft Nanoscience, Westf\"alische Wilhelms-Universit\"at M\"unster, 48149 M\"unster, Germany}

\begin{abstract}
One of the most important potential applications of self-propelled nano- and microdevices is targeted drug delivery. To realize this, biocompatible particles that can be guided collectively towards a target inside a patient's body are required. Acoustically propelled nano- and microparticles constitute a promising candidate for such biocompatible, artificial motile particles. The main remaining obstacle to targeted drug delivery by self-propelled nano- and microdevices is to also achieve a reliable and biocompatible method for guiding them collectively to their target. Here, we propose such a method. As we confirm by computer simulations, it allows for the remote guiding of large numbers of acoustically propelled particles to a prescribed target by combining a space- and time-dependent acoustic field and a time-dependent magnetic field. The method works without detailed knowledge about the particle positions  and for arbitrary initial particle distributions. With these features, it paves the way for the future application of self-propelled particles as vehicles for targeted drug delivery in nanomedicine.    
\end{abstract}
\maketitle

\section{\label{sec:introduction}Introduction}
The individual and collective properties of self-propelled nano- and microdevices, also called \ZT{active particles} \cite{BechingerdLLRVV2016}, form one of the most fascinating and fastest evolving areas of nanotechnology \cite{GuixMMM2014,EstebanFernandezdeAvilaALGZW2018,WangP2018,WangG2012,CeylanYKHS2019,SrivastavaCAB2019,ErkocYCYAS2019,SafdarKJ2018,KimGLF2018,LiEFdAGZW2017,GaoLLH2018,ReinisovaHP2019,GaoW2014a,ZareiZ2018,PengTW2017,KaturiMSS2016,LuoFWG2018,SanchezLJ2015,RaoLMZCW2015,VillaP2019,DongCYGR2018,VenugopalanEFdAPGW2020,GaoWYLMH2021,WangKNZ2021,VenugopalanEFdAPGW2020,GaoWYLMH2021,WangKNZ2021}. 
Benefiting from earlier progress in the fabrication of nanoparticles, research of the last two decades has resulted in a large number of realizations of artificial nano- and microparticles with different propulsion mechanisms \cite{BechingerdLLRVV2016,GuixMMM2014,KimGLF2018,WangP2018,GaoLLH2018,ReinisovaHP2019,LuoFWG2018,RaoLMZCW2015,VenugopalanEFdAPGW2020,WangKNZ2021}. Such particles are typically supplied with energy by fuels \cite{EstebanFernandezdeAvilaALGZW2018,SafdarKJ2018,PengTW2017,SanchezLJ2015,WangKNZ2021}, time-dependent electric \cite{GuixMMM2014,KimGLF2018,LuoFWG2018,WangKNZ2021} or magnetic \cite{GuixMMM2014,KimGLF2018,LuoFWG2018,WangKNZ2021} fields, or by irradiation with light \cite{JiangYS2010,BregullaYC2014,HeFHLGH2016,XuanSGWDH2018,VillaP2019}, X-rays \cite{XuCLFPLK2019}, or ultrasound \cite{WangCHM2012,GarciaGradillaEtAl2013,RaoLMZCW2015,SotoWGGGLKACW2016,ValdezGardunoLEOMSBSWGG2020,WangKNZ2021,ValdezGardunoLEOMSBSWGG2020} and can generate propulsion, e.g., via shape changes \cite{AhmedBJPDN2016}, chemical reactions \cite{EstebanFernandezdeAvilaALGZW2018,SafdarKJ2018,PengTW2017,SanchezLJ2015}, phase transitions in the surrounding medium \cite{VolpeBVKB2011,ButtinoniBKLBS2013,KuemmeltHWBEVLB2013,KuemmeltHWTBEVLB2014,tenHagenKWTLB2014}, thermophoresis \cite{JiangYS2010,BregullaYC2014,HeFHLGH2016,XuanSGWDH2018}, or self-acoustophoresis \cite{WangCHM2012,GarciaGradillaEtAl2013,RaoLMZCW2015,NadalL2014,SotoWGGGLKACW2016,CollisCS2017}. 
Even particles with multiple propulsion mechanisms have been realized \cite{ChenSKLW2019,TangEtAl2019}. 
Besides these various fully man-made self-propelled particles, there are several examples of \ZT{biohybrid} particles \cite{AlapanYYYES2019} that utilize microorganisms such as bacteria \cite{AkinSRSBRBMB2007,FelfoulEtAl2016,YazdiNSVDE2018} and algae \cite{YanEtAl2017,YasaEAS2018} for propulsion.  

Being intrinsically far from thermodynamic equilibrium, self-propelled nano- and microparticles are highly interesting from a nonequilibrium statistical physics perspective \cite{FodorNCTVvW2016}, while a somewhat less developed driver for research on active particles is their plethora of important potential applications. Among them are applications in environmental protection like environmental monitoring \cite{GaoW2014a,ZareiZ2018} and remediation \cite{GuixOGGSMEW2012,GaoW2014a,SanchezLJ2015,KaturiMSS2016,LiEFdAGZW2017,DongCYGR2018,SafdarKJ2018,SridharPS2018,ZareiZ2018,WangHHS2019}, applications in materials science like nanomotor lithography \cite{LiGDPSW2014}, active assembly of three-dimensional functional materials \cite{TasogluDGSD2014}, and realizations of active materials with fascinating novel properties \cite{BechingerdLLRVV2016,Marchetti2015,Saintillan2018}, as well as many applications in medicine \cite{EstebanFernandezdeAvilaALGZW2018,KimGLF2018,SafdarKJ2018,LiEFdAGZW2017,ReinisovaHP2019,ZareiZ2018,YanXZJVFNBZ2019,MedinaSanchezS2017,KaganEtAl2010,EstebanFernandezdeAvilaEtAl2017,EstebanFernandezdeAvilaMSLRCVMGZW2015,WangP2018,GaoLLH2018,ZhangEtAl2019,WangG2012,CeylanYKHS2019,PengTW2017,KaturiMSS2016,ErkocYCYAS2019,LuoFWG2018,HortelaoPPJBS2018,HeFHLGH2016,YasaEAS2018,BalasubramanianEtAl2011,GaoEFdAZW2018,HuHZHZJZ2018,BaylisEtAl2015,GaoDTLGZW2015,WuEtAl2018,WeiEtAl2019,GaoEtAl2012,KaganBCCEEW2012,HoopEtAl2018,ZhouHLYXYWW2019,HortelaoCMCPS2019,BozuyukYYCKS2018,ChoiLKH2018,XuEtAl2019,SridharPS2018,ChenEtAl2017b,CeylanYYTGS2019,MushtaqEtAl2019,XuanSGWDH2018,GarciaGradillaEtAl2013,SotoWGGGLKACW2016,EstebanFernandezdeAvilaARHSTDCZW2018,GaoLWWXH2019,SrivastavaCAB2019,YanEtAl2017,SchwarzKHNSMS2020,SotoWAD2020,WanEtAl2020a,XuMSMWS2020,WangKNZ2021,WanCWNXYZMS2019,SchmidtMSES2020,VenugopalanEFdAPGW2020,WangDWCR2020,YuZSWXCBJZ2021,SotoWAD2020,WangDWCR2020,WanEtAl2020a,XuMSMWS2020,SchmidtMSES2020,YuZSWXCBJZ2021,JinYDWWZ2021}. 
The latter include assisted fertilization \cite{SchwarzKHNSMS2020}, diagnostic sensing \cite{LiEFdAGZW2017,YanEtAl2017,KimGLF2018,WangP2018,ZareiZ2018,PengTW2017,EstebanFernandezdeAvilaMSLRCVMGZW2015,CeylanYKHS2019,SotoWAD2020,WangDWCR2020}, detoxification \cite{EstebanFernandezdeAvilaMSLRCVMGZW2015,LiEFdAGZW2017,ZhangEtAl2019,EstebanFernandezdeAvilaARHSTDCZW2018}, capture and isolation of circulating cancer cells \cite{BalasubramanianEtAl2011,GaoEFdAZW2018}, microbiopsy \cite{ErkocYCYAS2019,CeylanYKHS2019,SotoWAD2020}, thrombolysis \cite{PengTW2017,HuHZHZJZ2018,CeylanYKHS2019,WanEtAl2020a,XuMSMWS2020,WangKNZ2021}, tissue welding \cite{HeFHLGH2016}, and minimally invasive surgery in general \cite{LiEFdAGZW2017,ErkocYCYAS2019,CeylanYKHS2019,SotoWAD2020,WangKNZ2021}. 
Particularly great attention is payed to the area of drug delivery \cite{EstebanFernandezdeAvilaALGZW2018,SafdarKJ2018,ZhouHLYXYWW2019}, where self-propelled nano- and microparticles can enable a fast local distribution of drugs \cite{BaylisEtAl2015,GaoDTLGZW2015,EstebanFernandezdeAvilaALGZW2018,SafdarKJ2018,EstebanFernandezdeAvilaEtAl2017,WeiEtAl2019,ZhouHLYXYWW2019,HortelaoCMCPS2019,XuEtAl2019} or their targeted delivery to specific sites \cite{EstebanFernandezdeAvilaALGZW2018,ErkocYCYAS2019,SafdarKJ2018,KimGLF2018,LiEFdAGZW2017,GaoLLH2018,ReinisovaHP2019,LuoFWG2018,YanXZJVFNBZ2019,WangG2012,YasaEAS2018,HuHZHZJZ2018,WuEtAl2018,GaoEtAl2012,HoopEtAl2018,KaganEtAl2010,ZhouHLYXYWW2019,BozuyukYYCKS2018,SridharPS2018,ChenEtAl2017b,CeylanYYTGS2019,MushtaqEtAl2019,GarciaGradillaEtAl2013,SotoWGGGLKACW2016,GaoLWWXH2019,SrivastavaCAB2019,CeylanYKHS2019,SchwarzKHNSMS2020,ValdezGardunoLEOMSBSWGG2020,XuMSMWS2020,JinYDWWZ2021}. 
While enhanced local drug distribution has already been demonstrated \emph{in vivo} \cite{GaoDTLGZW2015,WeiEtAl2019,SrivastavaCAB2019,EstebanFernandezdeAvilaALGZW2018,LiEFdAGZW2017,WangP2018,EstebanFernandezdeAvilaEtAl2017,CeylanYKHS2019}, directed delivery of therapeutic and imaging agents to a distant target remains a big challenge \cite{ErkocYCYAS2019,KimGLF2018,LiEFdAGZW2017,GaoLLH2018,CeylanYYTGS2019,EstebanFernandezdeAvilaALGZW2018,YasaEAS2018,LuoFWG2018,CeylanYYTGS2019}. 

For a realization of targeted drug delivery by self-propelled nano- or microparticles, five basic requirements have to be met \cite{ReinisovaHP2019,ErkocYCYAS2019,KimGLF2018,LiEFdAGZW2017,WangP2018,GaoLLH2018,WangG2012,LuoFWG2018,MedinaSanchezS2017,SrivastavaCAB2019,YanXZJVFNBZ2019}:
\begin{enumerate}
\item Biocompatibility of the particles and their propulsion mechanism. In particular, the particles must not consist of a toxic material \cite{KimGLF2018,WangP2018,SrivastavaCAB2019,GaoLLH2018,ReinisovaHP2019,WangG2012,YanXZJVFNBZ2019,ZhouHLYXYWW2019,CeylanYYTGS2019,MedinaSanchezS2017,CeylanYKHS2019} or be propelled by a toxic fuel \cite{ErkocYCYAS2019,SafdarKJ2018,KimGLF2018,LiEFdAGZW2017,WangP2018,GaoLLH2018,ReinisovaHP2019,KaturiMSS2016,HortelaoPPJBS2018,MedinaSanchezS2017}. To exclude negative long-term effects of the particles, they should also be removable or biodegradable \cite{EstebanFernandezdeAvilaALGZW2018,ErkocYCYAS2019,KimGLF2018,LiEFdAGZW2017,WangP2018,ReinisovaHP2019,PengTW2017,CeylanYYTGS2019,LuoFWG2018,HortelaoPPJBS2018,YanXZJVFNBZ2019,ZhouHLYXYWW2019,MedinaSanchezS2017,CeylanYKHS2019}.
\item The ability of the particles to move actively, with sufficient speed, and over sufficiently long periods of time inside the body of a patient \cite{LuoFWG2018,LiEFdAGZW2017,HortelaoPPJBS2018,GaoLLH2018}. This excludes, e.g., particles that need to be illuminated for propulsion \cite{DaiWXZDLFT2016,ZhengDWXYLZWT2017,VillaP2019,DongCYGR2018,ChoiLKH2018,SridharPS2018,XuanSGWDH2018}. 
\item The particles must allow for the encapsulation and release of drug molecules \cite{ErkocYCYAS2019,KaganEtAl2010,ReinisovaHP2019,YanXZJVFNBZ2019,WangG2012}.
\item It should be possible to functionalize the particles \cite{WangG2012,LiEFdAGZW2017,KimGLF2018,PengTW2017} and to equip them with stealth features to go undetected by the immune system \cite{WangP2018,GaoLLH2018,ReinisovaHP2019,CeylanYKHS2019,PengTW2017}.
\item A robust and reliable method that allows for steering the particles to their target \cite{ReinisovaHP2019,ErkocYCYAS2019,KimGLF2018,LiEFdAGZW2017,GaoLLH2018,WangG2012,LuoFWG2018,HortelaoPPJBS2018,YanXZJVFNBZ2019,MedinaSanchezS2017}. 
\end{enumerate}
Fortunately, most of these requirements can already be fulfilled by using current technologies. The first problem can be solved by using ultrasound-propelled nano- or microparticles \cite{RaoLMZCW2015,WangKNZ2021}. These particles have a polar shape and move when exposed to ultrasound due to unbalanced hydrodynamic stresses that the surrounding liquid exerts at their surface. Since this propulsion mechanism works for a wide range of particle materials, this can be chosen to ensure biocompatibility \cite{WangLMAHM2014,EstebanFernandezdeAvilaMSLRCVMGZW2015,ReinisovaHP2019,GarciaGradillaSSKYWGW2014,ValdezGardunoLEOMSBSWGG2020} and biodegradability \cite{BozuyukYYCKS2018,WangQHTCHMWPN2018,WangQHTCHMWPN2018,EstebanFernandezdeAvilaALGZW2018,WangKNZ2021}. 
Furthermore, the acoustic propulsion mechanism itself is biocompatible \cite{RaoLMZCW2015,GarciaGradillaEtAl2013,KimGLF2018,WangKNZ2021,ValdezGardunoLEOMSBSWGG2020}. 
Ultrasound propulsion can yield considerable particle speeds already for ultrasound intensities that are typically used in sonography \cite{VossW2020,VossW2021,VossW2022b} and are harmless for patients. 
Such particles also fulfill the second requirement, since acoustic propulsion works in various liquids including biofluids \cite{GarciaGradillaEtAl2013,EstebanFernandezdeAvilaMSLRCVMGZW2015,WangLMAHM2014} and since the particles can be supplied with energy as long as necessary \emph{via} the ultrasound, in contrast to other types of self-propelled particles that eventually run out of fuel. 
The third requirement can be fulfilled by combining the particles with established techniques for encapsulation and release of drugs that are widely used in the context of nanocarriers \cite{BlancoSF2015,PeerKHFML2007,ReinisovaHP2019,GarciaGradillaSSKYWGW2014}.
Since the acoustic propulsion mechanism does not rely on a particular particle material, it is furthermore possible to use established techniques for functionalization of nanoparticles \cite{GarciaGradillaEtAl2013,EstebanFernandezdeAvilaMSLRCVMGZW2015,GarciaGradillaSSKYWGW2014} and equipping them with stealth features \cite{GaoLLH2018,XuanSGWDH2018} to fulfill the fourth requirement. 

Hence, the main remaining requirement is a method that allows for guiding acoustically propelled nano- and microparticles to a particular target \cite{KimGLF2018,LiEFdAGZW2017,GaoLLH2018,LuoFWG2018,YanXZJVFNBZ2019}. To be relevant for medical applications, this method must be biocompatible and reliable \cite{ErkocYCYAS2019,ReinisovaHP2019}. Furthermore, it must work for large numbers of particles \cite{LiEFdAGZW2017} with arbitrary (and unknown) initial distributions. It must also be robust over large distances \cite{KimGLF2018} and in complex environments with unknown structure, such as the vasculature of a patient \cite{LuoFWG2018}. Finally, it should not rely on tracking of the particle positions or orientations. 
Previous experimental studies have considered acoustically propelled particles that are steered by controlling their orientation using an external magnetic field \cite{GarciaGradillaEtAl2013,SotoWGGGLKACW2016,GaoLWWXH2019,GarciaGradillaSSKYWGW2014,WuEtAl2014,ValdezGardunoLEOMSBSWGG2020}. By observing the particles through a microscope and adjusting the orientation of the magnetic field (and thus of the particles) depending on their current position and a path along which they are supposed to move, it is possible to guide them along prescribed paths. However, this method requires tracking of the particles' positions in real time with a microscope, which is not possible in medical applications. Moreover, this method works only for single particles. When several particles need to be steered at the same time, like in a drug-delivery application \cite{LuoFWG2018,LiEFdAGZW2017,ErkocYCYAS2019,WangP2018}, the orientations of all particles are manipulated in the same way so that all particles move along similar trajectories with offsets that originate from their different initial positions, which prevents guiding all particles towards a common target. In a typical application, particles would furthermore distribute through a combination of convection by the blood flow and self-propulsion in the vasculature surrounding the target. Applying the presently existing guiding method would then in principle allow for the guiding of one particle precisely to the target, while other particles, especially if they start on the opposite side of the target, would move away from it. A guiding method for medical applications needs to overcome these limitations and allow for collective guiding to a common target without the need for particle tracking. 

In this article, we propose a method for collective guiding of acoustically propelled nano- and microparticles that meets all the aforementioned criteria and is therefore a promising candidate for application in medicine. 
The proposed method combines a space- and time-dependent ultrasound field that propels the particles with a time-dependent magnetic field that collectively aligns their propulsion directions. 
Using computer simulations, we demonstrate the feasibility of using this guiding method for potential applications in nanomedicine.

\section{\label{results}Results and discussion}
A method for guiding motile nano- and microparticles should be as simple as possible to facilitate its application.
Therefore, we do not pursue ideas to equip the particles with data processing units that control the propulsion of the particles and guide them to the target \cite{NelsonKA2010}, whose large-scale fabrication would be extremely challenging and expensive.  
Instead of such an internal guiding of the particles, we use a method that is based on external guiding.
This means that external fields are applied to influence the motion of the particles in such a way that they collectively move towards the common target.
For reasons of efficiency, the external fields should not simply pull or push the particles towards the target, but rather make use of the particles' propulsion to guide them in the right direction.

In principle, several types of external fields could be used to influence the motion of the particles.
However, fields that are relevant for applications have to fulfill some requirements.
First, they need to be able to run without significant absorption through biological tissue and to be harmless for the patient. 
This excludes electromagnetic waves, especially those with high frequencies whose spatial structure can be well controlled. Electrostatic fields are also excluded, since they would be screened by the ionic fluids inside the tissue.
Second, it must be possible to generate and control the external fields, which excludes the use of gravitational fields. External fields that fulfill these criteria are acoustic fields, when their intensity and frequency are not too large, and magnetic fields, when their flux density and temporal variation are sufficiently low. 
Therefore, we develop a method based on these two types of external fields.
By taking both fields into account simultaneously, we increase the number of available degrees of freedom compared to using only one of them, and this allows us to achieve a high degree of control using relatively simple fields, as discussed below. Furthermore, since both fields can in principle be space- and time-dependent, they provide a versatile method to control the particle motion. 
 
The simplicity of the field structure is a particularly important point, since it simplifies the experimental realization. In particular, using only a magnetic field could require the generation of a set of field lines that is not divergence-free and thus not possible to be generated.
It also needs to be taken into account that the fields need to be realized with realistic tools like a phased array transducer and magnetic field coils that are placed outside of the patient.
Since acoustic fields are easier to structure spatially than magnetic fields, it is reasonable to use the spatial degrees of freedom of the acoustic field and to keep the structure of the magnetic field simple.
Furthermore, the time dependence of the acoustic field can be large, whereas the magnetic field should change slowly in order to avoid effects of electromagnetic induction. 

Based on these considerations, we now propose a method for guiding ultrasound-propelled particles. 
It is based on the combination of a space- and time-dependent ultrasound field, a time-dependent magnetic field, and magnetic particles.
Making particles magnetic is typically possible by embedding smaller magnetic beads \cite{JinYDWWZ2021} or by coating with a magnetic shell. By using a suitable magnetic material such as magnetite, it is furthermore possible to retain biocompatibility of the particles.
When the magnetic particles are ferro- or ferrimagnetic, their orientation can be controlled by a homogeneous external magnetic field. For (super)paramagnetic particles, an inhomogeneous magnetic field can be used to control the particle orientation. 

The main idea of our method is to use a focused ultrasound beam to locally supply the particles with energy as well as a magnetic field to control the orientation of the particles. 
Since the propulsion speed  increases with the ultrasound intensity \cite{VossW2021}, only those particles that are within the focus of the beam are strongly propelled, whereas the other particles in the system are only weakly propelled or not at all. 
The magnetic field is then oriented so that it points from the focus towards the target, 
causing the fast particles to move towards the target,\footnote{We assume that the particles move in the direction of their orientation, as this is the case for many types of self-propelled particles \cite{GuixMMM2014,EstebanFernandezdeAvilaALGZW2018,WangP2018, GaoLLH2018,ZareiZ2018,PengTW2017,KaturiMSS2016,SanchezLJ2015,RaoLMZCW2015, DongCYGR2018,WangKNZ2021}. 
If this is not the case, the orientation of the magnetic field can easily be adapted to the relative orientation of the particle and its propulsion direction.} while the motion of the remaining particles is negligible due to their low propulsion speed. 
Depending on whether the particles are ferro-, ferri-, or (super)paramagnetic, the external magnetic field can be either homogeneous or inhomogeneous.
The focus of the ultrasound beam moves continuously through the system, while the magnetic field rotates with it so that it always points along the vector from the focus to the target. This makes the motile particles move towards the target, whereas for the out-of-focus (immotile) particles, only the orientation is changed. 
A possible trajectory for the focus is a spiral that starts in the outer regions of the system and ends with a cyclic motion along the surface of the target. 
This trajectory helps keeping the temporal variation of the magnetic field slow and thus to avoid effects of electromagnetic induction.
With this combination of two fields, the particles of the whole system move towards and collectively accumulate in the target, independent of their initial positions and orientations. The method furthermore does not require any knowledge of the initial particle distribution or the particle positions and orientations during the course of the trajectory.

This method fulfills all criteria mentioned in section \ref{sec:introduction}, with the nice additional feature that the target does not need to have a particular size or shape. 
 
\begin{figure*}[tbhp]
\centering
\includegraphics[width=\linewidth]{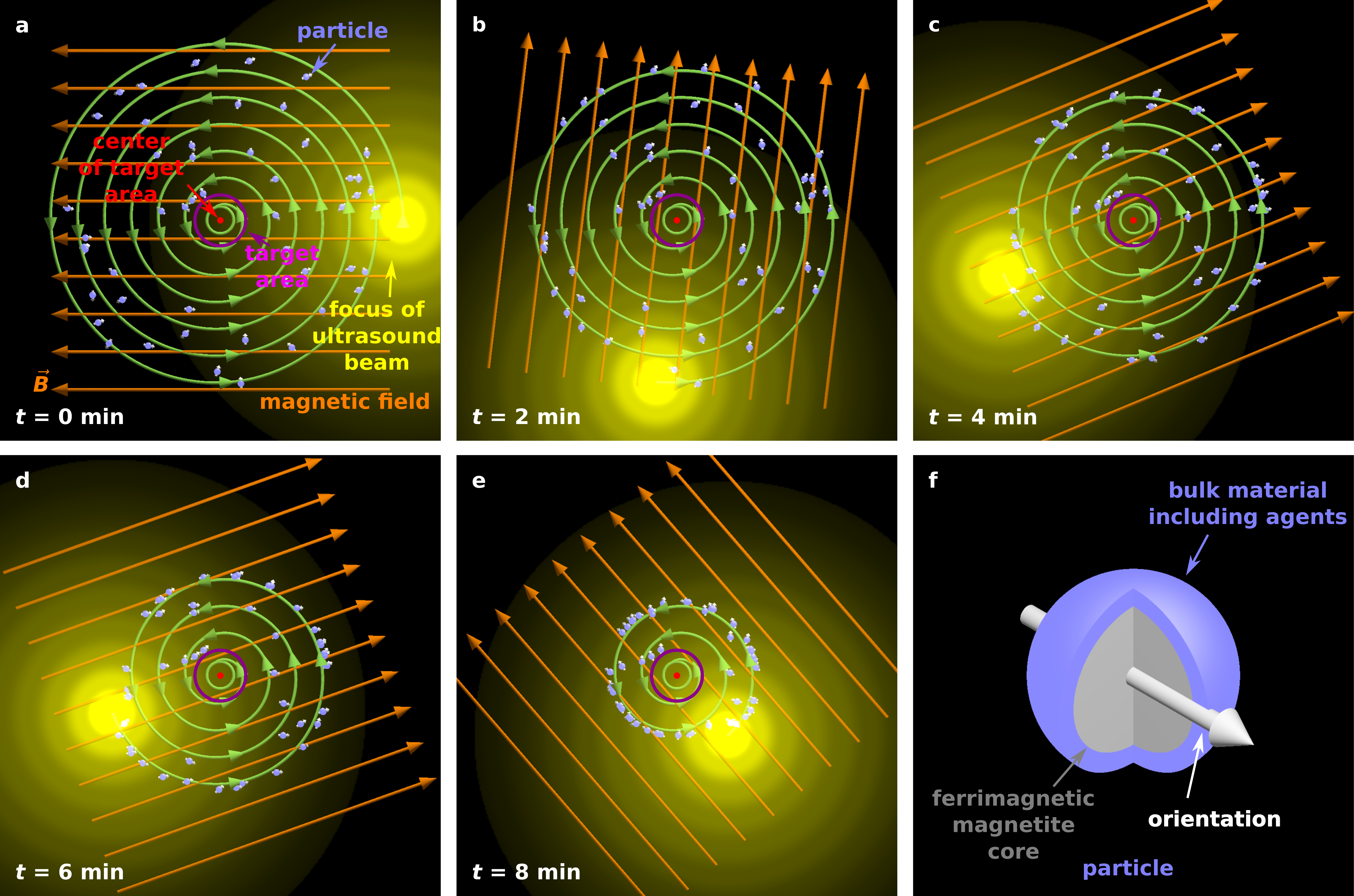}%
\caption{\label{fig:1}(\textbf{a})-(\textbf{e}) Illustration of the proposed particle guiding method. 
(\textbf{f}) Illustration of a single particle suitable for targeted drug delivery.}
\end{figure*}

Figure \ref{fig:1}a-e illustrates the proposed guiding method for 50 magnetic particles in the $x$-$y$ plane.
Although acoustically propelled particles can have various shapes \cite{WangCHM2012,AhmedBJPDN2016,SotoWGGGLKACW2016}, our model for convenience assumes spherical particles. We start from a random, homogeneous initial particle distribution within a disk-shaped region (Fig.\ \ref{fig:1}a) with a circular target area at its center. The ultrasound beam is taken to be antiparallel to the $z$-axis, and its focus moves in the $x$-$y$ plane along a spiral towards the target, while a homogeneous external magnetic field with flux density $\vec{B}$ rotates within the plane so that it is always parallel to the vector from the center of the focus of the ultrasound beam to the center of the target. 
Note that Fig.\ \ref{fig:1}a shows the initial particle orientation, where the particles are not yet aligned with the magnetic field. At later times (Fig.\ \ref{fig:1}b-e), the mean orientation of the particles is aligned with the magnetic field, but due to Brownian rotation the orientations of the individual particles can deviate from the orientation prescribed by the magnetic field.
In Fig.\ \ref{fig:1}f, a schematic example of a magnetic particle is shown, composed of a ferrimagnetic magnetite core and a shell of a bulk material with embedded therapeutic agents, similar to the particle design in Ref.\ \cite{JinYDWWZ2021}. If the bulk material is biocompatible and degrades slowly when it is in contact with biofluids, such a particle is suitable for targeted drug delivery. Furthermore, the core does not need to be massive, but can consist of a large number of smaller magnetic nanoparticles. 

In the following, we numerically demonstrate the reliability and robustness of the proposed method. 
To this end, we performed particle-based Brownian dynamics simulations of a system of magnetic ultrasound-propelled particles being guided towards a common target using the method described above. Further details on the computational implementation are given in the Methods. 

\begin{figure*}[htbp]
\centering
\includegraphics[width=1.0\linewidth]{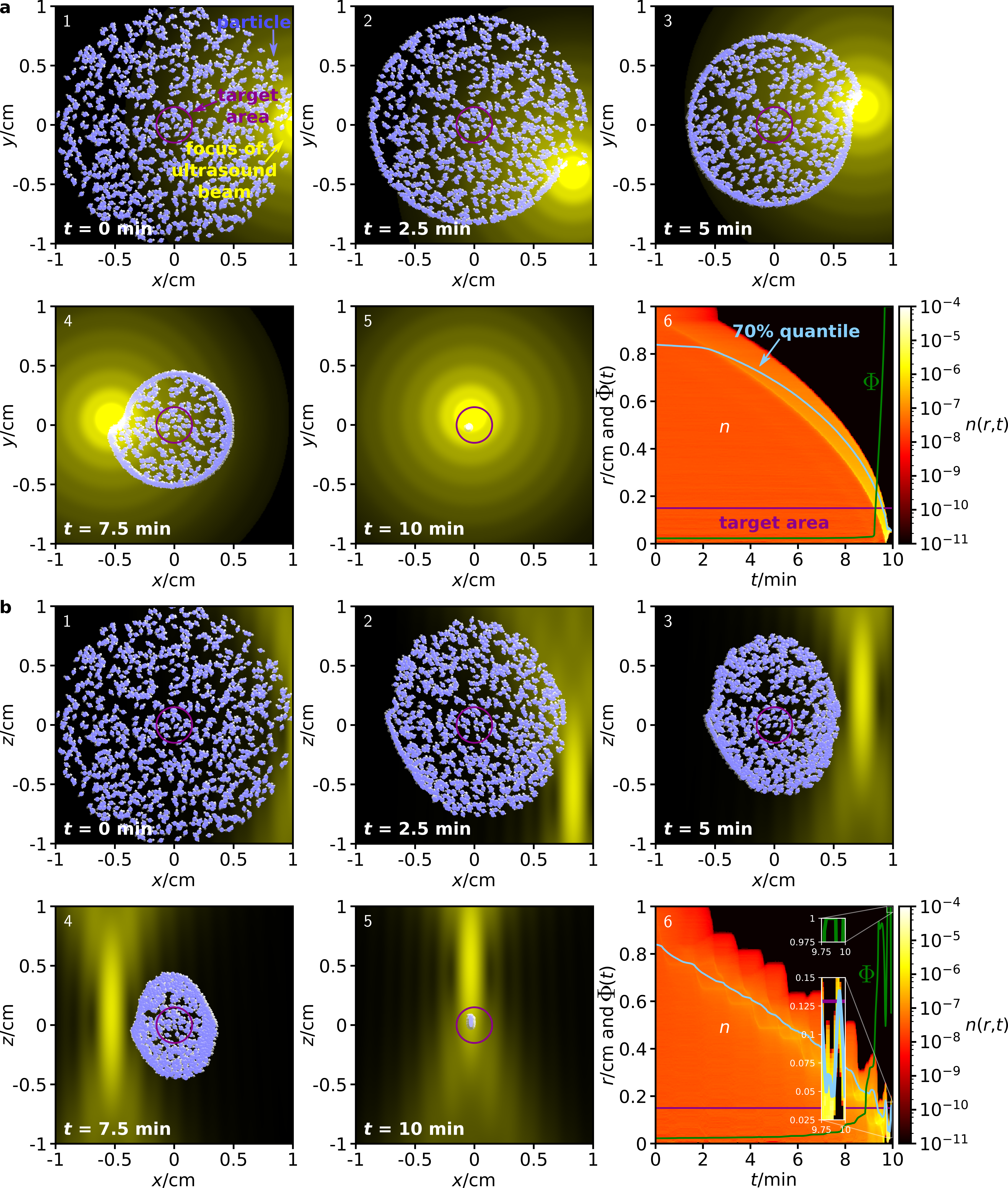}%
\caption{\label{fig:2}Time evolution of a distribution of $N_\mathrm{p}=1000$ particles with diameter $\sigma=100\,\mathrm{nm}$ in the (\textbf{a}) $x$-$y$ plane and (\textbf{b}) $x$-$z$ plane with a homogeneous environment. Plots 1-5 show the particle distribution for increasing times, where the particles are enlarged by a factor $2000$ to be visible. Plots 6 show the radial particle distribution $n(r,t)$ denoting the mean packing density at a distance $r$ from the target (bin width 50\,$\mu$m) at time $t$, the corresponding 70\% quantile, the boundary of the target area, and the fraction of particles $\Phi(t)$ that are within the target at time $t$, ensemble averaged for 100 simulations.}
\end{figure*}

First, we consider a system where the particles are guided through a homogeneous environment. 
Real-world examples of this are active particles to be guided within the eye, e.g., through the vitreous body to the retina \cite{WuEtAl2018}.
The system is similar to that shown in Fig.\ \ref{fig:1}, but now it contains $N_\mathrm{p}=1000$ particles, with diameter $100\,\mathrm{nm}$ that are initially distributed in a disk of diameter $2\,$cm, and guided towards a target area with a diameter of $3\,$mm. 
Figure \ref{fig:2} shows the time evolution of the particle distribution obtained from our simulations. 
To demonstrate that the method allows for full, 3-dimensional control of the particle motion, we now consider two separate initial particle distributions: one in the $x$-$y$ plane (Fig.\ \ref{fig:2}a) and one in the $x$-$z$ plane (Fig.\ \ref{fig:2}b).
To achieve particle guiding in a 3-dimensional system, one can either subsequently perform the procedures for particle guiding in respectively the $x$-$y$ plane and the $x$-$z$ plane, or combine them in different ways (e.g., by choosing a sequence of spherical spirals with decreasing radius for the trajectory of the focus).  
In both procedures, the ultrasound beam remains antiparallel to the $z$-axis, while its focus can move within the corresponding plane. Furthermore, the magnetic field can rotate within the plane. This could be realized in medicine with a single phased array transducer on one side of a patient and with three pairwise perpendicular pairs of static magnetic field coils with tunable flux densities or one pair of field coils that can rotate around its center. 
As is apparent from Fig.\ \ref{fig:2}, the proposed method works very well for both initial conditions, although the focus of the ultrasound beam has very different profiles in the two cross sections. Within 10 minutes, all particles are guided into the target area. 
Remarkably, this method works independently of the position or distance of the particles relative to the target and without any knowledge of the particle positions or orientations during the guiding procedure. 
Comparing the size of the initial particle distribution with the small particle size, some of the particles are guided over very long distances to the target. 
 
Next, we consider a system where the particles are in a complex environment. 
An example of such an environment is particles that are applied intravenously and need to be guided through the vascular system. Since the vasculature constitutes a complicated network of boundaries that confine the motion of the particles, this situation is very challenging for particle guiding. 
To test the proposed method for the case of a complex environment, we performed simulations where particles are confined within a network of channels resembling the human vasculature, having a constant diameter of 500\,$\mu$m. 
The simulations are otherwise analogous to those for a homogeneous environment described above. 
\begin{figure*}[htbp]
\centering
\includegraphics[width=0.93\linewidth]{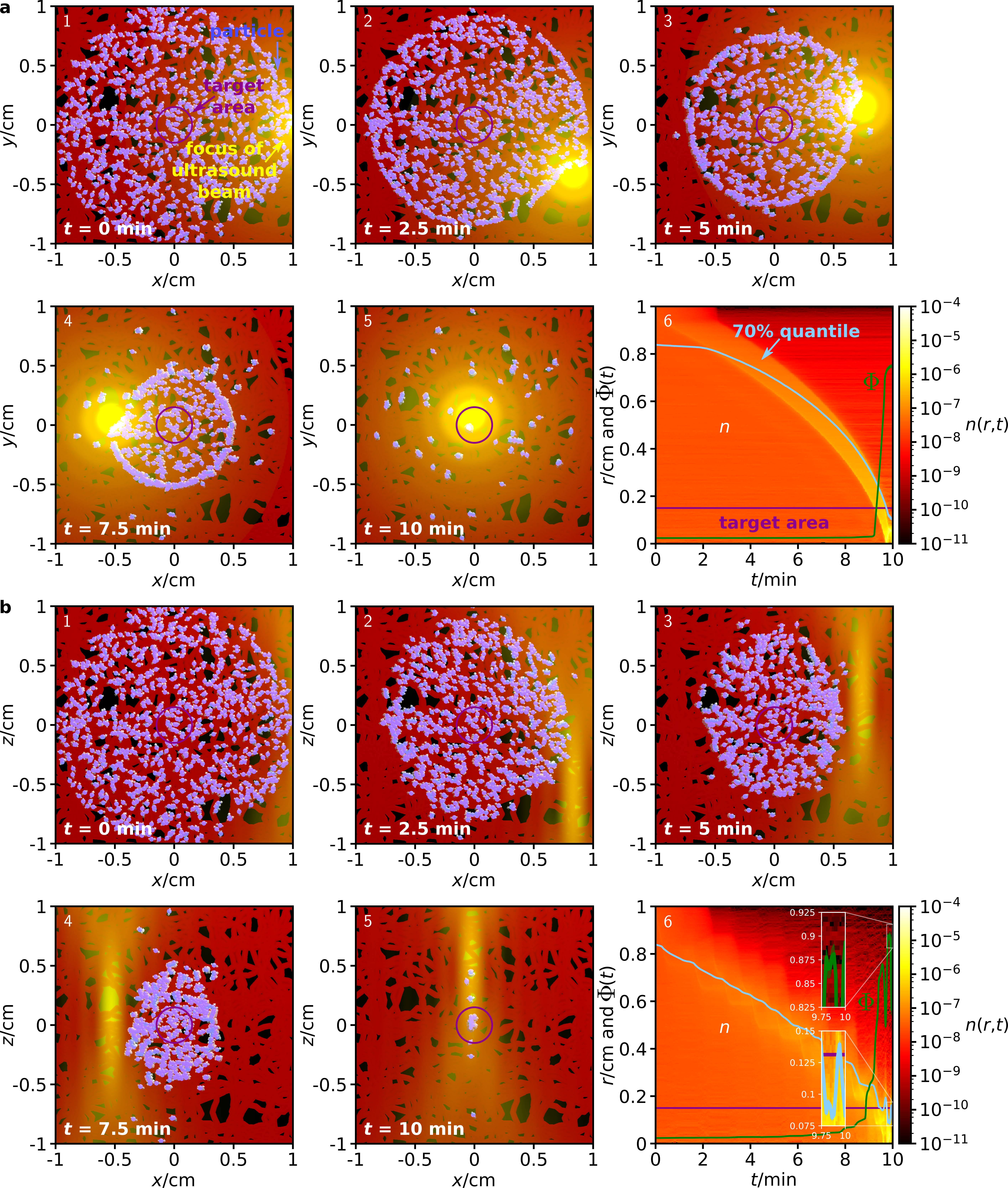}%
\caption{\label{fig:3}Analogous to Fig.\ \ref{fig:2}, but now for a complex environment given by a network of channels with diameter $500\,\mu$m to which the particles are confined.}
\end{figure*}
As can be seen from the results in Fig.\ \ref{fig:3}, the proposed method works surprisingly well even for particles in this very complex environment, although its efficiency is naturally slightly reduced compared to the fully homogeneous case. 
Due to the confinement, some of the particles get trapped in the network and cannot reach the target. 
For particles in an ensemble of 100 simulations in the $x$-$y$ plane, in average a fraction of $74.76\pm 0.70\%$ of the particles reaches the target within a $10$ minute period, while the corresponding fraction for particles in the $x$-$z$ plane is $89.28\pm 1.02\%$. 
While these figures are already surprisingly good, they could be further improved by repeated sweeps of the two fields. Hold times between two sweeps could potentially improve the results even further. 
During these hold times, the particles can rearrange (e.g., by diffusion), allowing them to reach a position from where they have higher success in being guided to the target by the next sweep.

\section{Conclusions}
In this article, we have proposed and computationally demonstrated a method for the collective guiding of acoustically propelled nano- and microparticles. This method is based on an experimentally realizable combination of a space- and time-dependent acoustic field and a time-dependent magnetic field.
Importantly, the method requires neither knowledge about the initial positions and orientations of the particles, nor any particle tracking during the guiding process.
Our results clearly show that the method allows for reliable transport of a large number of arbitrarily distributed particles over long distances to a prescribed target area where they accumulate.
The method is furthermore biocompatible, and the position, size, and shape of the target area can be freely chosen.
We also found that the method is relatively fast, requiring only a few minutes for particle accumulation in realistic situations, scalable with respect to the number of particles, and remains feasible when the particles are in a complex environment of unknown structure. 

These features all make the proposed method potentially applicable for controlling the collective dynamics of ultrasound-propelled active particles for future applications in, e.g., materials science and nanomedicine \cite{VossW2020,VossW2021}. 
In materials science, the method could be used to control the behavior of active materials and, e.g., induce the emergence of tailor-made persistent flow fields.
Modified by replacing the ultrasound field with a light field, the method can also be used for guiding light-propelled particles \cite{StenhammarWMC2016,JiangYS2010,BregullaYC2014,HeFHLGH2016,XuanSGWDH2018,VillaP2019}. 
In nanomedicine, the proposed method would be particularly useful, since medical applications go along with strict medical safety requirements that cannot be fulfilled by other approaches to particle guiding. 
In fact, the proposed guiding method solves the main problem that has so far prevented a wide application of active matter in medicine.
Thus it provides an important step towards the clinical realization of targeted drug delivery and other nanomedical treatments based on active particles.
The method is fully complementary to and can be synergistically combined with techniques from pharmacy and medicine that have been developed for pharmaceutical applications of nanoparticles (e.g., encapsulation and release of drugs, functionalization, stealth features).
For example, nanocarriers for targeted drug delivery, which have been intensively investigated in the last two decades, can lead to the accumulation of therapeutic agents, but a large fraction of the administered nanocarriers reaches off-target regions and causes serious adverse drug reactions \cite{LuoFWG2018}. By combining these particles with an ultrasound-propelled and magnetic unit, they would become motile and could be guided towards their target by our steering method. This should significantly increase targeting and reduce unwanted side effects. 
The ultrasound that is used for the propulsion of our particles fits synergistically to drug delivery, since it stimulates temporary permeability of cell membranes and capillaries which promotes drug delivery into a tumor \cite{SrivastavaCAB2019}. Besides particles for drug delivery, other nanoparticles, such as magnetic particles for thermal treatment of the target tissue \cite{ThiesenJ2008}, can be combined with our method. 
 
Since the acoustic and magnetic fields utilized can be generated by conventional phased array transducers and coils, a technical realization seems to be straightforward. Therefore, we hope that it will be demonstrated in experiments in the near future.
Compared to our proof-of-concept simulations, the performance of the proposed method can be further improved by optimization of the values of its various parameters. 
For example, the profile and trajectory of the ultrasound beam can be modified. One could consider also more complicated ultrasound fields, as long as they can be realized by conventional transducers or acoustic holography \cite{MeldeMQF2016}.

\section{\label{sec:methods}Methods}
The numerical results of this work were obtained by Brownian dynamics simulations, whose details we present in the following.

\subsection{Setup and parameters}
We consider a three-dimensional system that includes $N_\mathrm{p}$ particles, an ultrasound field, and a magnetic field. 
The system is either free of barriers or contains a channel system resembling a vasculature confining the particles. 
To show that the particle guiding works in all dimensions, we consider a particle distribution in the $x$-$y$ plane and a particle distribution in the $x$-$z$ plane. 
Initially, the particles are randomly and homogeneously distributed in a circular region of diameter 2\,cm. 
We simulate $N_\mathrm{p}=1000$ particles, which corresponds to an areal packing density of $1.96\cdot 10^{-8}$, realistic for drug delivery applications.
However, a variation of $N_\mathrm{p}$ should not alter the results significantly. 
The target is in the center of this region and given by a small circular region of diameter 3\,mm. 
Thus, the particles' distance from the center of the target is up to 1\,cm. This is much more than typical distances in the order of $50$-$100\,\mu\text m$ \cite{GarciaGradillaEtAl2013,SotoWGGGLKACW2016,KaganEtAl2010} over which particle guiding has been demonstrated in experiments so far. 
The particles are propelled by an ultrasound beam antiparallel to the $z$-axis, and oriented by a magnetic field. Both are described below.

\subsection{\label{sec:particledesign}Details about the particles}
The maximal particle speed needs to be larger than the speed of the blood flow in the capillaries (about $200\,\mu\mathrm{m/s}$ \cite{SchmidtLH2010} in the smallest capillaries), which constitute by far the largest fraction of the vasculature. For particular acoustically propelled particles, a speed of about 250\,$\mu$m/s has been reported \cite{GarciaGradillaEtAl2013}. Since the particle speed depends strongly on the shape \cite{CollisCS2017,VossW2021} and size \cite{SotoWGGGLKACW2016} of the particles and other parameters (including the frequency of the ultrasound), a strong enhancement of the particle speed by optimization of the parameters can likely be achieved \cite{VossW2021}. 
Furthermore, the particle speed can be enhanced by increasing the ultrasound intensity, since these quantities are approximately proportional to each other \cite{VossW2021}. 
In the experiments of Ref.\ \cite{WangCHM2012}, ultrasound-propelled rod-shaped particles were observed to have a propulsion speed of up to 200\,$\mu$m/s. In these experiments, it was also observed that the particles begin to levitate when the particles are exposed to pulsed ultrasound with a duty cycle of $0.04$ and a frequency of 3.7\,MHz. 
Equating the gravitational force of 0.027\,pN \cite{WangCHM2012} acting on the particles and the expression for the acoustic radiation force acting on a cylindrical particle \cite{Mitri2005} yields an acoustic energy density of 2.20\,J/m$^3$ that corresponds to the levitation threshold.
For their main experiments they used the same input voltage for the ultrasound transducer but continuous ultrasound (corresponding to a duty cycle of 1), i.e., an acoustic energy density of $(2.20/0.04)$\,J/m$^3=55$\,J/m$^3$.
If using the same ultrasound intensity in the focus of our ultrasound beam, the mean ultrasound intensity in the spherical region in which the particles are initially distributed in our simulations would be about 1.22\% of the ultrasound intensity in the focus, i.e., about 0.67\,J/m$^3$. 
According to the U.S.\ Food and Drug Administration, intensities of up to 720\,mW/cm$^2$, which corresponds to 4.8\,J/m$^3$ (assuming a realistic sound velocity of $c=1500$\,m/s in tissue \cite{Wells1975}), are considered as harmless and therefore allowed for diagnostic ultrasound \cite{BarnettTHZRDM2000}. 
Therefore, the ultrasound intensity could be increased by a factor of about $7$, which would enhance the particle speed from 200\,$\mu$m/s to 1.4\,mm/s.
Since already the optimization of the particles' shape and size and an increase of the ultrasound intensity allow for reaching much faster particle speeds than in Ref.\ \cite{WangCHM2012}, for our simulations we assume a particle speed of 1\,mm/s in the focus of the ultrasound beam. 
 
Since acoustically propelled particles can have very different shapes and the optimal particle shape still needs to be found \cite{WangCHM2012,AhmedBJPDN2016,SotoWGGGLKACW2016}, in our simulations we make the simple assumption of spherical particles. Furthermore, since the propulsion speed is an independent parameter in our simulations and since particle interactions are almost negligible due to the low packing density of the particles, the assumed particle shape should have no considerable effect on our simulation results. 

Similarly, there is no obvious choice of optimal particle size. The particles should be small, since smaller ultrasound-propelled particles can be faster than larger ones \cite{SotoWGGGLKACW2016}. The lower boundary for the particle size is at about 10\,nm, since smaller particles are filtered out by the kidney \cite{ReinisovaHP2019,DavisCS2008} when they circulate through the vascular system. On the other hand, the particles should be sufficiently small so that they do not clog the capillaries, whose minimal diameter is $5\,\mu\text m$ \cite{LuoFWG2018,WelterR2013}, and cause thromboses \cite{ErkocYCYAS2019}.
Furthermore, the particles should be able to pass the liver, which requires a size below 100-150\,nm \cite{DavisCS2008}.
Another important advantage of particles that have a size of 100\,nm or less is that they can pass the leaky vasculature of tumors and benefit from the enhanced permeability and retention effect, which leads to an accumulation within tumors \cite{ReinisovaHP2019,DavisCS2008}.
Therefore, we chose a particle size of $100\,\mathrm{nm}$ for our simulations. 

There are still many options for the particular design of the particles. 
An example design that is realistic for drug delivery applications is shown in Fig.\ \ref{fig:1}f. 
It has a magnetized ferrimagnetic core with a diameter of 80\,nm and a 10\,nm thick shell of a bulk material with embedded therapeutic agents.
In addition, the surface of the particle can be functionalized.
All materials used in the particle should be biocompatible, i.e., nontoxic and biodegradable.
The core of the particle does not need to be massive. It can consist of a large number of magnetic nanoparticles. 
A suitable material for the core is magnetite \cite{TassaSW2011,HarisinghaniBHDTHvdKdlRW2003,LiuSDZLGXGLZ2009}. 
As a particle with a size of about 80\,nm, magnetite is ferrimagnetic and has a remanent magnetic moment of $\mu = 1.394 \cdot 10^{-17}\,{\mathrm{J}}/{\mathrm{T}}$. This value follows from a remanent magnetic moment density of about 10\,emu/g \cite{LiKHOIO2017} and a mass density of $5.2\,\mathrm{g}/\mathrm{cm}^3$ \cite{AnthonyBBN2001}. 
The magnetization of the core allows to align the particle orientation by a homogeneous magnetic field. 
For convenience, we choose the same orientation for the particle, its propulsion, and its magnetization. 
The material of the shell should degrade slowly when it is in contact with biofluids.
For this purpose, one could choose Al, Ca, Mg, Si, or Zn or chemical compounds based on them, such as calcium carbonate, calcium sulphate, and calcium phosphate \cite{WangKNZ2021,LuALG2017,JinYDWWZ2021}. 
 
A surface functionalization of the particles would allow to improve the retention in the body, by equipping the particles with stealth features, and thus the success of a therapy \cite{LiEFdAGZW2017}.
This can also help to control the speed of degradation of the particles. 
For this purpose, the functionalization techniques that have been developed in pharmacy and medicine can be utilized.  
An option is to cover the particles by a suitable lipid (bi)layer (e.g., consisting of phospholipids), proteins (e.g., human serum albumin or gelatin), or hydroxyethyl starch.

\subsection{Details of the ultrasound field}
The focused ultrasound beam can be generated by a phased array transducer \cite{Thomenius1996} or a transducer consisting of an oscillating spherical cap \cite{Kennedy2005,MahoneyFMcDCH2001}. 
We consider here the second case in more detail, since it is relatively simple to realize. 
The ultrasound field can be calculated by solving the wave equation. 
For the acoustic pressure amplitude $p$ at position $\vec{r}=(x,y,z)^{\mathrm{T}}$ around a fixed transducer, one obtains the Rayleigh-Sommerfeld integral \cite{MahoneyFMcDCH2001} 
\begin{align}
p(\vec{r}) = \bigg| \rho u f \int_S \!\! \mathrm{d}^{2}r'\,
\frac{\mathrm e^{-(\mathrm ik+af)\norm{\vec{r}'-\vec{r}}}}{\norm{\vec{r}'-\vec{r}}} \bigg|
\label{eq:pressure}%
\end{align}
with the absolute value $\abs{\cdot}$, mass density of the surrounding medium (tissue) $\rho$, 
velocity amplitude of the oscillating cap $u$, and frequency of the ultrasound $f$.  
The integral is performed over the spherical-cap-shaped surface $S$ of the transducer, $\mathrm{i}$ is the imaginary unit,
$k = 2\pi f/c$ is the wave number of the ultrasound, $c$ is the sound velocity in tissue,
$a=1\,\mathrm{dB}/(\mathrm{cm\,MHz})=5\ln(10)/(\mathrm{m\,MHz})$ is the attenuation coefficient in tissue \cite{Wells1975}, and $\norm{\cdot}$ denotes the Euclidean norm.  
We assume that the radius of curvature and the diameter of the spherical-cap-shaped transducer are 20\,cm. 

The propulsion speed of the particles is approximately proportional to the sound intensity $I\propto p^2$ \cite{VossW2021}. 
Since we prescribe the propulsion speed of the particles in the center of the focus of the ultrasound beam by an independent parameter (see section \ref{sec:particledesign}), we are only interested in the spatial profile of $p^2(\vec{r})$, but not on the overall prefactor in Eq.\ \eqref{eq:pressure}.
Therefore, we normalize the maximum of $p(\vec{r})$ to $1$ and do not need to assume particular values for $\rho$ and $u$.
For the ultrasound frequency, we choose a value of $f=1\,\mathrm{MHz}$, since this value lies in the frequency range of therapeutically used ultrasound \cite{Kennedy2005} and leads to a good compromise between high efficiency and high resolution of the particle guiding. 
Smaller frequencies would have the advantage that they lead to a wider focus of the ultrasound beam so that more particles are propelled at the same time, but they would also reduce the spatial resolution for the particle guiding.
As a realistic value for the sound velocity in tissue, we choose $c=1500\,\mathrm{m}/\mathrm{s}$ \cite{Wells1975}.

In a simple setup, the focus of the ultrasound beam is moved by moving the transducer. 
For convenience, we assume that the ultrasound beam is always antiparallel to the $z$-axis but can be moved in all three directions. 
In contrast to the intensity profile for a phased array transducer, the profile of a spherical-cap-shaped transducer has a constant structure and is just shifted as a function of time. 
For the latter case, the intensity profile has an axis of rotational symmetry and its full-width-at-half-maximum shape is similar to a prolate spheroid with dimensions 1.585\,mm and 10.107\,mm. 
The normalized intensity profile can approximately be represented analytically by the function 
\begin{widetext}
\begin{align}
I_{\mathrm{n}}(x,y,z,t) = \begin{cases} \mathrm{sinc}^2\Big(\pi \sqrt{\big(\frac{x-x_0(t)}{b_{\mathrm{xy}}}\big)^2 + \big(\frac{y-y_0(t)} {b_{\mathrm{xy}}}\big)^2 + \big(\frac{z-z_0(t)}{b_{\mathrm{z}}}\big)^2} \Big)   &\mbox{ for } \big(\frac{x-x_0(t)} {b_{\mathrm{xy}}}\big)^2 + \big(\frac{y-y_0(t)} {b_{\mathrm{xy}}}\big)^2 + \big(\frac{z-z_0(t)}{b_{\mathrm{z}}}\big)^2 \le 1\;, \\
0 & \mbox{ for } \big(\frac{x-x_0(t)} {b_{\mathrm{xy}}}\big)^2 + \big(\frac{y-y_0(t)} {b_{\mathrm{xy}}}\big)^2 + \big(\frac{z-z_0(t)}{b_{\mathrm{z}}}\big)^2 > 1\;. \end{cases} 
\label{eq:fitfunsincrhoz}%
\end{align}
\end{widetext}
Here, $x_0(t)$, $y_0(t)$, and $z_0(t)$ are the time-dependent coordinates of the center of the focus of the ultrasound beam 
and $b_{\mathrm{xy}}=1.790\,\text{mm}$ and $b_{\mathrm{z}}=11.408\,\text{mm}$ 
are fit parameters that determine the aspect ratio $\chi=b_{\mathrm{z}}/b_{\mathrm{xy}}\approx 6.38$ of the spheroid.

\subsection{Details of the magnetic field}
For the magnetized particles described in section \ref{sec:particledesign}, a homogeneous magnetic field is sufficient to align them in any wanted direction. Such a magnetic field could be generated, e.g., by a pair of Helmholtz field coils. 
To reorient the magnetic field, one could use a pair of coils that can be rotated around its center or three pairs of static and pairwise perpendicular field coils whose flux densities can be tuned independently.
For the magnetic field, we use a constant flux density of $1\,\mathrm{mT}$. 
Since the magnetic field rotates with the orbital frequency of the ultrasound focus, which is below $0.1\,\mathrm{Hz}$, its flux density is harmless.
Taking into account that the extension of the magnetic field is about 2\,m in a realistic application, the electric field in a human body that can be induced by the rotation stays below $5\cdot 10^{-4}\,\mathrm{V/m}$, which does not cause health damages \cite{Boerner2011}. 
Rotating magnetic fields with a similar strength are also used in other research projects on medical applications \cite{HuHZHZJZ2018, WuEtAl2018,GaoEtAl2012,BozuyukYYCKS2018, ChenEtAl2017b, CeylanYYTGS2019,SchwarzKHNSMS2020}.

\subsection{Details on the simulations}
The simulations were performed using the software package Lammps \cite{Plimpton1995}.

\subsubsection{Equations of motion}
We consider a system of $N_\text{p}=1000$ spherical particles with diameter $\sigma=100\,\mathrm{nm}$ in two spatial dimensions, where the target is in the origin of coordinates.
The equations of motion of the particles are the overdamped Langevin equations \cite{BechingerdLLRVV2016}
\begin{align}
\dot{\vec{r}}_i &= \frac{1}{\gamma_\mathrm{t}}\vec{F}_i(\{\vec{r}_i\},\varphi_i,t) + \vec{\xi}_{\mathrm{t},i}\,,\label{eq:LangevinI}\\
\dot\varphi_i &= \frac{T_\mathrm{ext}(\varphi_i,t)}{\gamma_\mathrm{r}} +\xi_{\mathrm{r},i} \label{eq:LangevinII}
\end{align}
that describe the position $\vec{r}_i(t)$ and orientation $\varphi_i(t)$ of the $i$-th particle as functions of time $t$. 
Since the particles are spheres, their translational and rotational friction coefficients are $\gamma_\mathrm{t} = 3\pi\eta\sigma$ and $\gamma_\mathrm{r} = \pi\eta\sigma^3$, respectively, with the shear viscosity of blood plasma $\eta= 10^{-3}\,\mathrm{Pa\,s}$ \cite{KesmarkyKRT2008}.
$\vec{F}_i$ is the total force acting on particle $i$, $T_\mathrm{ext}$ is the external torque which the magnetic field exerts on the particles, and $\vec{\xi}_{\mathrm{t},i}$ and $\xi_{\mathrm{r},i}$ are zero-mean Gaussian white noise terms that take Brownian motion of particle $i$ into account.  

The force $\vec{F}_i$ is given by 
\begin{equation}
\vec{F}_i(\{\vec{r}_i\},\varphi_i,t) = \vec{F}_{\mathrm{int},i}(\{\vec{r}_i\})+\vec{F}_{\mathrm{p}}(\vec{r}_i,\varphi_i,t)
+ \vec{F}_{\mathrm{ext}}(\vec{r}_i) \;.
\label{eq:totalforce}%
\end{equation}
Its first contribution is the interaction force
\begin{widetext}
\begin{equation}
\vec{F}_{\mathrm{int},i}(\{\vec{r}_i\}) = \sum^{N_{\text{p}}}_{\begin{subarray}{c}j=1\\j\neq i\end{subarray}}
\begin{cases}24\dfrac{\varepsilon}{\sigma} \bigg(\! -2\bigg( \dfrac{\sigma}{\norm{\vec{r}_i-\vec{r}_j}} \bigg)^{13} + 
\bigg( \dfrac{\sigma}{\norm{\vec{r}_i-\vec{r}_j}} \bigg)^{7} \bigg)\dfrac{\vec{r}_j-\vec{r}_i}{\norm{\vec{r}_i-\vec{r}_j}} &\mbox{ for } \norm{\vec{r}_i-\vec{r}_j} < 2^{1/6}\sigma\;, \\
0 & \mbox{ else,} \end{cases} 
\label{eq:F_int}
\end{equation}
\end{widetext}
which is based on the assumption that the particles interact by the Weeks-Chandler-Andersen potential \cite{WeeksCA1971}
\begin{equation}
U(r) = \begin{cases}4\varepsilon \bigg(\Big( \dfrac{\sigma}{r} \Big)^{12} - \Big( \dfrac{\sigma}{r} \Big)^{6} \bigg) +\varepsilon &\mbox{ for } r < 2^{1/6}\sigma\;, \\
0 & \mbox{ else.} \end{cases}
\label{eq:V_int}\raisetag{3em}%
\end{equation}
Here, $\varepsilon$ is the characteristic interaction energy.
It is the potential energy of a pair of particles with distance $r=\sigma$ and here chosen as $\varepsilon=k_\mathrm{B}T_\mathrm{b}\approx 4.28\cdot 10^{-21}\,\mathrm{J}$ with the Boltzmann constant $k_\mathrm{B}$ and the body temperature $T_\mathrm{b}= 37^\circ\,\mathrm{C} \approx 310\,\mathrm{K}$. 
This choice ensures that nondriven particles that move only by Brownian motion keep a typical distance of $\sigma$ from each other.  
The second contribution to Eq.\ \eqref{eq:totalforce} is the acoustic propulsion force $\vec{F}_{\mathrm{p}}$ that acts on a particle. 
It is given by 
\begin{equation}
\vec{F}_{\mathrm{p}}(\vec{r},\varphi,t) = v_0\gamma_\mathrm{t} I_{\mathrm{n}}(\vec{r},t)\hat{u}(\varphi)
\label{eq:accoustic_force}%
\end{equation}
with the propulsion speed $v_0=1\,\mathrm{mm}/\mathrm{s}$ that we prescribe in the center of the focus and the orientational unit vector $\hat{u}(\varphi)=(\cos(\varphi),\sin(\varphi))^\mathrm{T}$ corresponding to the orientation angle $\varphi$. 
The third contribution to Eq.\ \eqref{eq:totalforce} is the external force $\vec{F}_{\mathrm{ext}}$.
It is zero when we consider a system without boundaries, but important in our simulations with the channel system. 
In the latter case, $\vec{F}_{\mathrm{ext}}(\vec{r})$ is the force that the channel walls exert on a particle at position $\vec{r}$.
For the interactions of a particle with the channel walls, we use again the Weeks-Chandler-Andersen potential \eqref{eq:V_int}. 

The external torque $T_\mathrm{ext}(\varphi,t)$ in Eq.\ \eqref{eq:LangevinII}, which the external magnetic field exerts on a particle with orientation $\varphi$, is given by
\begin{equation}
T_\mathrm{ext}(\varphi,t) = (\vec{\mu}(\varphi) \times \vec{B}(t) ) \cdot\hat{e}_z \;.
\label{eq:magnetic_torque}%
\end{equation}
Here, $\vec{\mu}(\varphi) = \mu \hat{u}(\varphi)$ is the vectorial and $\mu = 1.394 \cdot 10^{-17}\,{\mathrm{J}}/{\mathrm{T}}$ the scalar magnetic moment of a particle with orientation $\varphi$, 
$\times$ denotes the cross product,  
\begin{equation}
\vec{B}(t) = -B \frac{\vec{r}_\mathrm{f}(t)}{\norm{\vec{r}_\mathrm{f}(t)}} 
\end{equation}
is the vectorial flux density of the time-dependent magnetic field, 
and $\hat{e}_z$ is the unit vector in the $z$ direction. 
$\vec{r}_\mathrm{f}(t)$ is the position of the focus of the ultrasound beam at time $t$. 
The absolute flux density $B=\norm{\vec{B}}=1\,\mathrm{mT}$ of the magnetic field is sufficiently low to be harmless in medical applications but large enough to cause an alignment of the particles' orientation.

Finally, the zero-mean Gaussian white noises $\vec{\xi}_{\mathrm{t},i}(t)$ and $\xi_{\mathrm{r},i}(t)$ in Eqs.\ \eqref{eq:LangevinI} and \eqref{eq:LangevinII} are statistically independent. $\vec{\xi}_{\mathrm{t},i}(t)$ describes translational Brownian motion and $\xi_{\mathrm{r},i}(t)$ describes rotational Brownian motion. 
Their correlation is given by 
\begin{align}
\langle \vec{\xi}_{\mathrm{t},i}(t)\otimes \vec{\xi}_{\mathrm{t},j}(t') \rangle &= 2 D_\mathrm{t} \delta_{ij}\delta(t-t'){\boldsymbol 1}\;, \\
\langle \xi_{\mathrm{r},i}(t) \xi_{\mathrm{r},j}(t') \rangle &= 2 D_\mathrm{r} \delta_{ij}\delta(t-t')
\end{align}
with the ensemble average $\langle\cdot\rangle$, dyadic product $\otimes$, translational diffusion constant $D_\mathrm{t}=k_\mathrm{B}T_\mathrm{b}/\gamma_\mathrm{t}$, rotational diffusion constant $D_\mathrm{r}=k_\mathrm{B}T_\mathrm{b}/\gamma_\mathrm{r}$, Kronecker delta $\delta_{ij}$, delta distribution $\delta(t)$, and unit matrix $\boldsymbol{1}$.

The equations of motion \eqref{eq:LangevinI} and \eqref{eq:LangevinII} are solved by the Euler-Maruyama integration scheme with a time step size $\Delta t=0.3\,\mu\text{s}$, which is found to be needed to resolve the particle interaction correctly. 
We simulated a period of $t_\mathrm{s}=10\,\mathrm{min}$, which proved to be sufficient for guiding the particles to the target.

\subsubsection{Trajectory of the ultrasound beam}
The focus of the ultrasound beam moves along a spiral towards the center of the target, which is in the origin of coordinates. 
For the simulations in the $x$-$y$ plane, the spiral in circular, and for the simulations in the $x$-$z$ plane, it is ellipsoidal. 
We chose the distance of neighboring revolutions equal to the focus diameter. 
When the edge of the focus touches the center of the target, the ultrasound beam starts to move one last time around the center.
In this final orbit, the edge of the focus permanently touches the center of the target.
This last orbit increases the fraction of particles that reach the target. 

For the $x$-$y$ system, the trajectory of the focus is given by 
\begin{equation}
\vec{r}_\mathrm{f}(t) = r_\mathrm{f}(t) \hat{u}(\varphi_\mathrm{f}(t))
\label{eq:rfXY}%
\end{equation}
with the focus position $\vec{r}_\mathrm{f}(t)=(x_\mathrm{f}(t),y_\mathrm{f}(t))^\mathrm{T}$, the distance of the center of the focus from the origin of coordinates
\begin{equation}
r_\mathrm{f}(t) = R_\mathrm{f} + R_0 (1-s(t)) \;, 
\label{eq:rst}%
\end{equation}
and the angle 
\begin{equation}
\varphi_\mathrm{f}(t) = \pi \frac{R_0}{R_\mathrm{f}} s(t) \;.
\label{eq:phi_of_s}%
\end{equation}
Here, $R_\mathrm{f}=0.7927\,\mathrm{mm}$ is the radius of the focus in the $x$-$y$ plane, $R_0=1\,\mathrm{cm}$ is the radius of the region in which the particles are initially distributed, and 
\begin{equation}
s(t) = \frac{R_0+R_\mathrm{f}}{R_0} - \sqrt{\left(\frac{R_0+R_\mathrm{f}}{R_0}\right)^2- \frac{2v_\varphi R_\mathrm{f}}{\pi R_0^2}t}
\label{eq:s_of_t}%
\end{equation}
with $s(t)\in[0,1]$ is the trajectory parameter.
The constant $v_\varphi = 0.391\,\mathrm{mm}/\mathrm{s}$ is the tangential speed with which the focus moves around the center of the target. 
Its value is chosen so that the duration of a guiding run is $t_\mathrm{s}=10\,\mathrm{min}$. 
At time 
\begin{equation}
t_{1} = \frac{\pi R_0}{v_\varphi}\bigg( \frac{R_0}{2R_\mathrm{f}}+1 \bigg),
\end{equation}
$s$ is equal to $1$ and the focus touches the center. 
Then, the final circle of the trajectory starts. 
It has radius $R_\mathrm{f}$ and takes a period of
\begin{equation}
t_2=2\pi \frac{R_\mathrm{f}}{v_\varphi} \;.
\end{equation}
The total duration of a simulation run is therefore given by $t_\mathrm{s}=t_1+t_2$. 
 
For the $x$-$z$ system, the trajectory of the focus is, instead of Eq.\ \eqref{eq:rfXY}, given by the
focus position $\vec{r}_\mathrm{f}(t)=(x_\mathrm{f}(t),z_\mathrm{f}(t))^\mathrm{T}$ with
\begin{equation}
\vec{r}_\mathrm{f}(t) = r_\mathrm{f}(t) \hat{u}(\varphi_\mathrm{f}(t)) 
+ (\chi-1) R_\mathrm{f} \sin(\varphi_\mathrm{f}(t)) \hat{e}_z \;,
\label{eq:rf_ellip}
\end{equation}
where $\chi=6.38$ is the aspect ratio of the spheroidal focus region. 
Now, $r_\mathrm{f}(t)$ can no longer be interpreted as the distance of the center of the focus from the origin of coordinates, $v_\varphi$ is no longer the tangential speed of the focus, and the final orbit is an ellipse with aspect ratio $\chi$. 
It is described by Eq.\ \eqref{eq:rf_ellip} with $r_\mathrm{f} = R_\mathrm{f}$ and $\varphi_\mathrm{f}(t) = \pi R_0/R_\mathrm{f}+ (v_\varphi/R_\mathrm{f}) (t-t_1)$.

\subsubsection{Channel system}
Here, we describe the algorithm that we used to generate the channel systems. 
Each channel has width $w=500\,\mu\mathrm{m}$ \cite{AhmedBBLDN2017} and consists of two walls separated by distance $w$.
The walls are realized by a chain of small spherical wall particles. 
These particles have fixed positions, the same diameter of 100\,nm as the regular particles, and a center-to-center distance of 50\,nm from each other. 
The interaction of an ultrasound-propelled particle with a wall particle is the same as the regular particle-particle interaction given by Eq.\ \eqref{eq:F_int}. 

The channels are created by discrete random walks in a quadratic domain of size $l=2\,\mathrm{cm}$ with step size $w$.  
Each random walk starts at the boundary of the domain with a step towards the interior of the domain.
All steps are made in random directions that deviate by an angle $\Delta\alpha$ from the direction of the previous step, where the reference orientation of the initial step is perpendicular to the domain boundary. 
This random angle follows the probability distribution 
\begin{equation}
P(\Delta\alpha) = \frac{1}{\pi}(1+\cos(2\Delta\alpha))\;.
\end{equation}
To increase the persistence length of the random walk and thus to avoid loops of the random paths, the angle $\alpha$ is restricted to $\alpha\in [-\pi/4,\pi/4]$.
When a random walk reaches the boundary of the domain again, it is stopped. 
The starting points of the random walks are equally spaced with distance $d=1\,\mathrm{mm}$ along the full boundary of the domain. 

To create the channel system from the random walks, spheres with diameter $w$ are placed at each point of the random paths. 
Furthermore, cylinders with diameter $w$ and length $w$ are placed along the steps of the random walks such that the centers of their upper and lower bases coincide with two subsequent points of a random walk. 
Finally, the spheres and cylinders are merged, which yields a smooth channel system. 
The walls of the channel system are given by its surface within the domain.

For each simulation run, an individual channel system is generated so that our simulation results that originate from an averaging over simulation runs correspond to averaging over different initial particle distributions and different channel systems.

\begin{acknowledgments}
We thank Timo Betz, Astrid Br\"oker, Stephan Br\"oker, Julian Jeggle, Dennis Mulac, Johanna Schmeinck, and Johannes Vo\ss{} for helpful discussions. 
J.S.\ is supported by a Project Grant from the Swedish Research Council (grant number 2019-03718).
R.W.\ is funded by the Deutsche Forschungsgemeinschaft (DFG, German Research Foundation) -- WI 4170/3-1. 
The simulations for this work were performed on the computer cluster PALMA II of the University of M\"unster. 
\end{acknowledgments}

%\clearpage
\nocite{apsrev41Control}
\bibliographystyle{apsrev4-1}
\bibliography{control,refs}

%Control: author (-1) initials jnrlst
\begin{thebibliography}{129}%
\makeatletter
\providecommand \@ifxundefined [1]{%
 \@ifx{#1\undefined}
}%
\providecommand \@ifnum [1]{%
 \ifnum #1\expandafter \@firstoftwo
 \else \expandafter \@secondoftwo
 \fi
}%
\providecommand \@ifx [1]{%
 \ifx #1\expandafter \@firstoftwo
 \else \expandafter \@secondoftwo
 \fi
}%
\providecommand \natexlab [1]{#1}%
\providecommand \enquote  [1]{``#1''}%
\providecommand \bibnamefont  [1]{#1}%
\providecommand \bibfnamefont [1]{#1}%
\providecommand \citenamefont [1]{#1}%
\providecommand \href@noop [0]{\@secondoftwo}%
\providecommand \href [0]{\begingroup \@sanitize@url \@href}%
\providecommand \@href[1]{\@@startlink{#1}\@@href}%
\providecommand \@@href[1]{\endgroup#1\@@endlink}%
\providecommand \@sanitize@url [0]{\catcode `\\12\catcode `\$12\catcode
  `\&12\catcode `\#12\catcode `\^12\catcode `\_12\catcode `\%12\relax}%
\providecommand \@@startlink[1]{}%
\providecommand \@@endlink[0]{}%
\providecommand \url  [0]{\begingroup\@sanitize@url \@url }%
\providecommand \@url [1]{\endgroup\@href {#1}{\urlprefix }}%
\providecommand \urlprefix  [0]{URL }%
\providecommand \Eprint [0]{\href }%
\providecommand \doibase [0]{http://dx.doi.org/}%
\providecommand \selectlanguage [0]{\@gobble}%
\providecommand \bibinfo  [0]{\@secondoftwo}%
\providecommand \bibfield  [0]{\@secondoftwo}%
\providecommand \translation [1]{[#1]}%
\providecommand \BibitemOpen [0]{}%
\providecommand \bibitemStop [0]{}%
\providecommand \bibitemNoStop [0]{.\EOS\space}%
\providecommand \EOS [0]{\spacefactor3000\relax}%
\providecommand \BibitemShut  [1]{\csname bibitem#1\endcsname}%
\let\auto@bib@innerbib\@empty
%</preamble>
\bibitem [{\citenamefont {Bechinger}\ \emph {et~al.}(2016)\citenamefont
  {Bechinger}, \citenamefont {{Di Leonardo}}, \citenamefont {L{\"o}wen},
  \citenamefont {Reichhardt}, \citenamefont {Volpe},\ and\ \citenamefont
  {Volpe}}]{BechingerdLLRVV2016}%
  \BibitemOpen
  \bibfield  {author} {\bibinfo {author} {\bibfnamefont {C.}~\bibnamefont
  {Bechinger}}, \bibinfo {author} {\bibfnamefont {R.}~\bibnamefont {{Di
  Leonardo}}}, \bibinfo {author} {\bibfnamefont {H.}~\bibnamefont {L{\"o}wen}},
  \bibinfo {author} {\bibfnamefont {C.}~\bibnamefont {Reichhardt}}, \bibinfo
  {author} {\bibfnamefont {G.}~\bibnamefont {Volpe}}, \ and\ \bibinfo {author}
  {\bibfnamefont {G.}~\bibnamefont {Volpe}},\ }\bibfield  {title} {\enquote
  {\bibinfo {title} {Active particles in complex and crowded environments},}\
  }\href@noop {} {\bibfield  {journal} {\bibinfo  {journal} {Reviews of Modern
  Physics}\ }\textbf {\bibinfo {volume} {88}},\ \bibinfo {pages} {045006}
  (\bibinfo {year} {2016})}\BibitemShut {NoStop}%
\bibitem [{\citenamefont {Guix}\ \emph {et~al.}(2014)\citenamefont {Guix},
  \citenamefont {{Mayorga-Martinez}},\ and\ \citenamefont
  {Merko{\c{c}}i}}]{GuixMMM2014}%
  \BibitemOpen
  \bibfield  {author} {\bibinfo {author} {\bibfnamefont {M.}~\bibnamefont
  {Guix}}, \bibinfo {author} {\bibfnamefont {C.~C.}\ \bibnamefont
  {{Mayorga-Martinez}}}, \ and\ \bibinfo {author} {\bibfnamefont
  {A.}~\bibnamefont {Merko{\c{c}}i}},\ }\bibfield  {title} {\enquote {\bibinfo
  {title} {Nano/{M}icromotors in (bio)chemical science applications},}\
  }\href@noop {} {\bibfield  {journal} {\bibinfo  {journal} {Chemical Reviews}\
  }\textbf {\bibinfo {volume} {114}},\ \bibinfo {pages} {6285--6322} (\bibinfo
  {year} {2014})}\BibitemShut {NoStop}%
\bibitem [{\citenamefont {{Esteban-Fern{\'a}ndez de {\'A}vila}}\ \emph
  {et~al.}(2018{\natexlab{a}})\citenamefont {{Esteban-Fern{\'a}ndez de
  {\'A}vila}}, \citenamefont {Angsantikul}, \citenamefont {Li}, \citenamefont
  {Gao}, \citenamefont {Zhang},\ and\ \citenamefont
  {Wang}}]{EstebanFernandezdeAvilaALGZW2018}%
  \BibitemOpen
  \bibfield  {author} {\bibinfo {author} {\bibfnamefont {B.}~\bibnamefont
  {{Esteban-Fern{\'a}ndez de {\'A}vila}}}, \bibinfo {author} {\bibfnamefont
  {P.}~\bibnamefont {Angsantikul}}, \bibinfo {author} {\bibfnamefont
  {J.}~\bibnamefont {Li}}, \bibinfo {author} {\bibfnamefont {W.}~\bibnamefont
  {Gao}}, \bibinfo {author} {\bibfnamefont {L.}~\bibnamefont {Zhang}}, \ and\
  \bibinfo {author} {\bibfnamefont {J.}~\bibnamefont {Wang}},\ }\bibfield
  {title} {\enquote {\bibinfo {title} {Micromotors go in vivo: from test tubes
  to live animals},}\ }\href@noop {} {\bibfield  {journal} {\bibinfo  {journal}
  {Advanced Functional Materials}\ }\textbf {\bibinfo {volume} {28}},\ \bibinfo
  {pages} {1705640} (\bibinfo {year} {2018}{\natexlab{a}})}\BibitemShut
  {NoStop}%
\bibitem [{\citenamefont {Wang}\ and\ \citenamefont
  {Pumera}(2018)}]{WangP2018}%
  \BibitemOpen
  \bibfield  {author} {\bibinfo {author} {\bibfnamefont {H.}~\bibnamefont
  {Wang}}\ and\ \bibinfo {author} {\bibfnamefont {M.}~\bibnamefont {Pumera}},\
  }\bibfield  {title} {\enquote {\bibinfo {title} {Micro/{N}anomachines and
  living biosystems: from simple interactions to microcyborgs},}\ }\href@noop
  {} {\bibfield  {journal} {\bibinfo  {journal} {Advanced Functional
  Materials}\ }\textbf {\bibinfo {volume} {28}},\ \bibinfo {pages} {1705421}
  (\bibinfo {year} {2018})}\BibitemShut {NoStop}%
\bibitem [{\citenamefont {Wang}\ and\ \citenamefont {Gao}(2012)}]{WangG2012}%
  \BibitemOpen
  \bibfield  {author} {\bibinfo {author} {\bibfnamefont {J.}~\bibnamefont
  {Wang}}\ and\ \bibinfo {author} {\bibfnamefont {W.}~\bibnamefont {Gao}},\
  }\bibfield  {title} {\enquote {\bibinfo {title} {Nano/{M}icroscale motors:
  biomedical opportunities and challenges},}\ }\href@noop {} {\bibfield
  {journal} {\bibinfo  {journal} {ACS Nano}\ }\textbf {\bibinfo {volume} {6}},\
  \bibinfo {pages} {5745--5751} (\bibinfo {year} {2012})}\BibitemShut {NoStop}%
\bibitem [{\citenamefont {Ceylan}\ \emph
  {et~al.}(2019{\natexlab{a}})\citenamefont {Ceylan}, \citenamefont {Yasa},
  \citenamefont {Kilic}, \citenamefont {Hu},\ and\ \citenamefont
  {Sitti}}]{CeylanYKHS2019}%
  \BibitemOpen
  \bibfield  {author} {\bibinfo {author} {\bibfnamefont {H.}~\bibnamefont
  {Ceylan}}, \bibinfo {author} {\bibfnamefont {I.~C.}\ \bibnamefont {Yasa}},
  \bibinfo {author} {\bibfnamefont {U.}~\bibnamefont {Kilic}}, \bibinfo
  {author} {\bibfnamefont {W.}~\bibnamefont {Hu}}, \ and\ \bibinfo {author}
  {\bibfnamefont {M.}~\bibnamefont {Sitti}},\ }\bibfield  {title} {\enquote
  {\bibinfo {title} {Translational prospects of untethered medical
  microrobots},}\ }\href@noop {} {\bibfield  {journal} {\bibinfo  {journal}
  {Progress in Biomedical Engineering}\ }\textbf {\bibinfo {volume} {1}},\
  \bibinfo {pages} {012002} (\bibinfo {year} {2019}{\natexlab{a}})}\BibitemShut
  {NoStop}%
\bibitem [{\citenamefont {Srivastava}\ \emph {et~al.}(2019)\citenamefont
  {Srivastava}, \citenamefont {Clergeaud}, \citenamefont {Andresen},\ and\
  \citenamefont {Boisen}}]{SrivastavaCAB2019}%
  \BibitemOpen
  \bibfield  {author} {\bibinfo {author} {\bibfnamefont {S.~K.}\ \bibnamefont
  {Srivastava}}, \bibinfo {author} {\bibfnamefont {G.}~\bibnamefont
  {Clergeaud}}, \bibinfo {author} {\bibfnamefont {T.~L.}\ \bibnamefont
  {Andresen}}, \ and\ \bibinfo {author} {\bibfnamefont {A.}~\bibnamefont
  {Boisen}},\ }\bibfield  {title} {\enquote {\bibinfo {title} {Micromotors for
  drug delivery in vivo: the road ahead},}\ }\href@noop {} {\bibfield
  {journal} {\bibinfo  {journal} {Advanced Drug Delivery Reviews}\ }\textbf
  {\bibinfo {volume} {138}},\ \bibinfo {pages} {41--55} (\bibinfo {year}
  {2019})}\BibitemShut {NoStop}%
\bibitem [{\citenamefont {Erkoc}\ \emph {et~al.}(2019)\citenamefont {Erkoc},
  \citenamefont {Yasa}, \citenamefont {Ceylan}, \citenamefont {Yasa},
  \citenamefont {Alapan},\ and\ \citenamefont {Sitti}}]{ErkocYCYAS2019}%
  \BibitemOpen
  \bibfield  {author} {\bibinfo {author} {\bibfnamefont {P.}~\bibnamefont
  {Erkoc}}, \bibinfo {author} {\bibfnamefont {I.~C.}\ \bibnamefont {Yasa}},
  \bibinfo {author} {\bibfnamefont {H.}~\bibnamefont {Ceylan}}, \bibinfo
  {author} {\bibfnamefont {O.}~\bibnamefont {Yasa}}, \bibinfo {author}
  {\bibfnamefont {Y.}~\bibnamefont {Alapan}}, \ and\ \bibinfo {author}
  {\bibfnamefont {M.}~\bibnamefont {Sitti}},\ }\bibfield  {title} {\enquote
  {\bibinfo {title} {Mobile microrobots for active therapeutic delivery},}\
  }\href@noop {} {\bibfield  {journal} {\bibinfo  {journal} {Advanced
  Therapeutics}\ }\textbf {\bibinfo {volume} {2}},\ \bibinfo {pages} {1800064}
  (\bibinfo {year} {2019})}\BibitemShut {NoStop}%
\bibitem [{\citenamefont {Safdar}\ \emph {et~al.}(2018)\citenamefont {Safdar},
  \citenamefont {Khan},\ and\ \citenamefont {J{\"a}nis}}]{SafdarKJ2018}%
  \BibitemOpen
  \bibfield  {author} {\bibinfo {author} {\bibfnamefont {M.}~\bibnamefont
  {Safdar}}, \bibinfo {author} {\bibfnamefont {S.~U.}\ \bibnamefont {Khan}}, \
  and\ \bibinfo {author} {\bibfnamefont {J.}~\bibnamefont {J{\"a}nis}},\
  }\bibfield  {title} {\enquote {\bibinfo {title} {Progress toward catalytic
  micro- and nanomotors for biomedical and environmental applications},}\
  }\href@noop {} {\bibfield  {journal} {\bibinfo  {journal} {Advanced
  Materials}\ }\textbf {\bibinfo {volume} {30}},\ \bibinfo {pages} {1703660}
  (\bibinfo {year} {2018})}\BibitemShut {NoStop}%
\bibitem [{\citenamefont {Kim}\ \emph {et~al.}(2018)\citenamefont {Kim},
  \citenamefont {Guo}, \citenamefont {Liang},\ and\ \citenamefont
  {Fan}}]{KimGLF2018}%
  \BibitemOpen
  \bibfield  {author} {\bibinfo {author} {\bibfnamefont {K.}~\bibnamefont
  {Kim}}, \bibinfo {author} {\bibfnamefont {J.}~\bibnamefont {Guo}}, \bibinfo
  {author} {\bibfnamefont {Z.}~\bibnamefont {Liang}}, \ and\ \bibinfo {author}
  {\bibfnamefont {D.}~\bibnamefont {Fan}},\ }\bibfield  {title} {\enquote
  {\bibinfo {title} {Artificial micro/nanomachines for bioapplications:
  biochemical delivery and diagnostic sensing},}\ }\href@noop {} {\bibfield
  {journal} {\bibinfo  {journal} {Advanced Functional Materials}\ }\textbf
  {\bibinfo {volume} {28}},\ \bibinfo {pages} {1705867} (\bibinfo {year}
  {2018})}\BibitemShut {NoStop}%
\bibitem [{\citenamefont {Li}\ \emph {et~al.}(2017{\natexlab{a}})\citenamefont
  {Li}, \citenamefont {{Esteban-Fern{\'a}ndez de {\'A}vila}}, \citenamefont
  {Gao}, \citenamefont {Zhang},\ and\ \citenamefont {Wang}}]{LiEFdAGZW2017}%
  \BibitemOpen
  \bibfield  {author} {\bibinfo {author} {\bibfnamefont {J.}~\bibnamefont
  {Li}}, \bibinfo {author} {\bibfnamefont {B.}~\bibnamefont
  {{Esteban-Fern{\'a}ndez de {\'A}vila}}}, \bibinfo {author} {\bibfnamefont
  {W.}~\bibnamefont {Gao}}, \bibinfo {author} {\bibfnamefont {L.}~\bibnamefont
  {Zhang}}, \ and\ \bibinfo {author} {\bibfnamefont {J.}~\bibnamefont {Wang}},\
  }\bibfield  {title} {\enquote {\bibinfo {title} {Micro/{N}anorobots for
  biomedicine: delivery, surgery, sensing, and detoxification},}\ }\href@noop
  {} {\bibfield  {journal} {\bibinfo  {journal} {Science Robotics}\ }\textbf
  {\bibinfo {volume} {2}},\ \bibinfo {pages} {eaam6431} (\bibinfo {year}
  {2017}{\natexlab{a}})}\BibitemShut {NoStop}%
\bibitem [{\citenamefont {Gao}\ \emph {et~al.}(2018{\natexlab{a}})\citenamefont
  {Gao}, \citenamefont {Lin}, \citenamefont {Lin},\ and\ \citenamefont
  {He}}]{GaoLLH2018}%
  \BibitemOpen
  \bibfield  {author} {\bibinfo {author} {\bibfnamefont {C.}~\bibnamefont
  {Gao}}, \bibinfo {author} {\bibfnamefont {Z.}~\bibnamefont {Lin}}, \bibinfo
  {author} {\bibfnamefont {X.}~\bibnamefont {Lin}}, \ and\ \bibinfo {author}
  {\bibfnamefont {Q.}~\bibnamefont {He}},\ }\bibfield  {title} {\enquote
  {\bibinfo {title} {Cell membrane-camouflaged colloid motors for biomedical
  applications},}\ }\href@noop {} {\bibfield  {journal} {\bibinfo  {journal}
  {Advanced Therapeutics}\ }\textbf {\bibinfo {volume} {1}},\ \bibinfo {pages}
  {1800056} (\bibinfo {year} {2018}{\natexlab{a}})}\BibitemShut {NoStop}%
\bibitem [{\citenamefont {Reini{\v{s}}ov{\'a}}\ \emph
  {et~al.}(2019)\citenamefont {Reini{\v{s}}ov{\'a}}, \citenamefont
  {Hermanov{\'a}},\ and\ \citenamefont {Pumera}}]{ReinisovaHP2019}%
  \BibitemOpen
  \bibfield  {author} {\bibinfo {author} {\bibfnamefont {L.}~\bibnamefont
  {Reini{\v{s}}ov{\'a}}}, \bibinfo {author} {\bibfnamefont {S.}~\bibnamefont
  {Hermanov{\'a}}}, \ and\ \bibinfo {author} {\bibfnamefont {M.}~\bibnamefont
  {Pumera}},\ }\bibfield  {title} {\enquote {\bibinfo {title}
  {Micro/{N}anomachines: what is needed for them to become a real force in
  cancer therapy?}}\ }\href@noop {} {\bibfield  {journal} {\bibinfo  {journal}
  {Nanoscale}\ }\textbf {\bibinfo {volume} {11}},\ \bibinfo {pages}
  {6519--6532} (\bibinfo {year} {2019})}\BibitemShut {NoStop}%
\bibitem [{\citenamefont {Gao}\ and\ \citenamefont {Wang}(2014)}]{GaoW2014a}%
  \BibitemOpen
  \bibfield  {author} {\bibinfo {author} {\bibfnamefont {W.}~\bibnamefont
  {Gao}}\ and\ \bibinfo {author} {\bibfnamefont {J.}~\bibnamefont {Wang}},\
  }\bibfield  {title} {\enquote {\bibinfo {title} {The environmental impact of
  micro/nanomachines: a review},}\ }\href@noop {} {\bibfield  {journal}
  {\bibinfo  {journal} {ACS Nano}\ }\textbf {\bibinfo {volume} {8}},\ \bibinfo
  {pages} {3170--3180} (\bibinfo {year} {2014})}\BibitemShut {NoStop}%
\bibitem [{\citenamefont {Zarei}\ and\ \citenamefont
  {Zarei}(2018)}]{ZareiZ2018}%
  \BibitemOpen
  \bibfield  {author} {\bibinfo {author} {\bibfnamefont {M.}~\bibnamefont
  {Zarei}}\ and\ \bibinfo {author} {\bibfnamefont {M.}~\bibnamefont {Zarei}},\
  }\bibfield  {title} {\enquote {\bibinfo {title} {Self-propelled
  micro/nanomotors for sensing and environmental remediation},}\ }\href@noop {}
  {\bibfield  {journal} {\bibinfo  {journal} {Small}\ }\textbf {\bibinfo
  {volume} {14}},\ \bibinfo {pages} {1800912} (\bibinfo {year}
  {2018})}\BibitemShut {NoStop}%
\bibitem [{\citenamefont {Peng}\ \emph {et~al.}(2017)\citenamefont {Peng},
  \citenamefont {Tu},\ and\ \citenamefont {Wilson}}]{PengTW2017}%
  \BibitemOpen
  \bibfield  {author} {\bibinfo {author} {\bibfnamefont {F.}~\bibnamefont
  {Peng}}, \bibinfo {author} {\bibfnamefont {Y.}~\bibnamefont {Tu}}, \ and\
  \bibinfo {author} {\bibfnamefont {D.~A.}\ \bibnamefont {Wilson}},\ }\bibfield
   {title} {\enquote {\bibinfo {title} {Micro/{N}anomotors towards in vivo
  application: cell, tissue and biofluid},}\ }\href@noop {} {\bibfield
  {journal} {\bibinfo  {journal} {Chemical Society Reviews}\ }\textbf {\bibinfo
  {volume} {46}},\ \bibinfo {pages} {5289--5310} (\bibinfo {year}
  {2017})}\BibitemShut {NoStop}%
\bibitem [{\citenamefont {Katuri}\ \emph {et~al.}(2016)\citenamefont {Katuri},
  \citenamefont {Ma}, \citenamefont {Stanton},\ and\ \citenamefont
  {S{\'a}nchez}}]{KaturiMSS2016}%
  \BibitemOpen
  \bibfield  {author} {\bibinfo {author} {\bibfnamefont {J.}~\bibnamefont
  {Katuri}}, \bibinfo {author} {\bibfnamefont {X.}~\bibnamefont {Ma}}, \bibinfo
  {author} {\bibfnamefont {M.~M.}\ \bibnamefont {Stanton}}, \ and\ \bibinfo
  {author} {\bibfnamefont {S.}~\bibnamefont {S{\'a}nchez}},\ }\bibfield
  {title} {\enquote {\bibinfo {title} {Designing micro- and nanoswimmers for
  specific applications},}\ }\href@noop {} {\bibfield  {journal} {\bibinfo
  {journal} {Accounts of Chemical Research}\ }\textbf {\bibinfo {volume}
  {50}},\ \bibinfo {pages} {2--11} (\bibinfo {year} {2016})}\BibitemShut
  {NoStop}%
\bibitem [{\citenamefont {Luo}\ \emph {et~al.}(2018)\citenamefont {Luo},
  \citenamefont {Feng}, \citenamefont {Wang},\ and\ \citenamefont
  {Guan}}]{LuoFWG2018}%
  \BibitemOpen
  \bibfield  {author} {\bibinfo {author} {\bibfnamefont {M.}~\bibnamefont
  {Luo}}, \bibinfo {author} {\bibfnamefont {Y.}~\bibnamefont {Feng}}, \bibinfo
  {author} {\bibfnamefont {T.}~\bibnamefont {Wang}}, \ and\ \bibinfo {author}
  {\bibfnamefont {J.}~\bibnamefont {Guan}},\ }\bibfield  {title} {\enquote
  {\bibinfo {title} {Micro-/{N}anorobots at work in active drug delivery},}\
  }\href@noop {} {\bibfield  {journal} {\bibinfo  {journal} {Advanced
  Functional Materials}\ }\textbf {\bibinfo {volume} {28}},\ \bibinfo {pages}
  {1706100} (\bibinfo {year} {2018})}\BibitemShut {NoStop}%
\bibitem [{\citenamefont {S{\'a}nchez}\ \emph {et~al.}(2015)\citenamefont
  {S{\'a}nchez}, \citenamefont {Soler},\ and\ \citenamefont
  {Katuri}}]{SanchezLJ2015}%
  \BibitemOpen
  \bibfield  {author} {\bibinfo {author} {\bibfnamefont {S.}~\bibnamefont
  {S{\'a}nchez}}, \bibinfo {author} {\bibfnamefont {L.}~\bibnamefont {Soler}},
  \ and\ \bibinfo {author} {\bibfnamefont {J.}~\bibnamefont {Katuri}},\
  }\bibfield  {title} {\enquote {\bibinfo {title} {Chemically powered micro-
  and nanomotors},}\ }\href@noop {} {\bibfield  {journal} {\bibinfo  {journal}
  {Angewandte Chemie International Edition}\ }\textbf {\bibinfo {volume}
  {54}},\ \bibinfo {pages} {1414--1444} (\bibinfo {year} {2015})}\BibitemShut
  {NoStop}%
\bibitem [{\citenamefont {Rao}\ \emph {et~al.}(2015)\citenamefont {Rao},
  \citenamefont {Li}, \citenamefont {Meng}, \citenamefont {Zheng},
  \citenamefont {Cai},\ and\ \citenamefont {Wang}}]{RaoLMZCW2015}%
  \BibitemOpen
  \bibfield  {author} {\bibinfo {author} {\bibfnamefont {K.~J.}\ \bibnamefont
  {Rao}}, \bibinfo {author} {\bibfnamefont {F.}~\bibnamefont {Li}}, \bibinfo
  {author} {\bibfnamefont {L.}~\bibnamefont {Meng}}, \bibinfo {author}
  {\bibfnamefont {H.}~\bibnamefont {Zheng}}, \bibinfo {author} {\bibfnamefont
  {F.}~\bibnamefont {Cai}}, \ and\ \bibinfo {author} {\bibfnamefont
  {W.}~\bibnamefont {Wang}},\ }\bibfield  {title} {\enquote {\bibinfo {title}
  {A force to be reckoned with: a review of synthetic microswimmers powered by
  ultrasound},}\ }\href@noop {} {\bibfield  {journal} {\bibinfo  {journal}
  {Small}\ }\textbf {\bibinfo {volume} {11}},\ \bibinfo {pages} {2836--2846}
  (\bibinfo {year} {2015})}\BibitemShut {NoStop}%
\bibitem [{\citenamefont {Villa}\ and\ \citenamefont
  {Pumera}(2019)}]{VillaP2019}%
  \BibitemOpen
  \bibfield  {author} {\bibinfo {author} {\bibfnamefont {K.}~\bibnamefont
  {Villa}}\ and\ \bibinfo {author} {\bibfnamefont {M.}~\bibnamefont {Pumera}},\
  }\bibfield  {title} {\enquote {\bibinfo {title} {Fuel-free light-driven
  micro/nanomachines: artificial active matter mimicking nature},}\ }\href@noop
  {} {\bibfield  {journal} {\bibinfo  {journal} {Chemical Society Reviews}\
  }\textbf {\bibinfo {volume} {48}},\ \bibinfo {pages} {4966--4978} (\bibinfo
  {year} {2019})}\BibitemShut {NoStop}%
\bibitem [{\citenamefont {Dong}\ \emph {et~al.}(2018)\citenamefont {Dong},
  \citenamefont {Cai}, \citenamefont {Yang}, \citenamefont {Gao},\ and\
  \citenamefont {Ren}}]{DongCYGR2018}%
  \BibitemOpen
  \bibfield  {author} {\bibinfo {author} {\bibfnamefont {R.}~\bibnamefont
  {Dong}}, \bibinfo {author} {\bibfnamefont {Y.}~\bibnamefont {Cai}}, \bibinfo
  {author} {\bibfnamefont {Y.}~\bibnamefont {Yang}}, \bibinfo {author}
  {\bibfnamefont {W.}~\bibnamefont {Gao}}, \ and\ \bibinfo {author}
  {\bibfnamefont {B.}~\bibnamefont {Ren}},\ }\bibfield  {title} {\enquote
  {\bibinfo {title} {Photocatalytic micro/nanomotors: from construction to
  applications},}\ }\href@noop {} {\bibfield  {journal} {\bibinfo  {journal}
  {Accounts of Chemical Research}\ }\textbf {\bibinfo {volume} {51}},\ \bibinfo
  {pages} {1940--1947} (\bibinfo {year} {2018})}\BibitemShut {NoStop}%
\bibitem [{\citenamefont {Venugopalan}\ \emph {et~al.}(2020)\citenamefont
  {Venugopalan}, \citenamefont {{Esteban-Fern{\'a}ndez de {\'A}vila}},
  \citenamefont {Pal}, \citenamefont {Ghosh},\ and\ \citenamefont
  {Wang}}]{VenugopalanEFdAPGW2020}%
  \BibitemOpen
  \bibfield  {author} {\bibinfo {author} {\bibfnamefont {P.~L.}\ \bibnamefont
  {Venugopalan}}, \bibinfo {author} {\bibfnamefont {B.}~\bibnamefont
  {{Esteban-Fern{\'a}ndez de {\'A}vila}}}, \bibinfo {author} {\bibfnamefont
  {M.}~\bibnamefont {Pal}}, \bibinfo {author} {\bibfnamefont {A.}~\bibnamefont
  {Ghosh}}, \ and\ \bibinfo {author} {\bibfnamefont {J.}~\bibnamefont {Wang}},\
  }\bibfield  {title} {\enquote {\bibinfo {title} {Fantastic voyage of
  nanomotors into the cell},}\ }\href@noop {} {\bibfield  {journal} {\bibinfo
  {journal} {ACS Nano}\ }\textbf {\bibinfo {volume} {14}},\ \bibinfo {pages}
  {9423--9439} (\bibinfo {year} {2020})}\BibitemShut {NoStop}%
\bibitem [{\citenamefont {Gao}\ \emph {et~al.}(2021)\citenamefont {Gao},
  \citenamefont {Wang}, \citenamefont {Ye}, \citenamefont {Lin}, \citenamefont
  {Ma},\ and\ \citenamefont {He}}]{GaoWYLMH2021}%
  \BibitemOpen
  \bibfield  {author} {\bibinfo {author} {\bibfnamefont {C.}~\bibnamefont
  {Gao}}, \bibinfo {author} {\bibfnamefont {Y.}~\bibnamefont {Wang}}, \bibinfo
  {author} {\bibfnamefont {Z.}~\bibnamefont {Ye}}, \bibinfo {author}
  {\bibfnamefont {Z.}~\bibnamefont {Lin}}, \bibinfo {author} {\bibfnamefont
  {X.}~\bibnamefont {Ma}}, \ and\ \bibinfo {author} {\bibfnamefont
  {Q.}~\bibnamefont {He}},\ }\bibfield  {title} {\enquote {\bibinfo {title}
  {Biomedical micro-/nanomotors: from overcoming biological barriers to in vivo
  imaging},}\ }\href@noop {} {\bibfield  {journal} {\bibinfo  {journal}
  {Advanced Materials}\ }\textbf {\bibinfo {volume} {33}},\ \bibinfo {pages}
  {2000512} (\bibinfo {year} {2021})}\BibitemShut {NoStop}%
\bibitem [{\citenamefont {Wang}\ \emph {et~al.}(2021)\citenamefont {Wang},
  \citenamefont {Kostarelos}, \citenamefont {Nelson},\ and\ \citenamefont
  {Zhang}}]{WangKNZ2021}%
  \BibitemOpen
  \bibfield  {author} {\bibinfo {author} {\bibfnamefont {B.}~\bibnamefont
  {Wang}}, \bibinfo {author} {\bibfnamefont {K.}~\bibnamefont {Kostarelos}},
  \bibinfo {author} {\bibfnamefont {B.~J.}\ \bibnamefont {Nelson}}, \ and\
  \bibinfo {author} {\bibfnamefont {L.}~\bibnamefont {Zhang}},\ }\bibfield
  {title} {\enquote {\bibinfo {title} {Trends in micro-/nanorobotics: materials
  development, actuation, localization, and system integration for biomedical
  applications},}\ }\href@noop {} {\bibfield  {journal} {\bibinfo  {journal}
  {Advanced Materials}\ }\textbf {\bibinfo {volume} {33}},\ \bibinfo {pages}
  {2002047} (\bibinfo {year} {2021})}\BibitemShut {NoStop}%
\bibitem [{\citenamefont {{Jiang}}\ \emph {et~al.}(2010)\citenamefont
  {{Jiang}}, \citenamefont {{Yoshinaga}},\ and\ \citenamefont
  {{Sano}}}]{JiangYS2010}%
  \BibitemOpen
  \bibfield  {author} {\bibinfo {author} {\bibfnamefont {H.-R.}\ \bibnamefont
  {{Jiang}}}, \bibinfo {author} {\bibfnamefont {N.}~\bibnamefont
  {{Yoshinaga}}}, \ and\ \bibinfo {author} {\bibfnamefont {M.}~\bibnamefont
  {{Sano}}},\ }\bibfield  {title} {\enquote {\bibinfo {title} {Active motion of
  a {J}anus particle by self-thermophoresis in a defocused laser beam},}\
  }\href@noop {} {\bibfield  {journal} {\bibinfo  {journal} {Physical Review
  Letters}\ }\textbf {\bibinfo {volume} {105}},\ \bibinfo {pages} {268302}
  (\bibinfo {year} {2010})}\BibitemShut {NoStop}%
\bibitem [{\citenamefont {Bregulla}\ \emph {et~al.}(2014)\citenamefont
  {Bregulla}, \citenamefont {Yang},\ and\ \citenamefont
  {Cichos}}]{BregullaYC2014}%
  \BibitemOpen
  \bibfield  {author} {\bibinfo {author} {\bibfnamefont {A.~P.}\ \bibnamefont
  {Bregulla}}, \bibinfo {author} {\bibfnamefont {H.}~\bibnamefont {Yang}}, \
  and\ \bibinfo {author} {\bibfnamefont {F.}~\bibnamefont {Cichos}},\
  }\bibfield  {title} {\enquote {\bibinfo {title} {Stochastic localization of
  microswimmers by photon nudging},}\ }\href@noop {} {\bibfield  {journal}
  {\bibinfo  {journal} {ACS Nano}\ }\textbf {\bibinfo {volume} {8}},\ \bibinfo
  {pages} {6542--6550} (\bibinfo {year} {2014})}\BibitemShut {NoStop}%
\bibitem [{\citenamefont {He}\ \emph {et~al.}(2016)\citenamefont {He},
  \citenamefont {Frueh}, \citenamefont {Hu}, \citenamefont {Liu}, \citenamefont
  {Gai},\ and\ \citenamefont {He}}]{HeFHLGH2016}%
  \BibitemOpen
  \bibfield  {author} {\bibinfo {author} {\bibfnamefont {W.}~\bibnamefont
  {He}}, \bibinfo {author} {\bibfnamefont {J.}~\bibnamefont {Frueh}}, \bibinfo
  {author} {\bibfnamefont {N.}~\bibnamefont {Hu}}, \bibinfo {author}
  {\bibfnamefont {L.}~\bibnamefont {Liu}}, \bibinfo {author} {\bibfnamefont
  {M.}~\bibnamefont {Gai}}, \ and\ \bibinfo {author} {\bibfnamefont
  {Q.}~\bibnamefont {He}},\ }\bibfield  {title} {\enquote {\bibinfo {title}
  {Guidable thermophoretic {J}anus micromotors containing gold nanocolorifiers
  for infrared laser assisted tissue welding},}\ }\href@noop {} {\bibfield
  {journal} {\bibinfo  {journal} {Advanced Science}\ }\textbf {\bibinfo
  {volume} {3}},\ \bibinfo {pages} {1600206} (\bibinfo {year}
  {2016})}\BibitemShut {NoStop}%
\bibitem [{\citenamefont {Xuan}\ \emph {et~al.}(2018)\citenamefont {Xuan},
  \citenamefont {Shao}, \citenamefont {Gao}, \citenamefont {Wang},
  \citenamefont {Dai},\ and\ \citenamefont {He}}]{XuanSGWDH2018}%
  \BibitemOpen
  \bibfield  {author} {\bibinfo {author} {\bibfnamefont {M.}~\bibnamefont
  {Xuan}}, \bibinfo {author} {\bibfnamefont {J.}~\bibnamefont {Shao}}, \bibinfo
  {author} {\bibfnamefont {C.}~\bibnamefont {Gao}}, \bibinfo {author}
  {\bibfnamefont {W.}~\bibnamefont {Wang}}, \bibinfo {author} {\bibfnamefont
  {L.}~\bibnamefont {Dai}}, \ and\ \bibinfo {author} {\bibfnamefont
  {Q.}~\bibnamefont {He}},\ }\bibfield  {title} {\enquote {\bibinfo {title}
  {Self-propelled nanomotors for thermomechanically percolating cell
  membranes},}\ }\href@noop {} {\bibfield  {journal} {\bibinfo  {journal}
  {Angewandte Chemie International Edition}\ }\textbf {\bibinfo {volume}
  {57}},\ \bibinfo {pages} {12463--12467} (\bibinfo {year} {2018})}\BibitemShut
  {NoStop}%
\bibitem [{\citenamefont {Xu}\ \emph {et~al.}(2019{\natexlab{a}})\citenamefont
  {Xu}, \citenamefont {Chen}, \citenamefont {Lee}, \citenamefont {Feng},
  \citenamefont {Park}, \citenamefont {Lee},\ and\ \citenamefont
  {Kim}}]{XuCLFPLK2019}%
  \BibitemOpen
  \bibfield  {author} {\bibinfo {author} {\bibfnamefont {Z.}~\bibnamefont
  {Xu}}, \bibinfo {author} {\bibfnamefont {M.}~\bibnamefont {Chen}}, \bibinfo
  {author} {\bibfnamefont {H.}~\bibnamefont {Lee}}, \bibinfo {author}
  {\bibfnamefont {S.-P.}\ \bibnamefont {Feng}}, \bibinfo {author}
  {\bibfnamefont {J.~Y.}\ \bibnamefont {Park}}, \bibinfo {author}
  {\bibfnamefont {S.}~\bibnamefont {Lee}}, \ and\ \bibinfo {author}
  {\bibfnamefont {J.~T.}\ \bibnamefont {Kim}},\ }\bibfield  {title} {\enquote
  {\bibinfo {title} {X-ray-powered micromotors},}\ }\href@noop {} {\bibfield
  {journal} {\bibinfo  {journal} {ACS Applied Materials \& Interfaces}\
  }\textbf {\bibinfo {volume} {11}},\ \bibinfo {pages} {15727--15732} (\bibinfo
  {year} {2019}{\natexlab{a}})}\BibitemShut {NoStop}%
\bibitem [{\citenamefont {Wang}\ \emph {et~al.}(2012)\citenamefont {Wang},
  \citenamefont {Castro}, \citenamefont {Hoyos},\ and\ \citenamefont
  {Mallouk}}]{WangCHM2012}%
  \BibitemOpen
  \bibfield  {author} {\bibinfo {author} {\bibfnamefont {W.}~\bibnamefont
  {Wang}}, \bibinfo {author} {\bibfnamefont {L.}~\bibnamefont {Castro}},
  \bibinfo {author} {\bibfnamefont {M.}~\bibnamefont {Hoyos}}, \ and\ \bibinfo
  {author} {\bibfnamefont {T.~E.}\ \bibnamefont {Mallouk}},\ }\bibfield
  {title} {\enquote {\bibinfo {title} {Autonomous motion of metallic microrods
  propelled by ultrasound},}\ }\href@noop {} {\bibfield  {journal} {\bibinfo
  {journal} {ACS Nano}\ }\textbf {\bibinfo {volume} {6}},\ \bibinfo {pages}
  {6122--6132} (\bibinfo {year} {2012})}\BibitemShut {NoStop}%
\bibitem [{\citenamefont {{Garcia-Gradilla}}\ \emph {et~al.}(2013)\citenamefont
  {{Garcia-Gradilla}}, \citenamefont {Orozco}, \citenamefont
  {Sattayasamitsathit}, \citenamefont {Soto}, \citenamefont {Kuralay},
  \citenamefont {Pourazary}, \citenamefont {Katzenberg}, \citenamefont {Gao},
  \citenamefont {Shen},\ and\ \citenamefont {Wang}}]{GarciaGradillaEtAl2013}%
  \BibitemOpen
  \bibfield  {author} {\bibinfo {author} {\bibfnamefont {V.}~\bibnamefont
  {{Garcia-Gradilla}}}, \bibinfo {author} {\bibfnamefont {J.}~\bibnamefont
  {Orozco}}, \bibinfo {author} {\bibfnamefont {S.}~\bibnamefont
  {Sattayasamitsathit}}, \bibinfo {author} {\bibfnamefont {F.}~\bibnamefont
  {Soto}}, \bibinfo {author} {\bibfnamefont {F.}~\bibnamefont {Kuralay}},
  \bibinfo {author} {\bibfnamefont {A.}~\bibnamefont {Pourazary}}, \bibinfo
  {author} {\bibfnamefont {A.}~\bibnamefont {Katzenberg}}, \bibinfo {author}
  {\bibfnamefont {W.}~\bibnamefont {Gao}}, \bibinfo {author} {\bibfnamefont
  {Y.}~\bibnamefont {Shen}}, \ and\ \bibinfo {author} {\bibfnamefont
  {J.}~\bibnamefont {Wang}},\ }\bibfield  {title} {\enquote {\bibinfo {title}
  {Functionalized ultrasound-propelled magnetically guided nanomotors: toward
  practical biomedical applications},}\ }\href@noop {} {\bibfield  {journal}
  {\bibinfo  {journal} {ACS Nano}\ }\textbf {\bibinfo {volume} {7}},\ \bibinfo
  {pages} {9232--9240} (\bibinfo {year} {2013})}\BibitemShut {NoStop}%
\bibitem [{\citenamefont {Soto}\ \emph {et~al.}(2016)\citenamefont {Soto},
  \citenamefont {Wagner}, \citenamefont {{Garcia-Gradilla}}, \citenamefont
  {Gillespie}, \citenamefont {Lakshmipathy}, \citenamefont {Karshalev},
  \citenamefont {Angell}, \citenamefont {Chen},\ and\ \citenamefont
  {Wang}}]{SotoWGGGLKACW2016}%
  \BibitemOpen
  \bibfield  {author} {\bibinfo {author} {\bibfnamefont {F.}~\bibnamefont
  {Soto}}, \bibinfo {author} {\bibfnamefont {G.~L.}\ \bibnamefont {Wagner}},
  \bibinfo {author} {\bibfnamefont {V.}~\bibnamefont {{Garcia-Gradilla}}},
  \bibinfo {author} {\bibfnamefont {K.~T.}\ \bibnamefont {Gillespie}}, \bibinfo
  {author} {\bibfnamefont {D.~R.}\ \bibnamefont {Lakshmipathy}}, \bibinfo
  {author} {\bibfnamefont {E.}~\bibnamefont {Karshalev}}, \bibinfo {author}
  {\bibfnamefont {C.}~\bibnamefont {Angell}}, \bibinfo {author} {\bibfnamefont
  {Y.}~\bibnamefont {Chen}}, \ and\ \bibinfo {author} {\bibfnamefont
  {J.}~\bibnamefont {Wang}},\ }\bibfield  {title} {\enquote {\bibinfo {title}
  {Acoustically propelled nanoshells},}\ }\href@noop {} {\bibfield  {journal}
  {\bibinfo  {journal} {Nanoscale}\ }\textbf {\bibinfo {volume} {8}},\ \bibinfo
  {pages} {17788--17793} (\bibinfo {year} {2016})}\BibitemShut {NoStop}%
\bibitem [{\citenamefont {{Valdez-Gardu{\~n}o}}\ \emph
  {et~al.}(2020)\citenamefont {{Valdez-Gardu{\~n}o}}, \citenamefont
  {{Leal-Estrada}}, \citenamefont {{Oliveros-Mata}}, \citenamefont
  {{Sandoval-Bojorquez}}, \citenamefont {Soto}, \citenamefont {Wang},\ and\
  \citenamefont {{Garcia-Gradilla}}}]{ValdezGardunoLEOMSBSWGG2020}%
  \BibitemOpen
  \bibfield  {author} {\bibinfo {author} {\bibfnamefont {M.}~\bibnamefont
  {{Valdez-Gardu{\~n}o}}}, \bibinfo {author} {\bibfnamefont {M.}~\bibnamefont
  {{Leal-Estrada}}}, \bibinfo {author} {\bibfnamefont {E.~S.}\ \bibnamefont
  {{Oliveros-Mata}}}, \bibinfo {author} {\bibfnamefont {D.~I.}\ \bibnamefont
  {{Sandoval-Bojorquez}}}, \bibinfo {author} {\bibfnamefont {F.}~\bibnamefont
  {Soto}}, \bibinfo {author} {\bibfnamefont {J.}~\bibnamefont {Wang}}, \ and\
  \bibinfo {author} {\bibfnamefont {V.}~\bibnamefont {{Garcia-Gradilla}}},\
  }\bibfield  {title} {\enquote {\bibinfo {title} {Density asymmetry driven
  propulsion of ultrasound-powered {J}anus micromotor},}\ }\href@noop {}
  {\bibfield  {journal} {\bibinfo  {journal} {Advanced Functional Materials}\
  }\textbf {\bibinfo {volume} {30}},\ \bibinfo {pages} {2004043} (\bibinfo
  {year} {2020})}\BibitemShut {NoStop}%
\bibitem [{\citenamefont {Ahmed}\ \emph {et~al.}(2016)\citenamefont {Ahmed},
  \citenamefont {Baasch}, \citenamefont {Jang}, \citenamefont {Pane},
  \citenamefont {Dual},\ and\ \citenamefont {Nelson}}]{AhmedBJPDN2016}%
  \BibitemOpen
  \bibfield  {author} {\bibinfo {author} {\bibfnamefont {D.}~\bibnamefont
  {Ahmed}}, \bibinfo {author} {\bibfnamefont {T.}~\bibnamefont {Baasch}},
  \bibinfo {author} {\bibfnamefont {B.}~\bibnamefont {Jang}}, \bibinfo {author}
  {\bibfnamefont {S.}~\bibnamefont {Pane}}, \bibinfo {author} {\bibfnamefont
  {J.}~\bibnamefont {Dual}}, \ and\ \bibinfo {author} {\bibfnamefont {B.~J.}\
  \bibnamefont {Nelson}},\ }\bibfield  {title} {\enquote {\bibinfo {title}
  {Artificial swimmers propelled by acoustically activated flagella},}\
  }\href@noop {} {\bibfield  {journal} {\bibinfo  {journal} {Nano Letters}\
  }\textbf {\bibinfo {volume} {16}},\ \bibinfo {pages} {4968--4974} (\bibinfo
  {year} {2016})}\BibitemShut {NoStop}%
\bibitem [{\citenamefont {Volpe}\ \emph {et~al.}(2011)\citenamefont {Volpe},
  \citenamefont {Buttinoni}, \citenamefont {Vogt}, \citenamefont
  {K{\"u}mmerer},\ and\ \citenamefont {Bechinger}}]{VolpeBVKB2011}%
  \BibitemOpen
  \bibfield  {author} {\bibinfo {author} {\bibfnamefont {G.}~\bibnamefont
  {Volpe}}, \bibinfo {author} {\bibfnamefont {I.}~\bibnamefont {Buttinoni}},
  \bibinfo {author} {\bibfnamefont {D.}~\bibnamefont {Vogt}}, \bibinfo {author}
  {\bibfnamefont {H.}~\bibnamefont {K{\"u}mmerer}}, \ and\ \bibinfo {author}
  {\bibfnamefont {C.}~\bibnamefont {Bechinger}},\ }\bibfield  {title} {\enquote
  {\bibinfo {title} {Microswimmers in patterned environments},}\ }\href@noop {}
  {\bibfield  {journal} {\bibinfo  {journal} {Soft Matter}\ }\textbf {\bibinfo
  {volume} {7}},\ \bibinfo {pages} {8810--8815} (\bibinfo {year}
  {2011})}\BibitemShut {NoStop}%
\bibitem [{\citenamefont {{Buttinoni}}\ \emph {et~al.}(2013)\citenamefont
  {{Buttinoni}}, \citenamefont {{Bialk{\'e}}}, \citenamefont {{K{\"u}mmel}},
  \citenamefont {{L{\"o}wen}}, \citenamefont {{Bechinger}},\ and\ \citenamefont
  {{Speck}}}]{ButtinoniBKLBS2013}%
  \BibitemOpen
  \bibfield  {author} {\bibinfo {author} {\bibfnamefont {I.}~\bibnamefont
  {{Buttinoni}}}, \bibinfo {author} {\bibfnamefont {J.}~\bibnamefont
  {{Bialk{\'e}}}}, \bibinfo {author} {\bibfnamefont {F.}~\bibnamefont
  {{K{\"u}mmel}}}, \bibinfo {author} {\bibfnamefont {H.}~\bibnamefont
  {{L{\"o}wen}}}, \bibinfo {author} {\bibfnamefont {C.}~\bibnamefont
  {{Bechinger}}}, \ and\ \bibinfo {author} {\bibfnamefont {T.}~\bibnamefont
  {{Speck}}},\ }\bibfield  {title} {\enquote {\bibinfo {title} {Dynamical
  clustering and phase separation in suspensions of self-propelled colloidal
  particles},}\ }\href@noop {} {\bibfield  {journal} {\bibinfo  {journal}
  {Physical Review Letters}\ }\textbf {\bibinfo {volume} {110}},\ \bibinfo
  {pages} {238301} (\bibinfo {year} {2013})}\BibitemShut {NoStop}%
\bibitem [{\citenamefont {K{\"u}mmel}\ \emph {et~al.}(2013)\citenamefont
  {K{\"u}mmel}, \citenamefont {{ten Hagen}}, \citenamefont {Wittkowski},
  \citenamefont {Buttinoni}, \citenamefont {Eichhorn}, \citenamefont {Volpe},
  \citenamefont {L{\"o}wen},\ and\ \citenamefont
  {Bechinger}}]{KuemmeltHWBEVLB2013}%
  \BibitemOpen
  \bibfield  {author} {\bibinfo {author} {\bibfnamefont {F.}~\bibnamefont
  {K{\"u}mmel}}, \bibinfo {author} {\bibfnamefont {B.}~\bibnamefont {{ten
  Hagen}}}, \bibinfo {author} {\bibfnamefont {R.}~\bibnamefont {Wittkowski}},
  \bibinfo {author} {\bibfnamefont {I.}~\bibnamefont {Buttinoni}}, \bibinfo
  {author} {\bibfnamefont {R.}~\bibnamefont {Eichhorn}}, \bibinfo {author}
  {\bibfnamefont {G.}~\bibnamefont {Volpe}}, \bibinfo {author} {\bibfnamefont
  {H.}~\bibnamefont {L{\"o}wen}}, \ and\ \bibinfo {author} {\bibfnamefont
  {C.}~\bibnamefont {Bechinger}},\ }\bibfield  {title} {\enquote {\bibinfo
  {title} {Circular motion of asymmetric self-propelling particles},}\
  }\href@noop {} {\bibfield  {journal} {\bibinfo  {journal} {Physical Review
  Letters}\ }\textbf {\bibinfo {volume} {110}},\ \bibinfo {pages} {198302}
  (\bibinfo {year} {2013})}\BibitemShut {NoStop}%
\bibitem [{\citenamefont {K{\"u}mmel}\ \emph {et~al.}(2014)\citenamefont
  {K{\"u}mmel}, \citenamefont {{ten Hagen}}, \citenamefont {Wittkowski},
  \citenamefont {Takagi}, \citenamefont {Buttinoni}, \citenamefont {Eichhorn},
  \citenamefont {Volpe}, \citenamefont {L{\"o}wen},\ and\ \citenamefont
  {Bechinger}}]{KuemmeltHWTBEVLB2014}%
  \BibitemOpen
  \bibfield  {author} {\bibinfo {author} {\bibfnamefont {F.}~\bibnamefont
  {K{\"u}mmel}}, \bibinfo {author} {\bibfnamefont {B.}~\bibnamefont {{ten
  Hagen}}}, \bibinfo {author} {\bibfnamefont {R.}~\bibnamefont {Wittkowski}},
  \bibinfo {author} {\bibfnamefont {D.}~\bibnamefont {Takagi}}, \bibinfo
  {author} {\bibfnamefont {I.}~\bibnamefont {Buttinoni}}, \bibinfo {author}
  {\bibfnamefont {R.}~\bibnamefont {Eichhorn}}, \bibinfo {author}
  {\bibfnamefont {G.}~\bibnamefont {Volpe}}, \bibinfo {author} {\bibfnamefont
  {H.}~\bibnamefont {L{\"o}wen}}, \ and\ \bibinfo {author} {\bibfnamefont
  {C.}~\bibnamefont {Bechinger}},\ }\bibfield  {title} {\enquote {\bibinfo
  {title} {Reply to ``{C}omment on `{C}ircular motion of asymmetric
  self-propelling particles'"},}\ }\href@noop {} {\bibfield  {journal}
  {\bibinfo  {journal} {Physical Review Letters}\ }\textbf {\bibinfo {volume}
  {113}},\ \bibinfo {pages} {029802} (\bibinfo {year} {2014})}\BibitemShut
  {NoStop}%
\bibitem [{\citenamefont {{ten Hagen}}\ \emph {et~al.}(2014)\citenamefont {{ten
  Hagen}}, \citenamefont {K{\"u}mmel}, \citenamefont {Wittkowski},
  \citenamefont {Takagi}, \citenamefont {L{\"o}wen},\ and\ \citenamefont
  {Bechinger}}]{tenHagenKWTLB2014}%
  \BibitemOpen
  \bibfield  {author} {\bibinfo {author} {\bibfnamefont {B.}~\bibnamefont {{ten
  Hagen}}}, \bibinfo {author} {\bibfnamefont {F.}~\bibnamefont {K{\"u}mmel}},
  \bibinfo {author} {\bibfnamefont {R.}~\bibnamefont {Wittkowski}}, \bibinfo
  {author} {\bibfnamefont {D.}~\bibnamefont {Takagi}}, \bibinfo {author}
  {\bibfnamefont {H.}~\bibnamefont {L{\"o}wen}}, \ and\ \bibinfo {author}
  {\bibfnamefont {C.}~\bibnamefont {Bechinger}},\ }\bibfield  {title} {\enquote
  {\bibinfo {title} {Gravitaxis of asymmetric self-propelled colloidal
  particles},}\ }\href@noop {} {\bibfield  {journal} {\bibinfo  {journal}
  {Nature Communications}\ }\textbf {\bibinfo {volume} {5}},\ \bibinfo {pages}
  {4829} (\bibinfo {year} {2014})}\BibitemShut {NoStop}%
\bibitem [{\citenamefont {{Nadal}}\ and\ \citenamefont
  {{Lauga}}(2014)}]{NadalL2014}%
  \BibitemOpen
  \bibfield  {author} {\bibinfo {author} {\bibfnamefont {F.}~\bibnamefont
  {{Nadal}}}\ and\ \bibinfo {author} {\bibfnamefont {E.}~\bibnamefont
  {{Lauga}}},\ }\bibfield  {title} {\enquote {\bibinfo {title} {Asymmetric
  steady streaming as a mechanism for acoustic propulsion of rigid bodies},}\
  }\href@noop {} {\bibfield  {journal} {\bibinfo  {journal} {Physics of
  Fluids}\ }\textbf {\bibinfo {volume} {26}},\ \bibinfo {pages} {082001}
  (\bibinfo {year} {2014})}\BibitemShut {NoStop}%
\bibitem [{\citenamefont {Collis}\ \emph {et~al.}(2017)\citenamefont {Collis},
  \citenamefont {Chakraborty},\ and\ \citenamefont {Sader}}]{CollisCS2017}%
  \BibitemOpen
  \bibfield  {author} {\bibinfo {author} {\bibfnamefont {J.~F.}\ \bibnamefont
  {Collis}}, \bibinfo {author} {\bibfnamefont {D.}~\bibnamefont {Chakraborty}},
  \ and\ \bibinfo {author} {\bibfnamefont {J.~E.}\ \bibnamefont {Sader}},\
  }\bibfield  {title} {\enquote {\bibinfo {title} {Autonomous propulsion of
  nanorods trapped in an acoustic field},}\ }\href@noop {} {\bibfield
  {journal} {\bibinfo  {journal} {Journal of Fluid Mechanics}\ }\textbf
  {\bibinfo {volume} {825}},\ \bibinfo {pages} {29--48} (\bibinfo {year}
  {2017})}\BibitemShut {NoStop}%
\bibitem [{\citenamefont {Chen}\ \emph {et~al.}(2019)\citenamefont {Chen},
  \citenamefont {Soto}, \citenamefont {Karshalev}, \citenamefont {Li},\ and\
  \citenamefont {Wang}}]{ChenSKLW2019}%
  \BibitemOpen
  \bibfield  {author} {\bibinfo {author} {\bibfnamefont {C.}~\bibnamefont
  {Chen}}, \bibinfo {author} {\bibfnamefont {F.}~\bibnamefont {Soto}}, \bibinfo
  {author} {\bibfnamefont {E.}~\bibnamefont {Karshalev}}, \bibinfo {author}
  {\bibfnamefont {J.}~\bibnamefont {Li}}, \ and\ \bibinfo {author}
  {\bibfnamefont {J.}~\bibnamefont {Wang}},\ }\bibfield  {title} {\enquote
  {\bibinfo {title} {Hybrid nanovehicles: one machine, two engines},}\
  }\href@noop {} {\bibfield  {journal} {\bibinfo  {journal} {Advanced
  Functional Materials}\ }\textbf {\bibinfo {volume} {29}},\ \bibinfo {pages}
  {1806290} (\bibinfo {year} {2019})}\BibitemShut {NoStop}%
\bibitem [{\citenamefont {Tang}\ \emph {et~al.}(2019)\citenamefont {Tang} \emph
  {et~al.}}]{TangEtAl2019}%
  \BibitemOpen
  \bibfield  {author} {\bibinfo {author} {\bibfnamefont {S.}~\bibnamefont
  {Tang}} \emph {et~al.},\ }\bibfield  {title} {\enquote {\bibinfo {title}
  {Structure-dependent optical modulation of propulsion and collective behavior
  of acoustic/light-driven hybrid microbowls},}\ }\href@noop {} {\bibfield
  {journal} {\bibinfo  {journal} {Advanced Functional Materials}\ }\textbf
  {\bibinfo {volume} {29}},\ \bibinfo {pages} {1809003} (\bibinfo {year}
  {2019})}\BibitemShut {NoStop}%
\bibitem [{\citenamefont {Alapan}\ \emph {et~al.}(2019)\citenamefont {Alapan},
  \citenamefont {Yasa}, \citenamefont {Yigit}, \citenamefont {Yasa},
  \citenamefont {Erkoc},\ and\ \citenamefont {Sitti}}]{AlapanYYYES2019}%
  \BibitemOpen
  \bibfield  {author} {\bibinfo {author} {\bibfnamefont {Y.}~\bibnamefont
  {Alapan}}, \bibinfo {author} {\bibfnamefont {O.}~\bibnamefont {Yasa}},
  \bibinfo {author} {\bibfnamefont {B.}~\bibnamefont {Yigit}}, \bibinfo
  {author} {\bibfnamefont {I.~C.}\ \bibnamefont {Yasa}}, \bibinfo {author}
  {\bibfnamefont {P.}~\bibnamefont {Erkoc}}, \ and\ \bibinfo {author}
  {\bibfnamefont {M.}~\bibnamefont {Sitti}},\ }\bibfield  {title} {\enquote
  {\bibinfo {title} {Microrobotics and microorganisms: biohybrid autonomous
  cellular robots},}\ }\href@noop {} {\bibfield  {journal} {\bibinfo  {journal}
  {Annual Review of Control, Robotics, and Autonomous Systems}\ }\textbf
  {\bibinfo {volume} {2}},\ \bibinfo {pages} {205--230} (\bibinfo {year}
  {2019})}\BibitemShut {NoStop}%
\bibitem [{\citenamefont {Akin}\ \emph {et~al.}(2007)\citenamefont {Akin},
  \citenamefont {Sturgis}, \citenamefont {Ragheb}, \citenamefont {Sherman},
  \citenamefont {Burkholder}, \citenamefont {Robinson}, \citenamefont {Bhunia},
  \citenamefont {Mohammed},\ and\ \citenamefont {Bashir}}]{AkinSRSBRBMB2007}%
  \BibitemOpen
  \bibfield  {author} {\bibinfo {author} {\bibfnamefont {D.}~\bibnamefont
  {Akin}}, \bibinfo {author} {\bibfnamefont {J.}~\bibnamefont {Sturgis}},
  \bibinfo {author} {\bibfnamefont {K.}~\bibnamefont {Ragheb}}, \bibinfo
  {author} {\bibfnamefont {D.}~\bibnamefont {Sherman}}, \bibinfo {author}
  {\bibfnamefont {K.}~\bibnamefont {Burkholder}}, \bibinfo {author}
  {\bibfnamefont {J.~P.}\ \bibnamefont {Robinson}}, \bibinfo {author}
  {\bibfnamefont {A.~K.}\ \bibnamefont {Bhunia}}, \bibinfo {author}
  {\bibfnamefont {S.}~\bibnamefont {Mohammed}}, \ and\ \bibinfo {author}
  {\bibfnamefont {R.}~\bibnamefont {Bashir}},\ }\bibfield  {title} {\enquote
  {\bibinfo {title} {Bacteria-mediated delivery of nanoparticles and cargo into
  cells},}\ }\href@noop {} {\bibfield  {journal} {\bibinfo  {journal} {Nature
  Nanotechnology}\ }\textbf {\bibinfo {volume} {2}},\ \bibinfo {pages}
  {441--449} (\bibinfo {year} {2007})}\BibitemShut {NoStop}%
\bibitem [{\citenamefont {Felfoul}\ \emph {et~al.}(2016)\citenamefont {Felfoul}
  \emph {et~al.}}]{FelfoulEtAl2016}%
  \BibitemOpen
  \bibfield  {author} {\bibinfo {author} {\bibfnamefont {O.}~\bibnamefont
  {Felfoul}} \emph {et~al.},\ }\bibfield  {title} {\enquote {\bibinfo {title}
  {Magneto-aerotactic bacteria deliver drug-containing nanoliposomes to tumour
  hypoxic regions},}\ }\href@noop {} {\bibfield  {journal} {\bibinfo  {journal}
  {Nature Nanotechnology}\ }\textbf {\bibinfo {volume} {11}},\ \bibinfo {pages}
  {941--947} (\bibinfo {year} {2016})}\BibitemShut {NoStop}%
\bibitem [{\citenamefont {Yazdi}\ \emph {et~al.}(2018)\citenamefont {Yazdi},
  \citenamefont {Nosrati}, \citenamefont {Stevens}, \citenamefont {Vogel},
  \citenamefont {Davies},\ and\ \citenamefont {Escobedo}}]{YazdiNSVDE2018}%
  \BibitemOpen
  \bibfield  {author} {\bibinfo {author} {\bibfnamefont {S.~R.}\ \bibnamefont
  {Yazdi}}, \bibinfo {author} {\bibfnamefont {R.}~\bibnamefont {Nosrati}},
  \bibinfo {author} {\bibfnamefont {C.~A.}\ \bibnamefont {Stevens}}, \bibinfo
  {author} {\bibfnamefont {D.}~\bibnamefont {Vogel}}, \bibinfo {author}
  {\bibfnamefont {P.~L.}\ \bibnamefont {Davies}}, \ and\ \bibinfo {author}
  {\bibfnamefont {C.}~\bibnamefont {Escobedo}},\ }\bibfield  {title} {\enquote
  {\bibinfo {title} {Magnetotaxis enables magnetotactic bacteria to navigate in
  flow},}\ }\href@noop {} {\bibfield  {journal} {\bibinfo  {journal} {Small}\
  }\textbf {\bibinfo {volume} {14}},\ \bibinfo {pages} {1702982} (\bibinfo
  {year} {2018})}\BibitemShut {NoStop}%
\bibitem [{\citenamefont {Yan}\ \emph {et~al.}(2017)\citenamefont {Yan} \emph
  {et~al.}}]{YanEtAl2017}%
  \BibitemOpen
  \bibfield  {author} {\bibinfo {author} {\bibfnamefont {X.}~\bibnamefont
  {Yan}} \emph {et~al.},\ }\bibfield  {title} {\enquote {\bibinfo {title}
  {Multifunctional biohybrid magnetite microrobots for imaging-guided
  therapy},}\ }\href@noop {} {\bibfield  {journal} {\bibinfo  {journal}
  {Science Robotics}\ }\textbf {\bibinfo {volume} {2}},\ \bibinfo {pages}
  {eaaq1155} (\bibinfo {year} {2017})}\BibitemShut {NoStop}%
\bibitem [{\citenamefont {Yasa}\ \emph {et~al.}(2018)\citenamefont {Yasa},
  \citenamefont {Erkoc}, \citenamefont {Alapan},\ and\ \citenamefont
  {Sitti}}]{YasaEAS2018}%
  \BibitemOpen
  \bibfield  {author} {\bibinfo {author} {\bibfnamefont {O.}~\bibnamefont
  {Yasa}}, \bibinfo {author} {\bibfnamefont {P.}~\bibnamefont {Erkoc}},
  \bibinfo {author} {\bibfnamefont {Y.}~\bibnamefont {Alapan}}, \ and\ \bibinfo
  {author} {\bibfnamefont {M.}~\bibnamefont {Sitti}},\ }\bibfield  {title}
  {\enquote {\bibinfo {title} {Microalga-powered microswimmers toward active
  cargo delivery},}\ }\href@noop {} {\bibfield  {journal} {\bibinfo  {journal}
  {Advanced Materials}\ }\textbf {\bibinfo {volume} {30}},\ \bibinfo {pages}
  {1804130} (\bibinfo {year} {2018})}\BibitemShut {NoStop}%
\bibitem [{\citenamefont {Fodor}\ \emph {et~al.}(2016)\citenamefont {Fodor},
  \citenamefont {Nardini}, \citenamefont {Cates}, \citenamefont {Tailleur},
  \citenamefont {Visco},\ and\ \citenamefont {{van
  Wijland}}}]{FodorNCTVvW2016}%
  \BibitemOpen
  \bibfield  {author} {\bibinfo {author} {\bibfnamefont {E.}~\bibnamefont
  {Fodor}}, \bibinfo {author} {\bibfnamefont {C.}~\bibnamefont {Nardini}},
  \bibinfo {author} {\bibfnamefont {M.~E.}\ \bibnamefont {Cates}}, \bibinfo
  {author} {\bibfnamefont {J.}~\bibnamefont {Tailleur}}, \bibinfo {author}
  {\bibfnamefont {P.}~\bibnamefont {Visco}}, \ and\ \bibinfo {author}
  {\bibfnamefont {F.}~\bibnamefont {{van Wijland}}},\ }\bibfield  {title}
  {\enquote {\bibinfo {title} {How far from equilibrium is active matter},}\
  }\href@noop {} {\bibfield  {journal} {\bibinfo  {journal} {Physical Review
  Letters}\ }\textbf {\bibinfo {volume} {117}},\ \bibinfo {pages} {038103}
  (\bibinfo {year} {2016})}\BibitemShut {NoStop}%
\bibitem [{\citenamefont {Guix}\ \emph {et~al.}(2012)\citenamefont {Guix},
  \citenamefont {Orozco}, \citenamefont {Garc{\'\i}a}, \citenamefont {Gao},
  \citenamefont {Sattayasamitsathit}, \citenamefont {Merko{\c{c}}i},
  \citenamefont {Escarpa},\ and\ \citenamefont {Wang}}]{GuixOGGSMEW2012}%
  \BibitemOpen
  \bibfield  {author} {\bibinfo {author} {\bibfnamefont {M.}~\bibnamefont
  {Guix}}, \bibinfo {author} {\bibfnamefont {J.}~\bibnamefont {Orozco}},
  \bibinfo {author} {\bibfnamefont {M.}~\bibnamefont {Garc{\'\i}a}}, \bibinfo
  {author} {\bibfnamefont {W.}~\bibnamefont {Gao}}, \bibinfo {author}
  {\bibfnamefont {S.}~\bibnamefont {Sattayasamitsathit}}, \bibinfo {author}
  {\bibfnamefont {A.}~\bibnamefont {Merko{\c{c}}i}}, \bibinfo {author}
  {\bibfnamefont {A.}~\bibnamefont {Escarpa}}, \ and\ \bibinfo {author}
  {\bibfnamefont {J.}~\bibnamefont {Wang}},\ }\bibfield  {title} {\enquote
  {\bibinfo {title} {Superhydrophobic alkanethiol-coated microsubmarines for
  effective removal of oil},}\ }\href@noop {} {\bibfield  {journal} {\bibinfo
  {journal} {ACS Nano}\ }\textbf {\bibinfo {volume} {6}},\ \bibinfo {pages}
  {4445--4451} (\bibinfo {year} {2012})}\BibitemShut {NoStop}%
\bibitem [{\citenamefont {Sridhar}\ \emph {et~al.}(2018)\citenamefont
  {Sridhar}, \citenamefont {Park},\ and\ \citenamefont
  {Sitti}}]{SridharPS2018}%
  \BibitemOpen
  \bibfield  {author} {\bibinfo {author} {\bibfnamefont {V.}~\bibnamefont
  {Sridhar}}, \bibinfo {author} {\bibfnamefont {B.-W.}\ \bibnamefont {Park}}, \
  and\ \bibinfo {author} {\bibfnamefont {M.}~\bibnamefont {Sitti}},\ }\bibfield
   {title} {\enquote {\bibinfo {title} {Light-driven {J}anus hollow mesoporous
  {T}i{O}2-{A}u microswimmers},}\ }\href@noop {} {\bibfield  {journal}
  {\bibinfo  {journal} {Advanced Functional Materials}\ }\textbf {\bibinfo
  {volume} {28}},\ \bibinfo {pages} {1704902} (\bibinfo {year}
  {2018})}\BibitemShut {NoStop}%
\bibitem [{\citenamefont {Wang}\ \emph {et~al.}(2019)\citenamefont {Wang},
  \citenamefont {Hortel{\~a}o}, \citenamefont {Huang},\ and\ \citenamefont
  {S{\'a}nchez}}]{WangHHS2019}%
  \BibitemOpen
  \bibfield  {author} {\bibinfo {author} {\bibfnamefont {L.}~\bibnamefont
  {Wang}}, \bibinfo {author} {\bibfnamefont {A.~C.}\ \bibnamefont
  {Hortel{\~a}o}}, \bibinfo {author} {\bibfnamefont {X.}~\bibnamefont {Huang}},
  \ and\ \bibinfo {author} {\bibfnamefont {S.}~\bibnamefont {S{\'a}nchez}},\
  }\bibfield  {title} {\enquote {\bibinfo {title} {Lipase-powered mesoporous
  silica nanomotors for triglyceride degradation},}\ }\href@noop {} {\bibfield
  {journal} {\bibinfo  {journal} {Angewandte Chemie International Edition}\
  }\textbf {\bibinfo {volume} {58}},\ \bibinfo {pages} {7992--7996} (\bibinfo
  {year} {2019})}\BibitemShut {NoStop}%
\bibitem [{\citenamefont {Li}\ \emph {et~al.}(2014)\citenamefont {Li},
  \citenamefont {Gao}, \citenamefont {Dong}, \citenamefont {Pei}, \citenamefont
  {Sattayasamitsathit},\ and\ \citenamefont {Wang}}]{LiGDPSW2014}%
  \BibitemOpen
  \bibfield  {author} {\bibinfo {author} {\bibfnamefont {J.}~\bibnamefont
  {Li}}, \bibinfo {author} {\bibfnamefont {W.}~\bibnamefont {Gao}}, \bibinfo
  {author} {\bibfnamefont {R.}~\bibnamefont {Dong}}, \bibinfo {author}
  {\bibfnamefont {A.}~\bibnamefont {Pei}}, \bibinfo {author} {\bibfnamefont
  {S.}~\bibnamefont {Sattayasamitsathit}}, \ and\ \bibinfo {author}
  {\bibfnamefont {J.}~\bibnamefont {Wang}},\ }\bibfield  {title} {\enquote
  {\bibinfo {title} {Nanomotor lithography},}\ }\href@noop {} {\bibfield
  {journal} {\bibinfo  {journal} {Nature Communications}\ }\textbf {\bibinfo
  {volume} {5}},\ \bibinfo {pages} {5026} (\bibinfo {year} {2014})}\BibitemShut
  {NoStop}%
\bibitem [{\citenamefont {Tasoglu}\ \emph {et~al.}(2014)\citenamefont
  {Tasoglu}, \citenamefont {Diller}, \citenamefont {Guven}, \citenamefont
  {Sitti},\ and\ \citenamefont {Demirci}}]{TasogluDGSD2014}%
  \BibitemOpen
  \bibfield  {author} {\bibinfo {author} {\bibfnamefont {S.}~\bibnamefont
  {Tasoglu}}, \bibinfo {author} {\bibfnamefont {E.}~\bibnamefont {Diller}},
  \bibinfo {author} {\bibfnamefont {S.}~\bibnamefont {Guven}}, \bibinfo
  {author} {\bibfnamefont {M.}~\bibnamefont {Sitti}}, \ and\ \bibinfo {author}
  {\bibfnamefont {U.}~\bibnamefont {Demirci}},\ }\bibfield  {title} {\enquote
  {\bibinfo {title} {Untethered micro-robotic coding of three-dimensional
  material composition},}\ }\href@noop {} {\bibfield  {journal} {\bibinfo
  {journal} {Nature Communications}\ }\textbf {\bibinfo {volume} {5}},\
  \bibinfo {pages} {3124} (\bibinfo {year} {2014})}\BibitemShut {NoStop}%
\bibitem [{\citenamefont {{Marchetti}}(2015)}]{Marchetti2015}%
  \BibitemOpen
  \bibfield  {author} {\bibinfo {author} {\bibfnamefont {M.~C.}\ \bibnamefont
  {{Marchetti}}},\ }\bibfield  {title} {\enquote {\bibinfo {title}
  {Frictionless fluids from bacterial teamwork},}\ }\href@noop {} {\bibfield
  {journal} {\bibinfo  {journal} {Nature}\ }\textbf {\bibinfo {volume} {525}},\
  \bibinfo {pages} {37--39} (\bibinfo {year} {2015})}\BibitemShut {NoStop}%
\bibitem [{\citenamefont {Saintillan}(2018)}]{Saintillan2018}%
  \BibitemOpen
  \bibfield  {author} {\bibinfo {author} {\bibfnamefont {D.}~\bibnamefont
  {Saintillan}},\ }\bibfield  {title} {\enquote {\bibinfo {title} {Rheology of
  active fluids},}\ }\href@noop {} {\bibfield  {journal} {\bibinfo  {journal}
  {Annual Review of Fluid Mechanics}\ }\textbf {\bibinfo {volume} {50}},\
  \bibinfo {pages} {563--592} (\bibinfo {year} {2018})}\BibitemShut {NoStop}%
\bibitem [{\citenamefont {Yan}\ \emph {et~al.}(2019)\citenamefont {Yan},
  \citenamefont {Xu}, \citenamefont {Zhou}, \citenamefont {Jin}, \citenamefont
  {Vong}, \citenamefont {Feng}, \citenamefont {Ng}, \citenamefont {Bian},\ and\
  \citenamefont {Zhang}}]{YanXZJVFNBZ2019}%
  \BibitemOpen
  \bibfield  {author} {\bibinfo {author} {\bibfnamefont {X.}~\bibnamefont
  {Yan}}, \bibinfo {author} {\bibfnamefont {J.}~\bibnamefont {Xu}}, \bibinfo
  {author} {\bibfnamefont {Q.}~\bibnamefont {Zhou}}, \bibinfo {author}
  {\bibfnamefont {D.}~\bibnamefont {Jin}}, \bibinfo {author} {\bibfnamefont
  {C.~I.}\ \bibnamefont {Vong}}, \bibinfo {author} {\bibfnamefont
  {Q.}~\bibnamefont {Feng}}, \bibinfo {author} {\bibfnamefont {D.~H.~L.}\
  \bibnamefont {Ng}}, \bibinfo {author} {\bibfnamefont {L.}~\bibnamefont
  {Bian}}, \ and\ \bibinfo {author} {\bibfnamefont {L.}~\bibnamefont {Zhang}},\
  }\bibfield  {title} {\enquote {\bibinfo {title} {Molecular cargo delivery
  using multicellular magnetic microswimmers},}\ }\href@noop {} {\bibfield
  {journal} {\bibinfo  {journal} {Applied Materials Today}\ }\textbf {\bibinfo
  {volume} {15}},\ \bibinfo {pages} {242--251} (\bibinfo {year}
  {2019})}\BibitemShut {NoStop}%
\bibitem [{\citenamefont {{Medina-S{\'a}nchez}}\ and\ \citenamefont
  {Schmidt}(2017)}]{MedinaSanchezS2017}%
  \BibitemOpen
  \bibfield  {author} {\bibinfo {author} {\bibfnamefont {M.}~\bibnamefont
  {{Medina-S{\'a}nchez}}}\ and\ \bibinfo {author} {\bibfnamefont {O.~G.}\
  \bibnamefont {Schmidt}},\ }\bibfield  {title} {\enquote {\bibinfo {title}
  {Medical microbots need better imaging and control},}\ }\href@noop {}
  {\bibfield  {journal} {\bibinfo  {journal} {Nature}\ }\textbf {\bibinfo
  {volume} {545}},\ \bibinfo {pages} {406--408} (\bibinfo {year}
  {2017})}\BibitemShut {NoStop}%
\bibitem [{\citenamefont {Kagan}\ \emph {et~al.}(2010)\citenamefont {Kagan}
  \emph {et~al.}}]{KaganEtAl2010}%
  \BibitemOpen
  \bibfield  {author} {\bibinfo {author} {\bibfnamefont {D.}~\bibnamefont
  {Kagan}} \emph {et~al.},\ }\bibfield  {title} {\enquote {\bibinfo {title}
  {Rapid delivery of drug carriers propelled and navigated by catalytic
  nanoshuttles},}\ }\href@noop {} {\bibfield  {journal} {\bibinfo  {journal}
  {Small}\ }\textbf {\bibinfo {volume} {6}},\ \bibinfo {pages} {2741--2747}
  (\bibinfo {year} {2010})}\BibitemShut {NoStop}%
\bibitem [{\citenamefont {{Esteban-Fern{\'a}ndez de {\'A}vila}}\ \emph
  {et~al.}(2017)\citenamefont {{Esteban-Fern{\'a}ndez de {\'A}vila}} \emph
  {et~al.}}]{EstebanFernandezdeAvilaEtAl2017}%
  \BibitemOpen
  \bibfield  {author} {\bibinfo {author} {\bibfnamefont {B.}~\bibnamefont
  {{Esteban-Fern{\'a}ndez de {\'A}vila}}} \emph {et~al.},\ }\bibfield  {title}
  {\enquote {\bibinfo {title} {Micromotor-enabled active drug delivery for in
  vivo treatment of stomach infection},}\ }\href@noop {} {\bibfield  {journal}
  {\bibinfo  {journal} {Nature Communications}\ }\textbf {\bibinfo {volume}
  {8}},\ \bibinfo {pages} {272} (\bibinfo {year} {2017})}\BibitemShut {NoStop}%
\bibitem [{\citenamefont {{Esteban-Fern{\'a}ndez de {\'A}vila}}\ \emph
  {et~al.}(2015)\citenamefont {{Esteban-Fern{\'a}ndez de {\'A}vila}},
  \citenamefont {Mart{\'\i}n}, \citenamefont {Soto}, \citenamefont
  {{Lopez-Ramirez}}, \citenamefont {Campuzano}, \citenamefont
  {{V{\'a}squez-Machado}}, \citenamefont {Gao}, \citenamefont {Zhang},\ and\
  \citenamefont {Wang}}]{EstebanFernandezdeAvilaMSLRCVMGZW2015}%
  \BibitemOpen
  \bibfield  {author} {\bibinfo {author} {\bibfnamefont {B.}~\bibnamefont
  {{Esteban-Fern{\'a}ndez de {\'A}vila}}}, \bibinfo {author} {\bibfnamefont
  {A.}~\bibnamefont {Mart{\'\i}n}}, \bibinfo {author} {\bibfnamefont
  {F.}~\bibnamefont {Soto}}, \bibinfo {author} {\bibfnamefont {M.~A.}\
  \bibnamefont {{Lopez-Ramirez}}}, \bibinfo {author} {\bibfnamefont
  {S.}~\bibnamefont {Campuzano}}, \bibinfo {author} {\bibfnamefont {G.~M.}\
  \bibnamefont {{V{\'a}squez-Machado}}}, \bibinfo {author} {\bibfnamefont
  {W.}~\bibnamefont {Gao}}, \bibinfo {author} {\bibfnamefont {L.}~\bibnamefont
  {Zhang}}, \ and\ \bibinfo {author} {\bibfnamefont {J.}~\bibnamefont {Wang}},\
  }\bibfield  {title} {\enquote {\bibinfo {title} {Single cell real-time
  mi{RNA}s sensing based on nanomotors},}\ }\href@noop {} {\bibfield  {journal}
  {\bibinfo  {journal} {ACS Nano}\ }\textbf {\bibinfo {volume} {9}},\ \bibinfo
  {pages} {6756--6764} (\bibinfo {year} {2015})}\BibitemShut {NoStop}%
\bibitem [{\citenamefont {Zhang}\ \emph {et~al.}(2019)\citenamefont {Zhang}
  \emph {et~al.}}]{ZhangEtAl2019}%
  \BibitemOpen
  \bibfield  {author} {\bibinfo {author} {\bibfnamefont {F.}~\bibnamefont
  {Zhang}} \emph {et~al.},\ }\bibfield  {title} {\enquote {\bibinfo {title} {A
  macrophage-magnesium hybrid biomotor: fabrication and characterization},}\
  }\href@noop {} {\bibfield  {journal} {\bibinfo  {journal} {Advanced
  Materials}\ }\textbf {\bibinfo {volume} {31}},\ \bibinfo {pages} {1901828}
  (\bibinfo {year} {2019})}\BibitemShut {NoStop}%
\bibitem [{\citenamefont {Hortel{\~a}o}\ \emph {et~al.}(2018)\citenamefont
  {Hortel{\~a}o}, \citenamefont {Pati{\~n}o}, \citenamefont
  {{Perez-Jim{\'e}nez}}, \citenamefont {Blanco},\ and\ \citenamefont
  {S{\'a}nchez}}]{HortelaoPPJBS2018}%
  \BibitemOpen
  \bibfield  {author} {\bibinfo {author} {\bibfnamefont {A.~C.}\ \bibnamefont
  {Hortel{\~a}o}}, \bibinfo {author} {\bibfnamefont {T.}~\bibnamefont
  {Pati{\~n}o}}, \bibinfo {author} {\bibfnamefont {A.}~\bibnamefont
  {{Perez-Jim{\'e}nez}}}, \bibinfo {author} {\bibfnamefont
  {{\`A}.}~\bibnamefont {Blanco}}, \ and\ \bibinfo {author} {\bibfnamefont
  {S.}~\bibnamefont {S{\'a}nchez}},\ }\bibfield  {title} {\enquote {\bibinfo
  {title} {Enzyme-powered nanobots enhance anticancer drug delivery},}\
  }\href@noop {} {\bibfield  {journal} {\bibinfo  {journal} {Advanced
  Functional Materials}\ }\textbf {\bibinfo {volume} {28}},\ \bibinfo {pages}
  {1705086} (\bibinfo {year} {2018})}\BibitemShut {NoStop}%
\bibitem [{\citenamefont {Balasubramanian}\ \emph {et~al.}(2011)\citenamefont
  {Balasubramanian} \emph {et~al.}}]{BalasubramanianEtAl2011}%
  \BibitemOpen
  \bibfield  {author} {\bibinfo {author} {\bibfnamefont {S.}~\bibnamefont
  {Balasubramanian}} \emph {et~al.},\ }\bibfield  {title} {\enquote {\bibinfo
  {title} {Micromachine-enabled capture and isolation of cancer cells in
  complex media},}\ }\href@noop {} {\bibfield  {journal} {\bibinfo  {journal}
  {Angewandte Chemie International Edition}\ }\textbf {\bibinfo {volume}
  {50}},\ \bibinfo {pages} {4161--4164} (\bibinfo {year} {2011})}\BibitemShut
  {NoStop}%
\bibitem [{\citenamefont {Gao}\ \emph {et~al.}(2018{\natexlab{b}})\citenamefont
  {Gao}, \citenamefont {{Esteban-Fern{\'a}ndez de {\'A}vila}}, \citenamefont
  {Zhang},\ and\ \citenamefont {Wang}}]{GaoEFdAZW2018}%
  \BibitemOpen
  \bibfield  {author} {\bibinfo {author} {\bibfnamefont {W.}~\bibnamefont
  {Gao}}, \bibinfo {author} {\bibfnamefont {B.}~\bibnamefont
  {{Esteban-Fern{\'a}ndez de {\'A}vila}}}, \bibinfo {author} {\bibfnamefont
  {L.}~\bibnamefont {Zhang}}, \ and\ \bibinfo {author} {\bibfnamefont
  {J.}~\bibnamefont {Wang}},\ }\bibfield  {title} {\enquote {\bibinfo {title}
  {Targeting and isolation of cancer cells using micro/nanomotors},}\
  }\href@noop {} {\bibfield  {journal} {\bibinfo  {journal} {Advanced Drug
  Delivery Reviews}\ }\textbf {\bibinfo {volume} {125}},\ \bibinfo {pages}
  {94--101} (\bibinfo {year} {2018}{\natexlab{b}})}\BibitemShut {NoStop}%
\bibitem [{\citenamefont {Hu}\ \emph {et~al.}(2018)\citenamefont {Hu},
  \citenamefont {Huang}, \citenamefont {Zhu}, \citenamefont {Huang},
  \citenamefont {Zhao}, \citenamefont {Jin},\ and\ \citenamefont
  {ZhuGe}}]{HuHZHZJZ2018}%
  \BibitemOpen
  \bibfield  {author} {\bibinfo {author} {\bibfnamefont {J.}~\bibnamefont
  {Hu}}, \bibinfo {author} {\bibfnamefont {S.}~\bibnamefont {Huang}}, \bibinfo
  {author} {\bibfnamefont {L.}~\bibnamefont {Zhu}}, \bibinfo {author}
  {\bibfnamefont {W.}~\bibnamefont {Huang}}, \bibinfo {author} {\bibfnamefont
  {Y.}~\bibnamefont {Zhao}}, \bibinfo {author} {\bibfnamefont {K.}~\bibnamefont
  {Jin}}, \ and\ \bibinfo {author} {\bibfnamefont {Q.}~\bibnamefont {ZhuGe}},\
  }\bibfield  {title} {\enquote {\bibinfo {title} {Tissue plasminogen
  activator-porous magnetic microrods for targeted thrombolytic therapy after
  ischemic stroke},}\ }\href@noop {} {\bibfield  {journal} {\bibinfo  {journal}
  {ACS Applied Materials \& Interfaces}\ }\textbf {\bibinfo {volume} {10}},\
  \bibinfo {pages} {32988--32997} (\bibinfo {year} {2018})}\BibitemShut
  {NoStop}%
\bibitem [{\citenamefont {Baylis}\ \emph {et~al.}(2015)\citenamefont {Baylis},
  \citenamefont {Yeon}, \citenamefont {Thomson}, \citenamefont {Kazerooni},
  \citenamefont {Wang}, \citenamefont {{St. John}}, \citenamefont {Lim},
  \citenamefont {Chien}, \citenamefont {Lee}, \citenamefont {Zhang},
  \citenamefont {Piret}, \citenamefont {Machan}, \citenamefont {Burke},
  \citenamefont {White},\ and\ \citenamefont {Kastrup}}]{BaylisEtAl2015}%
  \BibitemOpen
  \bibfield  {author} {\bibinfo {author} {\bibfnamefont {J.~R.}\ \bibnamefont
  {Baylis}}, \bibinfo {author} {\bibfnamefont {J.~H.}\ \bibnamefont {Yeon}},
  \bibinfo {author} {\bibfnamefont {M.~H.}\ \bibnamefont {Thomson}}, \bibinfo
  {author} {\bibfnamefont {A.}~\bibnamefont {Kazerooni}}, \bibinfo {author}
  {\bibfnamefont {X.}~\bibnamefont {Wang}}, \bibinfo {author} {\bibfnamefont
  {A.~E.}\ \bibnamefont {{St. John}}}, \bibinfo {author} {\bibfnamefont
  {E.~B.}\ \bibnamefont {Lim}}, \bibinfo {author} {\bibfnamefont
  {D.}~\bibnamefont {Chien}}, \bibinfo {author} {\bibfnamefont
  {A.}~\bibnamefont {Lee}}, \bibinfo {author} {\bibfnamefont {J.~Q.}\
  \bibnamefont {Zhang}}, \bibinfo {author} {\bibfnamefont {J.~M.}\ \bibnamefont
  {Piret}}, \bibinfo {author} {\bibfnamefont {L.~S.}\ \bibnamefont {Machan}},
  \bibinfo {author} {\bibfnamefont {T.~F.}\ \bibnamefont {Burke}}, \bibinfo
  {author} {\bibfnamefont {N.~J.}\ \bibnamefont {White}}, \ and\ \bibinfo
  {author} {\bibfnamefont {C.~J.}\ \bibnamefont {Kastrup}},\ }\bibfield
  {title} {\enquote {\bibinfo {title} {Self-propelled particles that transport
  cargo through flowing blood and halt hemorrhage},}\ }\href@noop {} {\bibfield
   {journal} {\bibinfo  {journal} {Science Advances}\ }\textbf {\bibinfo
  {volume} {1}},\ \bibinfo {pages} {e1500379} (\bibinfo {year}
  {2015})}\BibitemShut {NoStop}%
\bibitem [{\citenamefont {Gao}\ \emph {et~al.}(2015)\citenamefont {Gao},
  \citenamefont {Dong}, \citenamefont {Thamphiwatana}, \citenamefont {Li},
  \citenamefont {Gao}, \citenamefont {Zhang},\ and\ \citenamefont
  {Wang}}]{GaoDTLGZW2015}%
  \BibitemOpen
  \bibfield  {author} {\bibinfo {author} {\bibfnamefont {W.}~\bibnamefont
  {Gao}}, \bibinfo {author} {\bibfnamefont {R.}~\bibnamefont {Dong}}, \bibinfo
  {author} {\bibfnamefont {S.}~\bibnamefont {Thamphiwatana}}, \bibinfo {author}
  {\bibfnamefont {J.}~\bibnamefont {Li}}, \bibinfo {author} {\bibfnamefont
  {W.}~\bibnamefont {Gao}}, \bibinfo {author} {\bibfnamefont {L.}~\bibnamefont
  {Zhang}}, \ and\ \bibinfo {author} {\bibfnamefont {J.}~\bibnamefont {Wang}},\
  }\bibfield  {title} {\enquote {\bibinfo {title} {Artificial micromotors in
  the mouse's stomach: a step toward in vivo use of synthetic motors},}\
  }\href@noop {} {\bibfield  {journal} {\bibinfo  {journal} {ACS Nano}\
  }\textbf {\bibinfo {volume} {9}},\ \bibinfo {pages} {117--123} (\bibinfo
  {year} {2015})}\BibitemShut {NoStop}%
\bibitem [{\citenamefont {Wu}\ \emph {et~al.}(2018)\citenamefont {Wu} \emph
  {et~al.}}]{WuEtAl2018}%
  \BibitemOpen
  \bibfield  {author} {\bibinfo {author} {\bibfnamefont {Z.}~\bibnamefont {Wu}}
  \emph {et~al.},\ }\bibfield  {title} {\enquote {\bibinfo {title} {A swarm of
  slippery micropropellers penetrates the vitreous body of the eye},}\
  }\href@noop {} {\bibfield  {journal} {\bibinfo  {journal} {Science Advances}\
  }\textbf {\bibinfo {volume} {4}},\ \bibinfo {pages} {eaat4388} (\bibinfo
  {year} {2018})}\BibitemShut {NoStop}%
\bibitem [{\citenamefont {Wei}\ \emph {et~al.}(2019)\citenamefont {Wei} \emph
  {et~al.}}]{WeiEtAl2019}%
  \BibitemOpen
  \bibfield  {author} {\bibinfo {author} {\bibfnamefont {X.}~\bibnamefont
  {Wei}} \emph {et~al.},\ }\bibfield  {title} {\enquote {\bibinfo {title}
  {Biomimetic micromotor enables active delivery of antigens for oral
  vaccination},}\ }\href@noop {} {\bibfield  {journal} {\bibinfo  {journal}
  {Nano Letters}\ }\textbf {\bibinfo {volume} {19}},\ \bibinfo {pages}
  {1914--1921} (\bibinfo {year} {2019})}\BibitemShut {NoStop}%
\bibitem [{\citenamefont {Gao}\ \emph {et~al.}(2012)\citenamefont {Gao} \emph
  {et~al.}}]{GaoEtAl2012}%
  \BibitemOpen
  \bibfield  {author} {\bibinfo {author} {\bibfnamefont {W.}~\bibnamefont
  {Gao}} \emph {et~al.},\ }\bibfield  {title} {\enquote {\bibinfo {title}
  {Cargo-towing fuel-free magnetic nanoswimmers for targeted drug delivery},}\
  }\href@noop {} {\bibfield  {journal} {\bibinfo  {journal} {Small}\ }\textbf
  {\bibinfo {volume} {8}},\ \bibinfo {pages} {460--467} (\bibinfo {year}
  {2012})}\BibitemShut {NoStop}%
\bibitem [{\citenamefont {Kagan}\ \emph {et~al.}(2012)\citenamefont {Kagan},
  \citenamefont {Benchimol}, \citenamefont {Claussen}, \citenamefont
  {{Chuluun-Erdene}}, \citenamefont {Esener},\ and\ \citenamefont
  {Wang}}]{KaganBCCEEW2012}%
  \BibitemOpen
  \bibfield  {author} {\bibinfo {author} {\bibfnamefont {D.}~\bibnamefont
  {Kagan}}, \bibinfo {author} {\bibfnamefont {M.~J.}\ \bibnamefont
  {Benchimol}}, \bibinfo {author} {\bibfnamefont {J.~C.}\ \bibnamefont
  {Claussen}}, \bibinfo {author} {\bibfnamefont {E.}~\bibnamefont
  {{Chuluun-Erdene}}}, \bibinfo {author} {\bibfnamefont {S.}~\bibnamefont
  {Esener}}, \ and\ \bibinfo {author} {\bibfnamefont {J.}~\bibnamefont
  {Wang}},\ }\bibfield  {title} {\enquote {\bibinfo {title} {Acoustic droplet
  vaporization and propulsion of perfluorocarbon-loaded microbullets for
  targeted tissue penetration and deformation},}\ }\href@noop {} {\bibfield
  {journal} {\bibinfo  {journal} {Angewandte Chemie International Edition}\
  }\textbf {\bibinfo {volume} {51}},\ \bibinfo {pages} {7519--7522} (\bibinfo
  {year} {2012})}\BibitemShut {NoStop}%
\bibitem [{\citenamefont {Hoop}\ \emph {et~al.}(2018)\citenamefont {Hoop} \emph
  {et~al.}}]{HoopEtAl2018}%
  \BibitemOpen
  \bibfield  {author} {\bibinfo {author} {\bibfnamefont {M.}~\bibnamefont
  {Hoop}} \emph {et~al.},\ }\bibfield  {title} {\enquote {\bibinfo {title}
  {Mobile magnetic nanocatalysts for bioorthogonal targeted cancer therapy},}\
  }\href@noop {} {\bibfield  {journal} {\bibinfo  {journal} {Advanced
  Functional Materials}\ }\textbf {\bibinfo {volume} {28}},\ \bibinfo {pages}
  {1705920} (\bibinfo {year} {2018})}\BibitemShut {NoStop}%
\bibitem [{\citenamefont {Zhou}\ \emph {et~al.}(2019)\citenamefont {Zhou},
  \citenamefont {Hou}, \citenamefont {Li}, \citenamefont {Yu}, \citenamefont
  {Xu}, \citenamefont {Yin}, \citenamefont {Wang},\ and\ \citenamefont
  {Wang}}]{ZhouHLYXYWW2019}%
  \BibitemOpen
  \bibfield  {author} {\bibinfo {author} {\bibfnamefont {M.}~\bibnamefont
  {Zhou}}, \bibinfo {author} {\bibfnamefont {T.}~\bibnamefont {Hou}}, \bibinfo
  {author} {\bibfnamefont {J.}~\bibnamefont {Li}}, \bibinfo {author}
  {\bibfnamefont {S.}~\bibnamefont {Yu}}, \bibinfo {author} {\bibfnamefont
  {Z.}~\bibnamefont {Xu}}, \bibinfo {author} {\bibfnamefont {M.}~\bibnamefont
  {Yin}}, \bibinfo {author} {\bibfnamefont {J.}~\bibnamefont {Wang}}, \ and\
  \bibinfo {author} {\bibfnamefont {X.}~\bibnamefont {Wang}},\ }\bibfield
  {title} {\enquote {\bibinfo {title} {Self-propelled and targeted drug
  delivery of poly(aspartic acid)/iron-zinc microrocket in the stomach},}\
  }\href@noop {} {\bibfield  {journal} {\bibinfo  {journal} {ACS Nano}\
  }\textbf {\bibinfo {volume} {13}},\ \bibinfo {pages} {1324--1332} (\bibinfo
  {year} {2019})}\BibitemShut {NoStop}%
\bibitem [{\citenamefont {Hortel{\~a}o}\ \emph {et~al.}(2019)\citenamefont
  {Hortel{\~a}o}, \citenamefont {Carrascosa}, \citenamefont
  {{Murillo-Cremaes}}, \citenamefont {Pati{\~n}o},\ and\ \citenamefont
  {S{\'a}nchez}}]{HortelaoCMCPS2019}%
  \BibitemOpen
  \bibfield  {author} {\bibinfo {author} {\bibfnamefont {A.~C.}\ \bibnamefont
  {Hortel{\~a}o}}, \bibinfo {author} {\bibfnamefont {R.}~\bibnamefont
  {Carrascosa}}, \bibinfo {author} {\bibfnamefont {N.}~\bibnamefont
  {{Murillo-Cremaes}}}, \bibinfo {author} {\bibfnamefont {T.}~\bibnamefont
  {Pati{\~n}o}}, \ and\ \bibinfo {author} {\bibfnamefont {S.}~\bibnamefont
  {S{\'a}nchez}},\ }\bibfield  {title} {\enquote {\bibinfo {title} {Targeting
  {3D} bladder cancer spheroids with urease-powered nanomotors},}\ }\href@noop
  {} {\bibfield  {journal} {\bibinfo  {journal} {ACS Nano}\ }\textbf {\bibinfo
  {volume} {13}},\ \bibinfo {pages} {429--439} (\bibinfo {year}
  {2019})}\BibitemShut {NoStop}%
\bibitem [{\citenamefont {Bozuyuk}\ \emph {et~al.}(2018)\citenamefont
  {Bozuyuk}, \citenamefont {Yasa}, \citenamefont {Yasa}, \citenamefont
  {Ceylan}, \citenamefont {Kizilel},\ and\ \citenamefont
  {Sitti}}]{BozuyukYYCKS2018}%
  \BibitemOpen
  \bibfield  {author} {\bibinfo {author} {\bibfnamefont {U.}~\bibnamefont
  {Bozuyuk}}, \bibinfo {author} {\bibfnamefont {O.}~\bibnamefont {Yasa}},
  \bibinfo {author} {\bibfnamefont {I.~C.}\ \bibnamefont {Yasa}}, \bibinfo
  {author} {\bibfnamefont {H.}~\bibnamefont {Ceylan}}, \bibinfo {author}
  {\bibfnamefont {S.}~\bibnamefont {Kizilel}}, \ and\ \bibinfo {author}
  {\bibfnamefont {M.}~\bibnamefont {Sitti}},\ }\bibfield  {title} {\enquote
  {\bibinfo {title} {Light-triggered drug release from {3D}-printed magnetic
  chitosan microswimmers},}\ }\href@noop {} {\bibfield  {journal} {\bibinfo
  {journal} {ACS Nano}\ }\textbf {\bibinfo {volume} {12}},\ \bibinfo {pages}
  {9617--9625} (\bibinfo {year} {2018})}\BibitemShut {NoStop}%
\bibitem [{\citenamefont {Choi}\ \emph {et~al.}(2018)\citenamefont {Choi},
  \citenamefont {Lee}, \citenamefont {Kim},\ and\ \citenamefont
  {Hahn}}]{ChoiLKH2018}%
  \BibitemOpen
  \bibfield  {author} {\bibinfo {author} {\bibfnamefont {H.}~\bibnamefont
  {Choi}}, \bibinfo {author} {\bibfnamefont {G.-H.}\ \bibnamefont {Lee}},
  \bibinfo {author} {\bibfnamefont {K.~S.}\ \bibnamefont {Kim}}, \ and\
  \bibinfo {author} {\bibfnamefont {S.~K.}\ \bibnamefont {Hahn}},\ }\bibfield
  {title} {\enquote {\bibinfo {title} {Light-guided nanomotor systems for
  autonomous photothermal cancer therapy},}\ }\href@noop {} {\bibfield
  {journal} {\bibinfo  {journal} {ACS Applied Materials \& Interfaces}\
  }\textbf {\bibinfo {volume} {10}},\ \bibinfo {pages} {2338--2346} (\bibinfo
  {year} {2018})}\BibitemShut {NoStop}%
\bibitem [{\citenamefont {Xu}\ \emph {et~al.}(2019{\natexlab{b}})\citenamefont
  {Xu} \emph {et~al.}}]{XuEtAl2019}%
  \BibitemOpen
  \bibfield  {author} {\bibinfo {author} {\bibfnamefont {D.}~\bibnamefont {Xu}}
  \emph {et~al.},\ }\bibfield  {title} {\enquote {\bibinfo {title} {Enzymatic
  micromotors as a mobile photosensitizer platform for highly efficient on-chip
  targeted antibacteria photodynamic therapy},}\ }\href@noop {} {\bibfield
  {journal} {\bibinfo  {journal} {Advanced Functional Materials}\ }\textbf
  {\bibinfo {volume} {29}},\ \bibinfo {pages} {1807727} (\bibinfo {year}
  {2019}{\natexlab{b}})}\BibitemShut {NoStop}%
\bibitem [{\citenamefont {Chen}\ \emph {et~al.}(2017)\citenamefont {Chen} \emph
  {et~al.}}]{ChenEtAl2017b}%
  \BibitemOpen
  \bibfield  {author} {\bibinfo {author} {\bibfnamefont {X.-Z.}\ \bibnamefont
  {Chen}} \emph {et~al.},\ }\bibfield  {title} {\enquote {\bibinfo {title}
  {Hybrid magnetoelectric nanowires for nanorobotic applications: fabrication,
  magnetoelectric coupling, and magnetically assisted in vitro targeted drug
  delivery},}\ }\href@noop {} {\bibfield  {journal} {\bibinfo  {journal}
  {Advanced Materials}\ }\textbf {\bibinfo {volume} {29}},\ \bibinfo {pages}
  {1605458} (\bibinfo {year} {2017})}\BibitemShut {NoStop}%
\bibitem [{\citenamefont {Ceylan}\ \emph
  {et~al.}(2019{\natexlab{b}})\citenamefont {Ceylan}, \citenamefont {Yasa},
  \citenamefont {Yasa}, \citenamefont {Tabak}, \citenamefont {Giltinan},\ and\
  \citenamefont {Sitti}}]{CeylanYYTGS2019}%
  \BibitemOpen
  \bibfield  {author} {\bibinfo {author} {\bibfnamefont {H.}~\bibnamefont
  {Ceylan}}, \bibinfo {author} {\bibfnamefont {I.~C.}\ \bibnamefont {Yasa}},
  \bibinfo {author} {\bibfnamefont {O.}~\bibnamefont {Yasa}}, \bibinfo {author}
  {\bibfnamefont {A.~F.}\ \bibnamefont {Tabak}}, \bibinfo {author}
  {\bibfnamefont {J.}~\bibnamefont {Giltinan}}, \ and\ \bibinfo {author}
  {\bibfnamefont {M.}~\bibnamefont {Sitti}},\ }\bibfield  {title} {\enquote
  {\bibinfo {title} {3{D}-printed biodegradable microswimmer for theranostic
  cargo delivery and release},}\ }\href@noop {} {\bibfield  {journal} {\bibinfo
   {journal} {ACS Nano}\ }\textbf {\bibinfo {volume} {13}},\ \bibinfo {pages}
  {3353--3362} (\bibinfo {year} {2019}{\natexlab{b}})}\BibitemShut {NoStop}%
\bibitem [{\citenamefont {Mushtaq}\ \emph {et~al.}(2019)\citenamefont {Mushtaq}
  \emph {et~al.}}]{MushtaqEtAl2019}%
  \BibitemOpen
  \bibfield  {author} {\bibinfo {author} {\bibfnamefont {F.}~\bibnamefont
  {Mushtaq}} \emph {et~al.},\ }\bibfield  {title} {\enquote {\bibinfo {title}
  {Motile piezoelectric nanoeels for targeted drug delivery},}\ }\href@noop {}
  {\bibfield  {journal} {\bibinfo  {journal} {Advanced Functional Materials}\
  }\textbf {\bibinfo {volume} {29}},\ \bibinfo {pages} {1808135} (\bibinfo
  {year} {2019})}\BibitemShut {NoStop}%
\bibitem [{\citenamefont {{Esteban-Fern{\'a}ndez de {\'A}vila}}\ \emph
  {et~al.}(2018{\natexlab{b}})\citenamefont {{Esteban-Fern{\'a}ndez de
  {\'A}vila}}, \citenamefont {Angsantikul}, \citenamefont
  {Ram{\'\i}rez-Herrera}, \citenamefont {Soto}, \citenamefont {Teymourian},
  \citenamefont {Dehaini}, \citenamefont {Chen}, \citenamefont {Zhang},\ and\
  \citenamefont {Wang}}]{EstebanFernandezdeAvilaARHSTDCZW2018}%
  \BibitemOpen
  \bibfield  {author} {\bibinfo {author} {\bibfnamefont {B.}~\bibnamefont
  {{Esteban-Fern{\'a}ndez de {\'A}vila}}}, \bibinfo {author} {\bibfnamefont
  {P.}~\bibnamefont {Angsantikul}}, \bibinfo {author} {\bibfnamefont {D.~E.}\
  \bibnamefont {Ram{\'\i}rez-Herrera}}, \bibinfo {author} {\bibfnamefont
  {F.}~\bibnamefont {Soto}}, \bibinfo {author} {\bibfnamefont {H.}~\bibnamefont
  {Teymourian}}, \bibinfo {author} {\bibfnamefont {D.}~\bibnamefont {Dehaini}},
  \bibinfo {author} {\bibfnamefont {Y.}~\bibnamefont {Chen}}, \bibinfo {author}
  {\bibfnamefont {L.}~\bibnamefont {Zhang}}, \ and\ \bibinfo {author}
  {\bibfnamefont {J.}~\bibnamefont {Wang}},\ }\bibfield  {title} {\enquote
  {\bibinfo {title} {Hybrid biomembrane-functionalized nanorobots for
  concurrent removal of pathogenic bacteria and toxins},}\ }\href@noop {}
  {\bibfield  {journal} {\bibinfo  {journal} {Science Robotics}\ }\textbf
  {\bibinfo {volume} {3}},\ \bibinfo {pages} {eaat0485} (\bibinfo {year}
  {2018}{\natexlab{b}})}\BibitemShut {NoStop}%
\bibitem [{\citenamefont {Gao}\ \emph {et~al.}(2019)\citenamefont {Gao},
  \citenamefont {Lin}, \citenamefont {Wang}, \citenamefont {Wu}, \citenamefont
  {Xie},\ and\ \citenamefont {He}}]{GaoLWWXH2019}%
  \BibitemOpen
  \bibfield  {author} {\bibinfo {author} {\bibfnamefont {C.}~\bibnamefont
  {Gao}}, \bibinfo {author} {\bibfnamefont {Z.}~\bibnamefont {Lin}}, \bibinfo
  {author} {\bibfnamefont {D.}~\bibnamefont {Wang}}, \bibinfo {author}
  {\bibfnamefont {Z.}~\bibnamefont {Wu}}, \bibinfo {author} {\bibfnamefont
  {H.}~\bibnamefont {Xie}}, \ and\ \bibinfo {author} {\bibfnamefont
  {Q.}~\bibnamefont {He}},\ }\bibfield  {title} {\enquote {\bibinfo {title}
  {Red blood cell-mimicking micromotor for active photodynamic cancer
  therapy},}\ }\href@noop {} {\bibfield  {journal} {\bibinfo  {journal} {ACS
  Applied Materials \& Interfaces}\ }\textbf {\bibinfo {volume} {11}},\
  \bibinfo {pages} {23392--23400} (\bibinfo {year} {2019})}\BibitemShut
  {NoStop}%
\bibitem [{\citenamefont {Schwarz}\ \emph {et~al.}(2020)\citenamefont
  {Schwarz}, \citenamefont {Karnaushenko}, \citenamefont {Hebenstreit},
  \citenamefont {Naumann}, \citenamefont {Schmidt},\ and\ \citenamefont
  {{Medina-S{\'a}nchez}}}]{SchwarzKHNSMS2020}%
  \BibitemOpen
  \bibfield  {author} {\bibinfo {author} {\bibfnamefont {L.}~\bibnamefont
  {Schwarz}}, \bibinfo {author} {\bibfnamefont {D.~D.}\ \bibnamefont
  {Karnaushenko}}, \bibinfo {author} {\bibfnamefont {F.}~\bibnamefont
  {Hebenstreit}}, \bibinfo {author} {\bibfnamefont {R.}~\bibnamefont
  {Naumann}}, \bibinfo {author} {\bibfnamefont {O.~G.}\ \bibnamefont
  {Schmidt}}, \ and\ \bibinfo {author} {\bibfnamefont {M.}~\bibnamefont
  {{Medina-S{\'a}nchez}}},\ }\bibfield  {title} {\enquote {\bibinfo {title} {A
  rotating spiral micromotor for noninvasive zygote transfer},}\ }\href@noop {}
  {\bibfield  {journal} {\bibinfo  {journal} {Advanced Science}\ }\textbf
  {\bibinfo {volume} {7}},\ \bibinfo {pages} {2000843} (\bibinfo {year}
  {2020})}\BibitemShut {NoStop}%
\bibitem [{\citenamefont {Soto}\ \emph {et~al.}(2020)\citenamefont {Soto},
  \citenamefont {Wang}, \citenamefont {Ahmed},\ and\ \citenamefont
  {Demirci}}]{SotoWAD2020}%
  \BibitemOpen
  \bibfield  {author} {\bibinfo {author} {\bibfnamefont {F.}~\bibnamefont
  {Soto}}, \bibinfo {author} {\bibfnamefont {J.}~\bibnamefont {Wang}}, \bibinfo
  {author} {\bibfnamefont {R.}~\bibnamefont {Ahmed}}, \ and\ \bibinfo {author}
  {\bibfnamefont {U.}~\bibnamefont {Demirci}},\ }\bibfield  {title} {\enquote
  {\bibinfo {title} {Medical micro/nanorobots in precision medicine},}\
  }\href@noop {} {\bibfield  {journal} {\bibinfo  {journal} {Advanced Science}\
  }\textbf {\bibinfo {volume} {7}},\ \bibinfo {pages} {2002203} (\bibinfo
  {year} {2020})}\BibitemShut {NoStop}%
\bibitem [{\citenamefont {Wan}\ \emph {et~al.}(2020)\citenamefont {Wan} \emph
  {et~al.}}]{WanEtAl2020a}%
  \BibitemOpen
  \bibfield  {author} {\bibinfo {author} {\bibfnamefont {M.}~\bibnamefont
  {Wan}} \emph {et~al.},\ }\bibfield  {title} {\enquote {\bibinfo {title}
  {Platelet-derived porous nanomotor for thrombus therapy},}\ }\href@noop {}
  {\bibfield  {journal} {\bibinfo  {journal} {Science Advances}\ }\textbf
  {\bibinfo {volume} {6}},\ \bibinfo {pages} {eaaz9014} (\bibinfo {year}
  {2020})}\BibitemShut {NoStop}%
\bibitem [{\citenamefont {Xu}\ \emph {et~al.}(2020)\citenamefont {Xu},
  \citenamefont {{Medina-S{\'a}nchez}}, \citenamefont {Maitz}, \citenamefont
  {Werner},\ and\ \citenamefont {Schmidt}}]{XuMSMWS2020}%
  \BibitemOpen
  \bibfield  {author} {\bibinfo {author} {\bibfnamefont {H.}~\bibnamefont
  {Xu}}, \bibinfo {author} {\bibfnamefont {M.}~\bibnamefont
  {{Medina-S{\'a}nchez}}}, \bibinfo {author} {\bibfnamefont {M.~F.}\
  \bibnamefont {Maitz}}, \bibinfo {author} {\bibfnamefont {C.}~\bibnamefont
  {Werner}}, \ and\ \bibinfo {author} {\bibfnamefont {O.~G.}\ \bibnamefont
  {Schmidt}},\ }\bibfield  {title} {\enquote {\bibinfo {title} {Sperm
  micromotors for cargo delivery through flowing blood},}\ }\href@noop {}
  {\bibfield  {journal} {\bibinfo  {journal} {ACS Nano}\ }\textbf {\bibinfo
  {volume} {14}},\ \bibinfo {pages} {2982--2993} (\bibinfo {year}
  {2020})}\BibitemShut {NoStop}%
\bibitem [{\citenamefont {Wan}\ \emph {et~al.}(2019)\citenamefont {Wan},
  \citenamefont {Chen}, \citenamefont {Wang}, \citenamefont {Niu},
  \citenamefont {Xu}, \citenamefont {Yu}, \citenamefont {Zhu}, \citenamefont
  {Mao},\ and\ \citenamefont {Shen}}]{WanCWNXYZMS2019}%
  \BibitemOpen
  \bibfield  {author} {\bibinfo {author} {\bibfnamefont {M.}~\bibnamefont
  {Wan}}, \bibinfo {author} {\bibfnamefont {H.}~\bibnamefont {Chen}}, \bibinfo
  {author} {\bibfnamefont {Q.}~\bibnamefont {Wang}}, \bibinfo {author}
  {\bibfnamefont {Q.}~\bibnamefont {Niu}}, \bibinfo {author} {\bibfnamefont
  {P.}~\bibnamefont {Xu}}, \bibinfo {author} {\bibfnamefont {Y.}~\bibnamefont
  {Yu}}, \bibinfo {author} {\bibfnamefont {T.}~\bibnamefont {Zhu}}, \bibinfo
  {author} {\bibfnamefont {C.}~\bibnamefont {Mao}}, \ and\ \bibinfo {author}
  {\bibfnamefont {J.}~\bibnamefont {Shen}},\ }\bibfield  {title} {\enquote
  {\bibinfo {title} {Bio-inspired nitric-oxide-driven nanomotor},}\ }\href@noop
  {} {\bibfield  {journal} {\bibinfo  {journal} {Nature Communications}\
  }\textbf {\bibinfo {volume} {10}},\ \bibinfo {pages} {966} (\bibinfo {year}
  {2019})}\BibitemShut {NoStop}%
\bibitem [{\citenamefont {Schmidt}\ \emph {et~al.}(2020)\citenamefont
  {Schmidt}, \citenamefont {{Medina-S{\'a}nchez}}, \citenamefont {Edmondson},\
  and\ \citenamefont {Schmidt}}]{SchmidtMSES2020}%
  \BibitemOpen
  \bibfield  {author} {\bibinfo {author} {\bibfnamefont {C.~K.}\ \bibnamefont
  {Schmidt}}, \bibinfo {author} {\bibfnamefont {M.}~\bibnamefont
  {{Medina-S{\'a}nchez}}}, \bibinfo {author} {\bibfnamefont {R.~J.}\
  \bibnamefont {Edmondson}}, \ and\ \bibinfo {author} {\bibfnamefont {O.~G.}\
  \bibnamefont {Schmidt}},\ }\bibfield  {title} {\enquote {\bibinfo {title}
  {Engineering microrobots for targeted cancer therapies from a medical
  perspective},}\ }\href@noop {} {\bibfield  {journal} {\bibinfo  {journal}
  {Nature Communications}\ }\textbf {\bibinfo {volume} {11}},\ \bibinfo {pages}
  {1--18} (\bibinfo {year} {2020})}\BibitemShut {NoStop}%
\bibitem [{\citenamefont {Wang}\ \emph {et~al.}(2020)\citenamefont {Wang},
  \citenamefont {Dong}, \citenamefont {Wu}, \citenamefont {Cai},\ and\
  \citenamefont {Ren}}]{WangDWCR2020}%
  \BibitemOpen
  \bibfield  {author} {\bibinfo {author} {\bibfnamefont {J.}~\bibnamefont
  {Wang}}, \bibinfo {author} {\bibfnamefont {R.}~\bibnamefont {Dong}}, \bibinfo
  {author} {\bibfnamefont {H.}~\bibnamefont {Wu}}, \bibinfo {author}
  {\bibfnamefont {Y.}~\bibnamefont {Cai}}, \ and\ \bibinfo {author}
  {\bibfnamefont {B.}~\bibnamefont {Ren}},\ }\bibfield  {title} {\enquote
  {\bibinfo {title} {A review on artificial micro/nanomotors for
  cancer‐targeted delivery, diagnosis, and therapy},}\ }\href@noop {}
  {\bibfield  {journal} {\bibinfo  {journal} {Nano-Micro Letters}\ }\textbf
  {\bibinfo {volume} {12}},\ \bibinfo {pages} {1--19} (\bibinfo {year}
  {2020})}\BibitemShut {NoStop}%
\bibitem [{\citenamefont {Yu}\ \emph {et~al.}(2021)\citenamefont {Yu},
  \citenamefont {Zhou}, \citenamefont {Sun}, \citenamefont {Wu}, \citenamefont
  {Xu}, \citenamefont {Chang}, \citenamefont {Bi}, \citenamefont {Jiang},\ and\
  \citenamefont {Zhu}}]{YuZSWXCBJZ2021}%
  \BibitemOpen
  \bibfield  {author} {\bibinfo {author} {\bibfnamefont {S.}~\bibnamefont
  {Yu}}, \bibinfo {author} {\bibfnamefont {Y.}~\bibnamefont {Zhou}}, \bibinfo
  {author} {\bibfnamefont {Y.}~\bibnamefont {Sun}}, \bibinfo {author}
  {\bibfnamefont {S.}~\bibnamefont {Wu}}, \bibinfo {author} {\bibfnamefont
  {T.}~\bibnamefont {Xu}}, \bibinfo {author} {\bibfnamefont {Y.-C.}\
  \bibnamefont {Chang}}, \bibinfo {author} {\bibfnamefont {S.}~\bibnamefont
  {Bi}}, \bibinfo {author} {\bibfnamefont {L.-P.}\ \bibnamefont {Jiang}}, \
  and\ \bibinfo {author} {\bibfnamefont {J.-J.}\ \bibnamefont {Zhu}},\
  }\bibfield  {title} {\enquote {\bibinfo {title} {Endogenous m{RNA} triggered
  {DNA}-{A}u nanomachine for in situ imaging and targeted multimodal
  synergistic cancer therapy},}\ }\href@noop {} {\bibfield  {journal} {\bibinfo
   {journal} {Angewandte Chemie International Edition}\ }\textbf {\bibinfo
  {volume} {60}},\ \bibinfo {pages} {5573--6184} (\bibinfo {year}
  {2021})}\BibitemShut {NoStop}%
\bibitem [{\citenamefont {Jin}\ \emph {et~al.}(2021)\citenamefont {Jin},
  \citenamefont {Yuan}, \citenamefont {Du}, \citenamefont {Wang}, \citenamefont
  {Wang},\ and\ \citenamefont {Zhang}}]{JinYDWWZ2021}%
  \BibitemOpen
  \bibfield  {author} {\bibinfo {author} {\bibfnamefont {D.}~\bibnamefont
  {Jin}}, \bibinfo {author} {\bibfnamefont {K.}~\bibnamefont {Yuan}}, \bibinfo
  {author} {\bibfnamefont {X.}~\bibnamefont {Du}}, \bibinfo {author}
  {\bibfnamefont {Q.}~\bibnamefont {Wang}}, \bibinfo {author} {\bibfnamefont
  {S.}~\bibnamefont {Wang}}, \ and\ \bibinfo {author} {\bibfnamefont
  {L.}~\bibnamefont {Zhang}},\ }\bibfield  {title} {\enquote {\bibinfo {title}
  {Domino reaction encoded heterogeneous colloidal microswarm with on-demand
  morphological adaptability},}\ }\href@noop {} {\bibfield  {journal} {\bibinfo
   {journal} {Advanced Materials}\ }\textbf {\bibinfo {volume} {33}},\ \bibinfo
  {pages} {2100070} (\bibinfo {year} {2021})}\BibitemShut {NoStop}%
\bibitem [{\citenamefont {Dai}\ \emph {et~al.}(2016)\citenamefont {Dai},
  \citenamefont {Wang}, \citenamefont {Xiong}, \citenamefont {Zhan},
  \citenamefont {Dai}, \citenamefont {Li}, \citenamefont {Feng},\ and\
  \citenamefont {Tang}}]{DaiWXZDLFT2016}%
  \BibitemOpen
  \bibfield  {author} {\bibinfo {author} {\bibfnamefont {B.}~\bibnamefont
  {Dai}}, \bibinfo {author} {\bibfnamefont {J.}~\bibnamefont {Wang}}, \bibinfo
  {author} {\bibfnamefont {Z.}~\bibnamefont {Xiong}}, \bibinfo {author}
  {\bibfnamefont {X.}~\bibnamefont {Zhan}}, \bibinfo {author} {\bibfnamefont
  {W.}~\bibnamefont {Dai}}, \bibinfo {author} {\bibfnamefont {C.}~\bibnamefont
  {Li}}, \bibinfo {author} {\bibfnamefont {S.}~\bibnamefont {Feng}}, \ and\
  \bibinfo {author} {\bibfnamefont {J.}~\bibnamefont {Tang}},\ }\bibfield
  {title} {\enquote {\bibinfo {title} {Programmable artificial phototactic
  microswimmer},}\ }\href@noop {} {\bibfield  {journal} {\bibinfo  {journal}
  {Nature Nanotechnology}\ }\textbf {\bibinfo {volume} {11}},\ \bibinfo {pages}
  {1087--1092} (\bibinfo {year} {2016})}\BibitemShut {NoStop}%
\bibitem [{\citenamefont {Zheng}\ \emph {et~al.}(2017)\citenamefont {Zheng},
  \citenamefont {Dai}, \citenamefont {Wang}, \citenamefont {Xiong},
  \citenamefont {Yang}, \citenamefont {Liu}, \citenamefont {Zhan},
  \citenamefont {Wan},\ and\ \citenamefont {Tang}}]{ZhengDWXYLZWT2017}%
  \BibitemOpen
  \bibfield  {author} {\bibinfo {author} {\bibfnamefont {J.}~\bibnamefont
  {Zheng}}, \bibinfo {author} {\bibfnamefont {B.}~\bibnamefont {Dai}}, \bibinfo
  {author} {\bibfnamefont {J.}~\bibnamefont {Wang}}, \bibinfo {author}
  {\bibfnamefont {Z.}~\bibnamefont {Xiong}}, \bibinfo {author} {\bibfnamefont
  {Y.}~\bibnamefont {Yang}}, \bibinfo {author} {\bibfnamefont {J.}~\bibnamefont
  {Liu}}, \bibinfo {author} {\bibfnamefont {X.}~\bibnamefont {Zhan}}, \bibinfo
  {author} {\bibfnamefont {Z.}~\bibnamefont {Wan}}, \ and\ \bibinfo {author}
  {\bibfnamefont {J.}~\bibnamefont {Tang}},\ }\bibfield  {title} {\enquote
  {\bibinfo {title} {Orthogonal navigation of multiple visible-light-driven
  artificial microswimmers},}\ }\href@noop {} {\bibfield  {journal} {\bibinfo
  {journal} {Nature Communications}\ }\textbf {\bibinfo {volume} {8}},\
  \bibinfo {pages} {1438} (\bibinfo {year} {2017})}\BibitemShut {NoStop}%
\bibitem [{\citenamefont {Wang}\ \emph {et~al.}(2014)\citenamefont {Wang},
  \citenamefont {Li}, \citenamefont {Mair}, \citenamefont {Ahmed},
  \citenamefont {Huang},\ and\ \citenamefont {Mallouk}}]{WangLMAHM2014}%
  \BibitemOpen
  \bibfield  {author} {\bibinfo {author} {\bibfnamefont {W.}~\bibnamefont
  {Wang}}, \bibinfo {author} {\bibfnamefont {S.}~\bibnamefont {Li}}, \bibinfo
  {author} {\bibfnamefont {L.}~\bibnamefont {Mair}}, \bibinfo {author}
  {\bibfnamefont {S.}~\bibnamefont {Ahmed}}, \bibinfo {author} {\bibfnamefont
  {T.~J.}\ \bibnamefont {Huang}}, \ and\ \bibinfo {author} {\bibfnamefont
  {T.~E.}\ \bibnamefont {Mallouk}},\ }\bibfield  {title} {\enquote {\bibinfo
  {title} {Acoustic propulsion of nanorod motors inside living cells},}\
  }\href@noop {} {\bibfield  {journal} {\bibinfo  {journal} {Angewandte Chemie
  International Edition}\ }\textbf {\bibinfo {volume} {53}},\ \bibinfo {pages}
  {3201--3204} (\bibinfo {year} {2014})}\BibitemShut {NoStop}%
\bibitem [{\citenamefont {{Garcia-Gradilla}}\ \emph {et~al.}(2014)\citenamefont
  {{Garcia-Gradilla}}, \citenamefont {Sattayasamitsathit}, \citenamefont
  {Soto}, \citenamefont {Kuralay}, \citenamefont {Yard{\i}mc{\i}},
  \citenamefont {Wiitala}, \citenamefont {Galarnyk},\ and\ \citenamefont
  {Wang}}]{GarciaGradillaSSKYWGW2014}%
  \BibitemOpen
  \bibfield  {author} {\bibinfo {author} {\bibfnamefont {V.}~\bibnamefont
  {{Garcia-Gradilla}}}, \bibinfo {author} {\bibfnamefont {S.}~\bibnamefont
  {Sattayasamitsathit}}, \bibinfo {author} {\bibfnamefont {F.}~\bibnamefont
  {Soto}}, \bibinfo {author} {\bibfnamefont {F.}~\bibnamefont {Kuralay}},
  \bibinfo {author} {\bibfnamefont {C.}~\bibnamefont {Yard{\i}mc{\i}}},
  \bibinfo {author} {\bibfnamefont {D.}~\bibnamefont {Wiitala}}, \bibinfo
  {author} {\bibfnamefont {M.}~\bibnamefont {Galarnyk}}, \ and\ \bibinfo
  {author} {\bibfnamefont {J.}~\bibnamefont {Wang}},\ }\bibfield  {title}
  {\enquote {\bibinfo {title} {Ultrasound-propelled nanoporous gold wire for
  efficient drug loading and release},}\ }\href@noop {} {\bibfield  {journal}
  {\bibinfo  {journal} {Small}\ }\textbf {\bibinfo {volume} {10}},\ \bibinfo
  {pages} {4154--4159} (\bibinfo {year} {2014})}\BibitemShut {NoStop}%
\bibitem [{\citenamefont {Wang}\ \emph {et~al.}(2018)\citenamefont {Wang},
  \citenamefont {Qin}, \citenamefont {Hu}, \citenamefont {Terzopoulou},
  \citenamefont {Chen}, \citenamefont {Huang}, \citenamefont {{Maniura-Weber}},
  \citenamefont {Pan{\'e}},\ and\ \citenamefont {Nelson}}]{WangQHTCHMWPN2018}%
  \BibitemOpen
  \bibfield  {author} {\bibinfo {author} {\bibfnamefont {X.}~\bibnamefont
  {Wang}}, \bibinfo {author} {\bibfnamefont {X.}~\bibnamefont {Qin}}, \bibinfo
  {author} {\bibfnamefont {C.}~\bibnamefont {Hu}}, \bibinfo {author}
  {\bibfnamefont {A.}~\bibnamefont {Terzopoulou}}, \bibinfo {author}
  {\bibfnamefont {X.}~\bibnamefont {Chen}}, \bibinfo {author} {\bibfnamefont
  {T.}~\bibnamefont {Huang}}, \bibinfo {author} {\bibfnamefont
  {K.}~\bibnamefont {{Maniura-Weber}}}, \bibinfo {author} {\bibfnamefont
  {S.}~\bibnamefont {Pan{\'e}}}, \ and\ \bibinfo {author} {\bibfnamefont
  {B.~J.}\ \bibnamefont {Nelson}},\ }\bibfield  {title} {\enquote {\bibinfo
  {title} {3{D} printed enzymatically biodegradable soft helical
  microswimmers},}\ }\href@noop {} {\bibfield  {journal} {\bibinfo  {journal}
  {Advanced Functional Materials}\ }\textbf {\bibinfo {volume} {28}},\ \bibinfo
  {pages} {1804107} (\bibinfo {year} {2018})}\BibitemShut {NoStop}%
\bibitem [{\citenamefont {Vo{\ss}}\ and\ \citenamefont
  {Wittkowski}(2020)}]{VossW2020}%
  \BibitemOpen
  \bibfield  {author} {\bibinfo {author} {\bibfnamefont {J.}~\bibnamefont
  {Vo{\ss}}}\ and\ \bibinfo {author} {\bibfnamefont {R.}~\bibnamefont
  {Wittkowski}},\ }\bibfield  {title} {\enquote {\bibinfo {title} {On the
  shape-dependent propulsion of nano- and microparticles by traveling
  ultrasound waves},}\ }\href@noop {} {\bibfield  {journal} {\bibinfo
  {journal} {Nanoscale Advances}\ }\textbf {\bibinfo {volume} {2}},\ \bibinfo
  {pages} {3890--3899} (\bibinfo {year} {2020})}\BibitemShut {NoStop}%
\bibitem [{\citenamefont {Vo{\ss}}\ and\ \citenamefont
  {Wittkowski}(2022)}]{VossW2021}%
  \BibitemOpen
  \bibfield  {author} {\bibinfo {author} {\bibfnamefont {J.}~\bibnamefont
  {Vo{\ss}}}\ and\ \bibinfo {author} {\bibfnamefont {R.}~\bibnamefont
  {Wittkowski}},\ }\bibfield  {title} {\enquote {\bibinfo {title} {Acoustically
  propelled nano- and microcones: fast forward and backward motion},}\
  }\href@noop {} {\bibfield  {journal} {\bibinfo  {journal} {Nanoscale
  Advances}\ }\textbf {\bibinfo {volume} {4}},\ \bibinfo {pages} {281--293}
  (\bibinfo {year} {2022})}\BibitemShut {NoStop}%
\bibitem [{\citenamefont {Vo\ss{}}\ and\ \citenamefont
  {Wittkowski}(2022)}]{VossW2022b}%
  \BibitemOpen
  \bibfield  {author} {\bibinfo {author} {\bibfnamefont {J.}~\bibnamefont
  {Vo\ss{}}}\ and\ \bibinfo {author} {\bibfnamefont {R.}~\bibnamefont
  {Wittkowski}},\ }\bibfield  {title} {\enquote {\bibinfo {title}
  {Orientation-dependent propulsion of cone-shaped nano- and microparticles by
  a traveling ultrasound wave},}\ }\href@noop {} {\bibfield  {journal}
  {\bibinfo  {journal} {submitted to ACS Nano}\ } (\bibinfo {year}
  {2022})}\BibitemShut {NoStop}%
\bibitem [{\citenamefont {Blanco}\ \emph {et~al.}(2015)\citenamefont {Blanco},
  \citenamefont {Shen},\ and\ \citenamefont {Ferrari}}]{BlancoSF2015}%
  \BibitemOpen
  \bibfield  {author} {\bibinfo {author} {\bibfnamefont {E.}~\bibnamefont
  {Blanco}}, \bibinfo {author} {\bibfnamefont {H.}~\bibnamefont {Shen}}, \ and\
  \bibinfo {author} {\bibfnamefont {M.}~\bibnamefont {Ferrari}},\ }\bibfield
  {title} {\enquote {\bibinfo {title} {Principles of nanoparticle design for
  overcoming biological barriers to drug delivery},}\ }\href@noop {} {\bibfield
   {journal} {\bibinfo  {journal} {Nature Biotechnology}\ }\textbf {\bibinfo
  {volume} {33}},\ \bibinfo {pages} {941--951} (\bibinfo {year}
  {2015})}\BibitemShut {NoStop}%
\bibitem [{\citenamefont {Peer}\ \emph {et~al.}(2007)\citenamefont {Peer},
  \citenamefont {Karp}, \citenamefont {Hong}, \citenamefont {Farokhzad},
  \citenamefont {Margalit},\ and\ \citenamefont {Langer}}]{PeerKHFML2007}%
  \BibitemOpen
  \bibfield  {author} {\bibinfo {author} {\bibfnamefont {D.}~\bibnamefont
  {Peer}}, \bibinfo {author} {\bibfnamefont {J.~M.}\ \bibnamefont {Karp}},
  \bibinfo {author} {\bibfnamefont {S.}~\bibnamefont {Hong}}, \bibinfo {author}
  {\bibfnamefont {O.~C.}\ \bibnamefont {Farokhzad}}, \bibinfo {author}
  {\bibfnamefont {R.}~\bibnamefont {Margalit}}, \ and\ \bibinfo {author}
  {\bibfnamefont {R.}~\bibnamefont {Langer}},\ }\bibfield  {title} {\enquote
  {\bibinfo {title} {Nanocarriers as an emerging platform for cancer
  therapy},}\ }\href@noop {} {\bibfield  {journal} {\bibinfo  {journal} {Nature
  Nanotechnology}\ }\textbf {\bibinfo {volume} {2}},\ \bibinfo {pages}
  {751--760} (\bibinfo {year} {2007})}\BibitemShut {NoStop}%
\bibitem [{\citenamefont {Wu}\ \emph {et~al.}(2014)\citenamefont {Wu} \emph
  {et~al.}}]{WuEtAl2014}%
  \BibitemOpen
  \bibfield  {author} {\bibinfo {author} {\bibfnamefont {Z.}~\bibnamefont {Wu}}
  \emph {et~al.},\ }\bibfield  {title} {\enquote {\bibinfo {title} {Turning
  erythrocytes into functional micromotors},}\ }\href@noop {} {\bibfield
  {journal} {\bibinfo  {journal} {ACS Nano}\ }\textbf {\bibinfo {volume} {8}},\
  \bibinfo {pages} {12041--12048} (\bibinfo {year} {2014})}\BibitemShut
  {NoStop}%
\bibitem [{\citenamefont {Nelson}\ \emph {et~al.}(2010)\citenamefont {Nelson},
  \citenamefont {Kaliakatsos},\ and\ \citenamefont {Abbott}}]{NelsonKA2010}%
  \BibitemOpen
  \bibfield  {author} {\bibinfo {author} {\bibfnamefont {B.~J.}\ \bibnamefont
  {Nelson}}, \bibinfo {author} {\bibfnamefont {I.~K.}\ \bibnamefont
  {Kaliakatsos}}, \ and\ \bibinfo {author} {\bibfnamefont {J.~J.}\ \bibnamefont
  {Abbott}},\ }\bibfield  {title} {\enquote {\bibinfo {title} {Microrobots for
  minimally invasive medicine},}\ }\href@noop {} {\bibfield  {journal}
  {\bibinfo  {journal} {Annual Review of Biomedical Engineering}\ }\textbf
  {\bibinfo {volume} {12}},\ \bibinfo {pages} {55--85} (\bibinfo {year}
  {2010})}\BibitemShut {NoStop}%
\bibitem [{\citenamefont {Stenhammar}\ \emph {et~al.}(2016)\citenamefont
  {Stenhammar}, \citenamefont {Wittkowski}, \citenamefont {Marenduzzo},\ and\
  \citenamefont {Cates}}]{StenhammarWMC2016}%
  \BibitemOpen
  \bibfield  {author} {\bibinfo {author} {\bibfnamefont {J.}~\bibnamefont
  {Stenhammar}}, \bibinfo {author} {\bibfnamefont {R.}~\bibnamefont
  {Wittkowski}}, \bibinfo {author} {\bibfnamefont {D.}~\bibnamefont
  {Marenduzzo}}, \ and\ \bibinfo {author} {\bibfnamefont {M.~E.}\ \bibnamefont
  {Cates}},\ }\bibfield  {title} {\enquote {\bibinfo {title} {Light-induced
  self-assembly of active rectification devices},}\ }\href@noop {} {\bibfield
  {journal} {\bibinfo  {journal} {Science Advances}\ }\textbf {\bibinfo
  {volume} {2}},\ \bibinfo {pages} {e1501850} (\bibinfo {year}
  {2016})}\BibitemShut {NoStop}%
\bibitem [{\citenamefont {Thiesen}\ and\ \citenamefont
  {Jordan}(2008)}]{ThiesenJ2008}%
  \BibitemOpen
  \bibfield  {author} {\bibinfo {author} {\bibfnamefont {B.}~\bibnamefont
  {Thiesen}}\ and\ \bibinfo {author} {\bibfnamefont {A.}~\bibnamefont
  {Jordan}},\ }\bibfield  {title} {\enquote {\bibinfo {title} {Clinical
  applications of magnetic nanoparticles for hyperthermia},}\ }\href@noop {}
  {\bibfield  {journal} {\bibinfo  {journal} {International Journal of
  Hyperthermia}\ }\textbf {\bibinfo {volume} {24}},\ \bibinfo {pages}
  {467--474} (\bibinfo {year} {2008})}\BibitemShut {NoStop}%
\bibitem [{\citenamefont {Melde}\ \emph {et~al.}(2016)\citenamefont {Melde},
  \citenamefont {Mark}, \citenamefont {Qiu},\ and\ \citenamefont
  {Fischer}}]{MeldeMQF2016}%
  \BibitemOpen
  \bibfield  {author} {\bibinfo {author} {\bibfnamefont {K.}~\bibnamefont
  {Melde}}, \bibinfo {author} {\bibfnamefont {A.~G.}\ \bibnamefont {Mark}},
  \bibinfo {author} {\bibfnamefont {T.}~\bibnamefont {Qiu}}, \ and\ \bibinfo
  {author} {\bibfnamefont {P.}~\bibnamefont {Fischer}},\ }\bibfield  {title}
  {\enquote {\bibinfo {title} {Holograms for acoustics},}\ }\href@noop {}
  {\bibfield  {journal} {\bibinfo  {journal} {Nature}\ }\textbf {\bibinfo
  {volume} {537}},\ \bibinfo {pages} {518--522} (\bibinfo {year}
  {2016})}\BibitemShut {NoStop}%
\bibitem [{\citenamefont {Schmidt}\ \emph {et~al.}(2010)\citenamefont
  {Schmidt}, \citenamefont {Lang},\ and\ \citenamefont
  {Heckmann}}]{SchmidtLH2010}%
  \BibitemOpen
  \bibfield  {author} {\bibinfo {author} {\bibfnamefont {R.~F.}\ \bibnamefont
  {Schmidt}}, \bibinfo {author} {\bibfnamefont {F.}~\bibnamefont {Lang}}, \
  and\ \bibinfo {author} {\bibfnamefont {M.}~\bibnamefont {Heckmann}},\
  }\href@noop {} {\emph {\bibinfo {title} {Physiologie des {M}enschen: mit
  {P}athophysiologie}}},\ \bibinfo {edition} {31st}\ ed.\ (\bibinfo
  {publisher} {Springer Medizin Verlag},\ \bibinfo {address} {Heidelberg},\
  \bibinfo {year} {2010})\BibitemShut {NoStop}%
\bibitem [{\citenamefont {Mitri}(2005)}]{Mitri2005}%
  \BibitemOpen
  \bibfield  {author} {\bibinfo {author} {\bibfnamefont {F.~G.}\ \bibnamefont
  {Mitri}},\ }\bibfield  {title} {\enquote {\bibinfo {title} {Theoretical
  calculation of the acoustic radiation force acting on elastic and
  viscoelastic cylinders placed in a plane standing or quasistanding wave
  field},}\ }\href@noop {} {\bibfield  {journal} {\bibinfo  {journal} {European
  Physical Journal B}\ }\textbf {\bibinfo {volume} {44}},\ \bibinfo {pages}
  {71--78} (\bibinfo {year} {2005})}\BibitemShut {NoStop}%
\bibitem [{\citenamefont {Wells}(1975)}]{Wells1975}%
  \BibitemOpen
  \bibfield  {author} {\bibinfo {author} {\bibfnamefont {P.~N.~T.}\
  \bibnamefont {Wells}},\ }\bibfield  {title} {\enquote {\bibinfo {title}
  {Absorption and dispersion of ultrasound in biological tissue},}\ }\href@noop
  {} {\bibfield  {journal} {\bibinfo  {journal} {Ultrasound in Medicine \&
  Biology}\ }\textbf {\bibinfo {volume} {1}},\ \bibinfo {pages} {369--376}
  (\bibinfo {year} {1975})}\BibitemShut {NoStop}%
\bibitem [{\citenamefont {Barnett}\ \emph {et~al.}(2000)\citenamefont
  {Barnett}, \citenamefont {{Ter Haar}}, \citenamefont {Ziskin}, \citenamefont
  {Rott}, \citenamefont {Duck},\ and\ \citenamefont
  {Maeda}}]{BarnettTHZRDM2000}%
  \BibitemOpen
  \bibfield  {author} {\bibinfo {author} {\bibfnamefont {S.~B.}\ \bibnamefont
  {Barnett}}, \bibinfo {author} {\bibfnamefont {G.~R.}\ \bibnamefont {{Ter
  Haar}}}, \bibinfo {author} {\bibfnamefont {M.~C.}\ \bibnamefont {Ziskin}},
  \bibinfo {author} {\bibfnamefont {H.}~\bibnamefont {Rott}}, \bibinfo {author}
  {\bibfnamefont {F.~A.}\ \bibnamefont {Duck}}, \ and\ \bibinfo {author}
  {\bibfnamefont {K.}~\bibnamefont {Maeda}},\ }\bibfield  {title} {\enquote
  {\bibinfo {title} {International recommendations and guidelines for the safe
  use of diagnostic ultrasound in medicine},}\ }\href@noop {} {\bibfield
  {journal} {\bibinfo  {journal} {Ultrasound in Medicine \& Biology}\ }\textbf
  {\bibinfo {volume} {26}},\ \bibinfo {pages} {355--366} (\bibinfo {year}
  {2000})}\BibitemShut {NoStop}%
\bibitem [{\citenamefont {Davis}\ \emph {et~al.}(2008)\citenamefont {Davis},
  \citenamefont {Chen},\ and\ \citenamefont {Shin}}]{DavisCS2008}%
  \BibitemOpen
  \bibfield  {author} {\bibinfo {author} {\bibfnamefont {M.~E.}\ \bibnamefont
  {Davis}}, \bibinfo {author} {\bibfnamefont {Z.}~\bibnamefont {Chen}}, \ and\
  \bibinfo {author} {\bibfnamefont {D.~M.}\ \bibnamefont {Shin}},\ }\bibfield
  {title} {\enquote {\bibinfo {title} {Nanoparticle therapeutics: an emerging
  treatment modality for cancer},}\ }\href@noop {} {\bibfield  {journal}
  {\bibinfo  {journal} {Nature Reviews Drug Discovery}\ }\textbf {\bibinfo
  {volume} {7}},\ \bibinfo {pages} {771--782} (\bibinfo {year}
  {2008})}\BibitemShut {NoStop}%
\bibitem [{\citenamefont {Welter}\ and\ \citenamefont
  {Rieger}(2013)}]{WelterR2013}%
  \BibitemOpen
  \bibfield  {author} {\bibinfo {author} {\bibfnamefont {M.}~\bibnamefont
  {Welter}}\ and\ \bibinfo {author} {\bibfnamefont {H.}~\bibnamefont
  {Rieger}},\ }\bibfield  {title} {\enquote {\bibinfo {title} {Interstitial
  fluid flow and drug delivery in vascularized tumors: a computational
  model},}\ }\href@noop {} {\bibfield  {journal} {\bibinfo  {journal} {PLOS
  ONE}\ }\textbf {\bibinfo {volume} {8}},\ \bibinfo {pages} {e70395} (\bibinfo
  {year} {2013})}\BibitemShut {NoStop}%
\bibitem [{\citenamefont {Tassa}\ \emph {et~al.}(2011)\citenamefont {Tassa},
  \citenamefont {Shaw},\ and\ \citenamefont {Weissleder}}]{TassaSW2011}%
  \BibitemOpen
  \bibfield  {author} {\bibinfo {author} {\bibfnamefont {C.}~\bibnamefont
  {Tassa}}, \bibinfo {author} {\bibfnamefont {S.~Y.}\ \bibnamefont {Shaw}}, \
  and\ \bibinfo {author} {\bibfnamefont {R.}~\bibnamefont {Weissleder}},\
  }\bibfield  {title} {\enquote {\bibinfo {title} {Dextran-coated iron oxide
  nanoparticles: a versatile platform for targeted molecular imaging, molecular
  diagnostics, and therapy},}\ }\href@noop {} {\bibfield  {journal} {\bibinfo
  {journal} {Accounts of Chemical Research}\ }\textbf {\bibinfo {volume}
  {44}},\ \bibinfo {pages} {842--852} (\bibinfo {year} {2011})}\BibitemShut
  {NoStop}%
\bibitem [{\citenamefont {Harisinghani}\ \emph {et~al.}(2003)\citenamefont
  {Harisinghani}, \citenamefont {Barentsz}, \citenamefont {Hahn}, \citenamefont
  {Deserno}, \citenamefont {Tabatabaei}, \citenamefont {{Hulsbergen van de
  Kaa}}, \citenamefont {{de la Rosette}},\ and\ \citenamefont
  {Weissleder}}]{HarisinghaniBHDTHvdKdlRW2003}%
  \BibitemOpen
  \bibfield  {author} {\bibinfo {author} {\bibfnamefont {M.~G.}\ \bibnamefont
  {Harisinghani}}, \bibinfo {author} {\bibfnamefont {J.}~\bibnamefont
  {Barentsz}}, \bibinfo {author} {\bibfnamefont {P.~F.}\ \bibnamefont {Hahn}},
  \bibinfo {author} {\bibfnamefont {W.~M.}\ \bibnamefont {Deserno}}, \bibinfo
  {author} {\bibfnamefont {S.}~\bibnamefont {Tabatabaei}}, \bibinfo {author}
  {\bibfnamefont {C.}~\bibnamefont {{Hulsbergen van de Kaa}}}, \bibinfo
  {author} {\bibfnamefont {J.}~\bibnamefont {{de la Rosette}}}, \ and\ \bibinfo
  {author} {\bibfnamefont {R.}~\bibnamefont {Weissleder}},\ }\bibfield  {title}
  {\enquote {\bibinfo {title} {Noninvasive detection of clinically occult
  lymph-node metastases in prostate cancer},}\ }\href@noop {} {\bibfield
  {journal} {\bibinfo  {journal} {New England Journal of Medicine}\ }\textbf
  {\bibinfo {volume} {348}},\ \bibinfo {pages} {2491--2499} (\bibinfo {year}
  {2003})}\BibitemShut {NoStop}%
\bibitem [{\citenamefont {Liu}\ \emph {et~al.}(2009)\citenamefont {Liu} \emph
  {et~al.}}]{LiuSDZLGXGLZ2009}%
  \BibitemOpen
  \bibfield  {author} {\bibinfo {author} {\bibfnamefont {J.}~\bibnamefont
  {Liu}} \emph {et~al.},\ }\bibfield  {title} {\enquote {\bibinfo {title}
  {Highly water-dispersible biocompatible magnetite particles with low
  cytotoxicity stabilized by citrate groups},}\ }\href@noop {} {\bibfield
  {journal} {\bibinfo  {journal} {Angewandte Chemie International Edition}\
  }\textbf {\bibinfo {volume} {48}},\ \bibinfo {pages} {5875--5879} (\bibinfo
  {year} {2009})}\BibitemShut {NoStop}%
\bibitem [{\citenamefont {Li}\ \emph {et~al.}(2017{\natexlab{b}})\citenamefont
  {Li}, \citenamefont {Kartikowati}, \citenamefont {Horie}, \citenamefont
  {Ogi}, \citenamefont {Iwaki},\ and\ \citenamefont {Okuyama}}]{LiKHOIO2017}%
  \BibitemOpen
  \bibfield  {author} {\bibinfo {author} {\bibfnamefont {Q.}~\bibnamefont
  {Li}}, \bibinfo {author} {\bibfnamefont {C.~W.}\ \bibnamefont {Kartikowati}},
  \bibinfo {author} {\bibfnamefont {S.}~\bibnamefont {Horie}}, \bibinfo
  {author} {\bibfnamefont {T.}~\bibnamefont {Ogi}}, \bibinfo {author}
  {\bibfnamefont {T.}~\bibnamefont {Iwaki}}, \ and\ \bibinfo {author}
  {\bibfnamefont {K.}~\bibnamefont {Okuyama}},\ }\bibfield  {title} {\enquote
  {\bibinfo {title} {Correlation between particle size/domain structure and
  magnetic properties of highly crystalline {F}e$_3${O}$_4$ nanoparticles},}\
  }\href@noop {} {\bibfield  {journal} {\bibinfo  {journal} {Scientific
  Reports}\ }\textbf {\bibinfo {volume} {7}},\ \bibinfo {pages} {9894}
  (\bibinfo {year} {2017}{\natexlab{b}})}\BibitemShut {NoStop}%
\bibitem [{\citenamefont {Anthony}\ \emph {et~al.}(2001)\citenamefont
  {Anthony}, \citenamefont {Bideaux}, \citenamefont {Bladh},\ and\
  \citenamefont {Nichols}}]{AnthonyBBN2001}%
  \BibitemOpen
  \bibfield  {author} {\bibinfo {author} {\bibfnamefont {J.~W.}\ \bibnamefont
  {Anthony}}, \bibinfo {author} {\bibfnamefont {R.~A.}\ \bibnamefont
  {Bideaux}}, \bibinfo {author} {\bibfnamefont {K.~W.}\ \bibnamefont {Bladh}},
  \ and\ \bibinfo {author} {\bibfnamefont {M.~C.}\ \bibnamefont {Nichols}},\
  }\href@noop {} {\emph {\bibinfo {title} {Handbook of Mineralogy}}}\ (\bibinfo
   {publisher} {Mineral Data Publishing},\ \bibinfo {address} {Chantilly,
  Virginia, USA},\ \bibinfo {year} {2001})\BibitemShut {NoStop}%
\bibitem [{\citenamefont {Lu}\ \emph {et~al.}(2017)\citenamefont {Lu},
  \citenamefont {Aimetti}, \citenamefont {Langer},\ and\ \citenamefont
  {Gu}}]{LuALG2017}%
  \BibitemOpen
  \bibfield  {author} {\bibinfo {author} {\bibfnamefont {Y.}~\bibnamefont
  {Lu}}, \bibinfo {author} {\bibfnamefont {A.~A.}\ \bibnamefont {Aimetti}},
  \bibinfo {author} {\bibfnamefont {R.}~\bibnamefont {Langer}}, \ and\ \bibinfo
  {author} {\bibfnamefont {Z.}~\bibnamefont {Gu}},\ }\bibfield  {title}
  {\enquote {\bibinfo {title} {Bioresponsive materials},}\ }\href@noop {}
  {\bibfield  {journal} {\bibinfo  {journal} {Nature Reviews Materials}\
  }\textbf {\bibinfo {volume} {2}},\ \bibinfo {pages} {1--17} (\bibinfo {year}
  {2017})}\BibitemShut {NoStop}%
\bibitem [{\citenamefont {Thomenius}(1996)}]{Thomenius1996}%
  \BibitemOpen
  \bibfield  {author} {\bibinfo {author} {\bibfnamefont {K.~E.}\ \bibnamefont
  {Thomenius}},\ }\bibfield  {title} {\enquote {\bibinfo {title} {Evolution of
  ultrasound beamformers},}\ }in\ \href@noop {} {\emph {\bibinfo {booktitle}
  {Proceedings of the IEEE Ultrasonics Symposium}}},\ Vol.~\bibinfo {volume}
  {2}\ (\bibinfo {organization} {IEEE},\ \bibinfo {address} {San Antonio,
  Texas, USA},\ \bibinfo {year} {1996})\ pp.\ \bibinfo {pages}
  {1615--1622}\BibitemShut {NoStop}%
\bibitem [{\citenamefont {Kennedy}(2005)}]{Kennedy2005}%
  \BibitemOpen
  \bibfield  {author} {\bibinfo {author} {\bibfnamefont {J.~E.}\ \bibnamefont
  {Kennedy}},\ }\bibfield  {title} {\enquote {\bibinfo {title} {High-intensity
  focused ultrasound in the treatment of solid tumours},}\ }\href@noop {}
  {\bibfield  {journal} {\bibinfo  {journal} {Nature Reviews Cancer}\ }\textbf
  {\bibinfo {volume} {5}},\ \bibinfo {pages} {321--327} (\bibinfo {year}
  {2005})}\BibitemShut {NoStop}%
\bibitem [{\citenamefont {Mahoney}\ \emph {et~al.}(2001)\citenamefont
  {Mahoney}, \citenamefont {Fjield}, \citenamefont {McDannold}, \citenamefont
  {Clement},\ and\ \citenamefont {Hynynen}}]{MahoneyFMcDCH2001}%
  \BibitemOpen
  \bibfield  {author} {\bibinfo {author} {\bibfnamefont {K.}~\bibnamefont
  {Mahoney}}, \bibinfo {author} {\bibfnamefont {T.}~\bibnamefont {Fjield}},
  \bibinfo {author} {\bibfnamefont {N.}~\bibnamefont {McDannold}}, \bibinfo
  {author} {\bibfnamefont {G.}~\bibnamefont {Clement}}, \ and\ \bibinfo
  {author} {\bibfnamefont {K.}~\bibnamefont {Hynynen}},\ }\bibfield  {title}
  {\enquote {\bibinfo {title} {Comparison of modelled and observed in vivo
  temperature elevations induced by focused ultrasound: implications for
  treatment planning},}\ }\href@noop {} {\bibfield  {journal} {\bibinfo
  {journal} {Physics in Medicine \& Biology}\ }\textbf {\bibinfo {volume}
  {47}},\ \bibinfo {pages} {1785–1798} (\bibinfo {year} {2001})}\BibitemShut
  {NoStop}%
\bibitem [{\citenamefont {B{\"o}rner}(2011)}]{Boerner2011}%
  \BibitemOpen
  \bibfield  {author} {\bibinfo {author} {\bibfnamefont {F.}~\bibnamefont
  {B{\"o}rner}},\ }\href@noop {} {\emph {\bibinfo {title} {Elektromagnetische
  {F}elder an {A}nlagen, {M}aschinen und {G}er{\"a}ten}}},\ Vol.~\bibinfo
  {volume} {5}\ (\bibinfo  {publisher} {Deutsche Gesetzliche
  Unfallversicherung},\ \bibinfo {address} {Berlin},\ \bibinfo {year}
  {2011})\BibitemShut {NoStop}%
\bibitem [{\citenamefont {Plimpton}(1995)}]{Plimpton1995}%
  \BibitemOpen
  \bibfield  {author} {\bibinfo {author} {\bibfnamefont {S.}~\bibnamefont
  {Plimpton}},\ }\bibfield  {title} {\enquote {\bibinfo {title} {Fast parallel
  algorithms for short-range molecular dynamics},}\ }\href@noop {} {\bibfield
  {journal} {\bibinfo  {journal} {Journal of Computational Physics}\ }\textbf
  {\bibinfo {volume} {117}},\ \bibinfo {pages} {1--19} (\bibinfo {year}
  {1995})}\BibitemShut {NoStop}%
\bibitem [{\citenamefont {K{\'e}sm{\'a}rky}\ \emph {et~al.}(2008)\citenamefont
  {K{\'e}sm{\'a}rky}, \citenamefont {Kenyeres}, \citenamefont {R{\'a}bai},\
  and\ \citenamefont {T{\'o}th}}]{KesmarkyKRT2008}%
  \BibitemOpen
  \bibfield  {author} {\bibinfo {author} {\bibfnamefont {G.}~\bibnamefont
  {K{\'e}sm{\'a}rky}}, \bibinfo {author} {\bibfnamefont {P.}~\bibnamefont
  {Kenyeres}}, \bibinfo {author} {\bibfnamefont {M.}~\bibnamefont {R{\'a}bai}},
  \ and\ \bibinfo {author} {\bibfnamefont {K.}~\bibnamefont {T{\'o}th}},\
  }\bibfield  {title} {\enquote {\bibinfo {title} {Plasma viscosity: a
  forgotten variable},}\ }\href@noop {} {\bibfield  {journal} {\bibinfo
  {journal} {Clinical Hemorheology and Microcirculation}\ }\textbf {\bibinfo
  {volume} {39}},\ \bibinfo {pages} {243} (\bibinfo {year} {2008})}\BibitemShut
  {NoStop}%
\bibitem [{\citenamefont {Weeks}\ \emph {et~al.}(1971)\citenamefont {Weeks},
  \citenamefont {Chandler},\ and\ \citenamefont {Andersen}}]{WeeksCA1971}%
  \BibitemOpen
  \bibfield  {author} {\bibinfo {author} {\bibfnamefont {J.~D.}\ \bibnamefont
  {Weeks}}, \bibinfo {author} {\bibfnamefont {D.}~\bibnamefont {Chandler}}, \
  and\ \bibinfo {author} {\bibfnamefont {H.~C.}\ \bibnamefont {Andersen}},\
  }\bibfield  {title} {\enquote {\bibinfo {title} {Role of repulsive forces in
  determining the equilibrium structure of simple liquids},}\ }\href@noop {}
  {\bibfield  {journal} {\bibinfo  {journal} {Journal of Chemical Physics}\
  }\textbf {\bibinfo {volume} {54}},\ \bibinfo {pages} {5237--5247} (\bibinfo
  {year} {1971})}\BibitemShut {NoStop}%
\bibitem [{\citenamefont {Ahmed}\ \emph {et~al.}(2017)\citenamefont {Ahmed},
  \citenamefont {Baasch}, \citenamefont {Blondel}, \citenamefont {L{\"a}ubli},
  \citenamefont {Dual},\ and\ \citenamefont {Nelson}}]{AhmedBBLDN2017}%
  \BibitemOpen
  \bibfield  {author} {\bibinfo {author} {\bibfnamefont {D.}~\bibnamefont
  {Ahmed}}, \bibinfo {author} {\bibfnamefont {T.}~\bibnamefont {Baasch}},
  \bibinfo {author} {\bibfnamefont {N.}~\bibnamefont {Blondel}}, \bibinfo
  {author} {\bibfnamefont {N.}~\bibnamefont {L{\"a}ubli}}, \bibinfo {author}
  {\bibfnamefont {J.}~\bibnamefont {Dual}}, \ and\ \bibinfo {author}
  {\bibfnamefont {B.~J.}\ \bibnamefont {Nelson}},\ }\bibfield  {title}
  {\enquote {\bibinfo {title} {Neutrophil-inspired propulsion in a combined
  acoustic and magnetic field},}\ }\href@noop {} {\bibfield  {journal}
  {\bibinfo  {journal} {Nature Communications}\ }\textbf {\bibinfo {volume}
  {8}},\ \bibinfo {pages} {770} (\bibinfo {year} {2017})}\BibitemShut {NoStop}%
\end{thebibliography}%

\end{document}